\documentclass[a4paper,11pt]{article}
\pdfoutput=1
\usepackage{jcappub}
\usepackage[T1]{fontenc}
\usepackage[compat=1.1.0]{tikz-feynman}
\usepackage{tikzsymbols}
\usepackage{upgreek}
\usepackage{natbib}
\usepackage{appendix}
\usepackage{mathtools,slashed}
\usepackage{subcaption}
\usepackage{tikz}
\usepackage{color, colortbl}
\usepackage{bigfoot}
\usepackage{soul}
\usepackage{orcidlink}
\DeclareNewFootnote[para]{math}[fnsymbol]
\MakeSortedPerPage{footnotemath}
\definecolor{deepmagenta}{rgb}{0.8, 0.0, 0.8}
\definecolor{mediumtealblue}{rgb}{0.0, 0.33, 0.71}
\definecolor{warmblack}{rgb}{0.0, 0.26, 0.26}
\definecolor{bostonuniversityred}{rgb}{0.8, 0.0, 0.0}
\definecolor{junglegreen}{rgb}{0.16, 0.67, 0.53}
\definecolor{lightcornflowerblue}{rgb}{0.6, 0.81, 0.93}
\definecolor{mypink1}{rgb}{0.858, 0.188, 0.478}
\definecolor{mypink2}{RGB}{219, 48, 122}
\definecolor{mypink3}{cmyk}{0, 0.7808, 0.4429, 0.1412}
\definecolor{mygray}{gray}{0.2}
\definecolor{ForestGreen}{RGB}{34,139,34}
\usepackage[font=small]{caption}
\usepackage[font={small},labelfont={bf}]{caption}
\usepackage{url}
\definecolor{MyDarkBlue}{rgb}{0.1, 0.1, 0.8}
\definecolor{SBlue}{rgb}{0.2, 0.4, 0.7} 
\definecolor{MyLightBlue}{rgb}{0.22,0.51,0.9}
\definecolor{MyGreen}{rgb}{0.0, 0.5, 0.0}
\definecolor{BrickRed}{rgb}{0.8, 0.25, 0.33}
\usepackage{hyperref}
\hypersetup{colorlinks, citecolor=BrickRed,linkcolor=MyDarkBlue, urlcolor=MyGreen}
\title{Two-component Dark Matter and low scale Thermal Leptogenesis}
\author[a]{Subhaditya Bhattacharya\orcidlink{0000-0002-8841-603X}, }
\author[a,b]{Devabrat Mahanta\orcidlink{0000-0003-4330-2052}, }
\author[a]{Niloy Mondal\orcidlink{0009-0006-5837-9772}, }
\author[a]{Dipankar Pradhan\orcidlink{0000-0002-2450-6677} }
\affiliation[a]{Department of Physics, Indian Institute of Technology Guwahati,\\North Guwahati, Assam-781039, India.}
\affiliation[b]{Department of Physics, Pragjyotish College,  Guwahati, Assam-781009, India.}
\emailAdd{subhab@iitg.ac.in}
\emailAdd{devabrat@pragjyotishcollege.ac.in}
\emailAdd{niloy18@iitg.ac.in}
\emailAdd{d.pradhan@iitg.ac.in}
\abstract{The observable cosmos exhibits sizable baryon asymmetry, small active neutrino masses, and the presence of dark matter (DM). To address these phenomena 
together, we propose a two component DM scenario in an extension of Scotogenic model, imposing $\mathbb{Z}_2 \otimes \mathbb{Z}_2^{\prime}$ symmetry.  
 The electroweak sphaleron process converts the $\rm Y_{B-L}^{}$ yield, generated through the Leptogenesis mechanism, into the baryon asymmetry ($\rm Y_{\Delta B}^{}$) at \textcolor{black}{ $\rm T_{\rm sph}\sim 130$ GeV}, the sphalerons decoupling temperature. In this framework, the CP asymmetry as well as the radiative 
 neutrino mass generation explicitly involve the two DM particles, thus establishing a correlation between the baryon asymmetry, DM and observed active neutrino masses.
We study in details the allowed parameter space available after considering all the constraints from the three phenomena as well as from the collider search limits, and outline the region which could potentially be tested in future DM detection experiments through direct or indirect detection searches, lepton flavor-violating decays, etc.}
\begin{document} 
\makeatletter
\gdef\@fpheader{}
\makeatother
\maketitle
\flushbottom
\section{Introduction}
Particle physics is at a crucial juncture where there are at least three major issues that prompt us to probe physics beyond the Standard Model (SM); 
baryon asymmetry of the Universe (BAU), tiny but non-zero neutrino masses and the existence of a dark matter (DM). 
But with no experimental evidences of New Physics (NP) so far, the efforts are directed to several possibilities. Addressing all these issues together in an 
extension of SM has been done extensively in the literature, and ours is another effort to that direction. 
Why and how this is different and interesting will be elaborated as we go along. 
   
The presence of baryon asymmetry in the Universe (BAU), or matter domination over antimatter, has been a long-standing problem in astroparticle physics. 
The baryon asymmetry is quantified in terms of the baryon-to-photon ratio, and the current measured value is \cite{Planck:2018vyg},
\begin{eqnarray}
\eta_{B}^{} & = & \dfrac{n_{b}-n_{\bar{b}}}{n_{\gamma}}= \left(6.16 \pm 0.16 \right)\times 10^{-10}\,.
\label{eq:bau}
\end{eqnarray}
Here, $n_{b}$, $n_{\bar{b}}$, and $n_{\gamma}$ are the number densities of baryons, anti-baryons, and photons in the present Universe, respectively. The conditions to dynamically generate a baryon asymmetry out of a baryon symmetric Universe were first proposed by Sakharov \cite{Sakharov:1967dj}. Although achieving these conditions within the SM is possible, but never adequate. Baryogenesis through Leptogenesis, first proposed in \cite{Fukugita:1986hr}, is a promising and viable idea that requires physics beyond the 
SM (BSM). When the temperature of the Universe drops below the \textcolor{black}{$ T_{\rm sph}\sim 130$ GeV} \cite{DOnofrio:2014rug}, sphalerons 
start to decouple from the thermal bath and the asymmetry conversion ceases; for some excellent reviews, see \cite{Buchmuller:2004tu, Buchmuller:2004nz, Davidson:2008bu, Pilaftsis:2009pk}. While Leptogenesis is generally a high-scale phenomenon and lies beyond the reach of current and near-future experiments, there are ways of achieving low-scale Leptogenesis, see for example, \cite{Pilaftsis:2003gt, Hugle:2018qbw, Alanne:2018brf, PhysRevD.102.055009, Borah:2020ivi}. Our work also addresses such a possibility. 

The presence of a non-luminous dark matter (DM), as hinted from several astrophysical and cosmological experiments, has also been a puzzle since long. Anisotropies in CMBR 
suggests that DM is present in huge amount ($\sim$27\%), while the energy content of the visible Universe is tiny ($\sim$5\%) !! The abundance of DM is quantified in terms of the 
DM number density to the total critical energy density of the Universe, $\Omega_{\rm DM}=\rho_{\rm DM}/\rho_{\rm crit}$, and the recent measured value is \cite{Planck:2018vyg}
\begin{equation}
\rm\Omega_{DM}h^{2}=0.1200\pm 0.0012.
\end{equation}
Here $h$ is the reduced Hubble parameter $\rm H/100$ $\rm km~s^{-1} Mpc^{-1}$ with $\rm H = 67.4\pm 0.5$ $\rm km~s^{-1} Mpc^{-1}$ being the current Hubble constant. What constitutes DM is not known. No particle from the SM can describe the properties of a DM. Weakly interacting massive particle (WIMP) \cite{PhysRevD.42.3310, Jungman:1995df}, feebly interacting massive particle (FIMP) \cite{Hall:2009bx} provide some elegant explanations for producing DM in the early Universe. Although one neutral massive particle can explain the observed DM characteristics, the DM can comprise of multiple fundamental particles.  This is especially motivating when looking at the large variety of fundamental particles in the visible sector that comprises only about $5\%$ of the Universe's energy density. The interplay of DM-DM interactions in such cases provide interesting phenomenology, see,  \cite{Profumo:2009tb,Drozd:2011aa,Cao:2007fy,Feldman:2010wy,Heeck:2012bz,Tomozawa:2008zq,Malekjani:2007pv,Daikoku:2011mq,Biswas:2013nn,Aoki:2013gzs,Gu:2013iy,Bhattacharya:2013hva}.

It is intriguing to note further that $\Omega_{\rm DM}h^{2} \simeq 5\Omega_{\rm B}h^{2}$. While it may be a numerical coincidence, and the production of DM 
and BAU may stem from different NP, it is often interesting to look for a common origin.
In the Scotogenic model \cite{Ma:2006km}, baryon asymmetry can be generated using the leptogenesis mechanism at the TeV scale,  
left-handed neutrinos acquire small Majorana masses via a one-loop radiative seesaw mechanism, while the neutral component of the 
inert doublet can be a viable DM candidate. However, the lepton asymmetry generation remains decoupled from the DM production, 
as both the scenarios are effective at different scales. There are mainly two ways to connect BAU and DM, known as co-genesis scenarios. In one case, 
the DM is assumed asymmetric, similar to the visible sector, and the asymmetry in both visible and dark sectors is generated from the out-of-equilibrium decay 
of a heavy particle \cite{Kaplan:2009ag, Davoudiasl:2012uw, Zurek:2013wia, Barman:2021ost, Cui:2020dly, Davoudiasl:2010am, Falkowski:2011xh, Petraki:2013wwa}. The other class of the co-genesis scenario generates a baryon asymmetry from the annihilation of DM \cite{Cui:2011ab, Bernal:2012gv, Kumar:2013uca, Racker:2014uga, Dasgupta:2016odo, Borah:2018uci, Dasgupta:2019lha}.

In this work, we propose a two-component DM model where both DMs are directly responsible for the baryon asymmetry production in the Universe while 
being consistent with the neutrino oscillation data. To the best of our knowledge, this is attempted for the first time. 
Although there are efforts where DM particle generates the CP asymmetry by entering into the one-loop diagrams \cite{Hugle:2018qbw, Croon:2022gwq}, or co-genesis with 
two component DM \cite{Borah:2019epq}, but the DM parameter space mostly remains independent of the leptogenesis. In the proposed model, the same DM coupling 
that generates diagrams for the CP asymmetry also determines the DM relic density and thus are correlated.
The particle content of our model is an extension of the minimal scotogenic model that includes two inert doublets, two right-handed neutrinos (RHNs), and a real scalar.
The low-scale leptogenesis requires the introduction of two new interaction vertices, along with the minimal scotogenic vertex, which doesn't affect 
neutrino mass generation, giving rise to RHN mass relaxation. These new vertices can contain DMs after ensuring the particles' stability by appropriate
$\mathbb{Z}_2\otimes\mathbb{Z}_2^{\prime}$ symmetry assignments. A similar scenario was studied in \cite{LeDall:2014too, Alanne:2018brf}, 
where the CP asymmetry is generated by introducing a real scalar particle (not a DM) in type-I seesaw model. Additionally, few more studies \cite{Mahanta:2019gfe, Abe:2021mfy, Ghosh:2024mpz} known as $N_2$ leptogenesis, assumes 
the decay of heavy RHN ($m_{N_2}>m_{N_1}$) relevant for asymmetry generation, but without direct involvement of DM. 
However, in our framework, the asymmetry generated by the tree and 1-loop decays of the lightest RHN, contain two DMs, which are the lightest stable particles 
under each discrete symmetries $\mathbb{Z}_2\otimes\mathbb{Z}_2^{\prime}$. By this choice, we were also able to prohibit the lepton number violating (LNV) 
decays involving the SM Higgs and only allow the symmetry singlet terms.
For simplicity, we assume that all the model parameters are real, except for the Yukawa couplings related to the tree-level LNV decay to explain the observed BAU.
In this way, the asymmetry generation directly involves the masses of two DMs and the couplings associated with interaction with DMs, which are also responsible 
for the DM analysis. Therefore, the BAU satisfied points are constrained by the DM bounds. 
Apart, the radiative diagrams for neutrino mass generation also have a strong connection with the DM, thus connecting all the 
three phenomena together.

There is also a phenomenological advantage for the presence of two DMs in this model. In an inert doublet model, the DM is tightly constrained and can't produce the correct relic density unless the DM mass is around the Higgs mass or approximately $\rm 600~GeV$. Introducing an additional DM component shares the remaining relic density and enhances the allowed parameter space via DM-DM conversion.

The paper is organized as follows. In Section\,.~\ref{sec:model}, we provide a general discussion of the model and its motivations. In Section\,.~\ref{sec:constrnt}, we explore the collider and lepton flavor constraints on the model parameters. The thermal leptogenesis and dark matter phenomenology, including DM relic density and direct (indirect) detection prospects, are discussed in detail in Sections\,.~\ref{sec:thermal-lepto} and \ref{sec:dm}, respectively. Finally, we summarize and conclude in Section\,.~\ref{sec:summary}. Several appendices provide the details of the relevant calculations. 
\section{The Model and motivations}
\label{sec:model}

The minimal scotogenic model \cite{Ma:2006km}, proposed to generate neutrino mass at one loop level, successfully addresses the BAU and DM in the universe. The one-loop neutrino mass generation relaxes the constraints coming from the neutrino masses to Leptogenesis. The well-known Davidson Ibarra (DI) bound lowers down to around $m^{}_{N_1^{}}\sim 10$ TeV \cite{Hugle:2018qbw, Bhattacharya:2023kws, Ghosh:2024mpz} from $m^{}_{N_1^{}}\sim10^{10}$ GeV in the type-I seesaw model \cite{Davidson:2002qv}. However, with two right-handed neutrinos, the scale of Leptogenesis again pushes to $m^{}_{N_2^{}}\gtrsim 10^{10}$ GeV \cite{Hugle:2018qbw, Mahanta:2019gfe}. Depending on the lightest $\mathbb{Z}_{2}$ odd state, the lightest neutral component of the inert doublet or the lightest right-handed neutrino can be a DM candidate in the scotogenic model. In the case of the scalar DM scenario, three RHNs participate in Leptogenesis. In such a case leptogenesis is possible at the TeV scale independent of DM parameters. In the case of the fermionic DM scenario, two RHNs participate in leptogenesis and the observed asymmetry can be generated around $m_{N_2^{}}^{} \sim 10^{10}$ GeV. With two RHNs responsible for CP asymmetry generation, the lightest RHN $N_{1}$, being DM only enters the leptogenesis scenario by scattering washouts \cite{Mahanta:2019gfe}. The scattering washouts are insignificant compared to the huge washout from the inverse decay of $N_{2}$. Therefore the leptogenesis parameter space remains independent of DM parameter space. In this work, we propose an extension of the scotogenic model where a DM particle directly generates the CP asymmetry in LNV decays. A two-component DM scenario naturally emerges in the model.    

We extend the SM with two copies of RHNs ($N_{1,2}$), two scalar inert doublets ($\eta_{1,2}$), and a real singlet scalar ($\phi$).
All the SM fields are even under an imposed $\mathbb{Z}_{2}\otimes \mathbb{Z}_{2}^{\prime}$ symmetry while the new fields transform non-trivially as shown in tab\,.~\ref{tab:tab1}.
The neutrinos get mass by the scotogenic mechanism \cite{Ma:2006km} as shown in fig\,.~\ref{fig:NuMass} of appendix\,.~\ref{sec:Numass}. 
We keep the right-handed neutrinos $N_{i}$ to be heavier than the doublet scalars $\eta_{i}$ and the singlet scalar $\phi$.
A net lepton asymmetry can be generated from the out-of-equilibrium decay of the lightest RHN ($N_{1}\longrightarrow l_{\alpha} \eta_{1}$ ). Due to the imposed $\mathbb{Z}_{2}\otimes \mathbb{Z}_{2}^{\prime}$ symmetry the singlet scalar enters the vertex correction diagram generating the required CP asymmetry.
The additional $\mathbb{Z}_{2}$ symmetry naturally leads to a two-component DM scenario.
Under these circumstances, the lightest of the inert doublets could be a viable WIMP due to its gauge portal interaction. At the same time, the singlet scalar $\phi$ can be any DM, depending on the strength of its Higgs portal interaction and its interaction rate with the WIMP.
Finally, we could get a WIMP-WIMP \cite{Bhattacharya:2016ysw}, WIMP-FIMP \cite{Bhattacharya:2021rwh}, WIMP-pFIMP \cite{Bhattacharya:2022dco,Bhattacharya:2022vxm,Bhattacharya:2024nla}, etc. However, this article focuses solely on the WIMP-WIMP scenario.
\begin{table}[htb!]
\centering
\begin{tabular}{|c|ccc|}
\hline
\rowcolor{teal!30}Fields &$ SU(3)_c \otimes SU(2)_L\otimes U(1)_Y$ & $\mathbb{Z}_{2}$ & $\mathbb{Z}_{2}^{\prime}$\\\hline
\rowcolor{cyan!15}$N_{1}$ & (1,1,0) & -1 & ~1 \\
\rowcolor{cyan!10}$N_{2}$ & (1,1,0) & ~1 & -1 \\
\rowcolor{green!15}$\eta_1$ & (1,2,1/2) & -1 & ~1\\
\rowcolor{green!10}$\eta_2$ & (1,2,1/2) & ~1 & -1\\
\rowcolor{lime!20}$\phi$ & (1,1,0) &-1 & -1\\\hline
\end{tabular}
\caption{Particle content of the extended model and their corresponding charges.}
\label{tab:tab1}
\end{table}
\begin{align}
\mathcal{L}=\mathcal{L}_{\rm SM}+\dfrac{1}{2}|\partial_{\mu}\phi|^2+\sum_{k=1,2}\left(i\overline{N}_k\slashed{\partial}N_k+|\mathcal{D}_{\mu}\eta_k|^2\right)+\mathcal{L}_{Yuk}-\mathcal{V}\,.
\end{align}
\noindent Here the Lagrangian $\mathcal{L}_{Yuk}$ contain the Yukawa couplings as well as the Majorana mass term for the RHNs and is given by,
\begin{eqnarray}
\mathcal{L}_{Yuk} & = &- \sum_{i=1,2}\left(h_{ii\alpha}\overline{L}_{\alpha}\Tilde{\eta}_iN_{i} + h.c\right)-\dfrac{1}{2}\sum_{i=1,2}m_{N_i}\overline{N_{i}^{c}}N_{i}.
\end{eqnarray}
\noindent The scalar potential $\mathcal{V}$ is given by
\begin{align}
\nonumber\mathcal{V} =& \mu_{\eta_i}^2\left(\eta^{\dagger}_{i}\eta_i\right)+\dfrac{1}{4}\lambda_{ij}\left|\eta^{\dagger}_{i}\eta_j+\eta^{\dagger}_{j}\eta_i\right|^{2} + \dfrac{1}{2}\mu_{\phi}^{2}\phi^2+\dfrac{1}{4!}\lambda_{\phi}\phi^{4} + \lambda_{iiH}\left(H^{\dagger}\eta_i\right) \left(\eta^{\dagger}_{i} H\right) \\\nonumber +&  \lambda^{\prime}_{iiH}\left(\eta^{\dagger}_{i}\eta_i\right) \left( H^{\dagger}H \right)+\dfrac{1}{2}\left[\lambda^{\prime\prime}_{iiH}\left( \eta^{\dagger}_iH \right)\left( \eta^{\dagger}_iH \right)+h.c. \right] + \dfrac{1}{2}\lambda_{\phi H} \phi^2\left(H^{\dagger}H \right) \\ +& \dfrac{1}{2}\lambda_{ii\phi}\left(  \eta^{\dagger}_i\eta_i\right)\phi^{2}+\dfrac{1}{2}\sum_{i\neq j}\left(y_{ij\phi} \overline{N_{i}^{c}}N_{j} \phi+\mu_{ij\phi}\eta_{i}^{\dagger}\eta_j\phi +h.c.\right)\,.
\label{eq:poten}
\end{align}
where, we assumed real $y_{ij\phi}$ and inert doublets ($\eta_js$) has the following forms,
\begin{equation}
\eta_j=
\dfrac{1}{\sqrt{2}}\begin{pmatrix}
\sqrt{2}\eta^{+}_j \\ \eta_{R_j}^0 + i \eta_{I_j}^0
\end{pmatrix}\,.
\end{equation}
After the electroweak symmetry breaking, the masses of the physical scalars would be, 
\begin{gather}
m_{\phi}^2=\mu_{\phi}^2+\dfrac{1}{2}\lambda_{\phi H}v^2\,,\\
m_{\eta^+_i}^2\footnotemarkmath=\mu_{\eta_i}^2+\lambda^{\prime}_{iiH}v^2\,,\\
m_{\eta^0_{R_i}}^2=m_{\eta^+_i}^2+\dfrac{1}{2}\left(\lambda_{iiH}+\lambda^{\prime\prime}_{iiH}\right)v^2\,,\\
m_{\eta^0_{I_i}}^2=m_{\eta^+_i}^2+\dfrac{1}{2}\left(\lambda_{iiH}-\lambda^{\prime\prime}_{iiH}\right)v^2\,.
\end{gather}\footnotetextmath{$\mu_{\eta_i}^2>0\implies m_{\eta^+_i}^2>\lambda_{iiH}^{\prime}v^2$ required for the stability of inert doublets.}
\section{Constraints on model parameters}\label{sec:constrnt}
\subsection{Collider constraints}
LHC and LEP experiments put constraints on the decay of SM gauge bosons \cite{ATLAS:2023ynf, CMS:2022ett, L3:1993jix}. One such constraint on the model appears from $Z\longrightarrow \eta_{R}\eta_{I}$  requiring $m_{Z}^{}< m_{\eta_{R}}+m_{\eta_{I}}$.
\begin{eqnarray}
\begin{split}
\rm\Gamma_{Z\to invisible}
<  \begin{cases}
\rm 506\pm 13~MeV~~~(ATLAS)\,,\\
\rm523\pm 16~MeV~~~( CMS)\,,\\
\rm498\pm 17 MeV~~~(L3)\,.\end{cases}
\end{split}\label{Z_invisible_decay}
\end{eqnarray}
In the parameter space $m_{h}/2> m_{\eta_{R}},m_{\eta_{I}}$ the constraints from the Higgs invisible decay is applied.
The observed (expected) upper limit on the invisible branching fraction of the Higgs boson corresponds to an integrated luminosity of
$138 ~fb^{-1}$, at $95\%$ confidence level \cite{ATLAS:2023tkt,CMS:2023sdw} with total decay width of 125.1 GeV Higgs is $3.2^{+2.8}_{-2.2}\rm~ MeV$ \cite{ParticleDataGroup:2022pth},
\begin{eqnarray}\begin{split}
\rm \mathcal{B}_{h\to\rm{invisible}}< \begin{cases}0.107~(0.077)~~~\rm( ATLAS)\,,\\0.15~(0.08)\rm~~~( CMS)\,.\end{cases}
\end{split}\label{higgs_invisible_decay}
\end{eqnarray}
This constrain our model parameters $(\lambda_{iiH}+\lambda_{iiH}^{\prime}\pm \lambda_{iiH}^{''}),\lambda_{\phi H}$ to be less than around $10^{-3}$ in the regime $m_{\eta_{R}},m_{\eta_{I}},m_{\phi}< m_{h}/2$.\textcolor{black}{~Additionally LEP-II precision data ruled out some parameter space \cite{Lundstrom:2008ai, Swiezewska:2012eh, Arhrib:2012ia, Belanger:2015kga} of inert doublet model (IDM), which can be summarised as follows: $m_{\eta_j^{\pm}}<70$ GeV (based on the LEP-II chargino search from $e^+e^-\rightarrow\tilde{H}^+\tilde{H}^-$), $\text{max}\big(m_{\eta_{R}},m_{\eta_{I}}\big)< 110$ GeV (based on the LEP-II neutralino search), and $m_{\eta_{R}}-m_{\eta_{I}}>8$ GeV~\cite{Abouabid:2020eik}. Also, recent analyses by ATLAS using $\sqrt{s}=13$ TeV LHC data have set limits on the masses of charginos and neutralinos of the minimal supersymmetric model (MSSM) by examining events with a pair of boosted hadronically decaying bosons and missing transverse momentum in pp collisions~\cite{ATLAS:2021yqv}. A similar collider signature can be produced from the IDM. Ref.~\cite{Banerjee:2021oxc} conducted a full recast study for IDM using~\cite{ATLAS:2021yqv}. Based on this recast study authors of Ref.~\cite{Ghosh:2021noq} find out that the Higgs portal DM in the IDM scenario remains viable with hierarchical heavy scalars $m_{\eta_{R}},~m_{\eta_i^{\pm}}\gtrsim$ $123$ GeV. Using the CMS Run II data for invisible Higgs decays via vector boson fusion~\cite{CMS:2018yfx}, a recast study on IDM parameter space was done by Ref.~\cite{Dercks:2018wch,Belanger:2021lwd}, where the rigions with $(\lambda_{iiH}+\lambda_{iiH}^{'}+\lambda_{iiH}^{''})>1(3)$ for $m_{\eta^{0}_{R_{i}}}\sim65(70)$ GeV and regions with  $(\lambda_{iiH}+\lambda_{iiH}^{'}+\lambda_{iiH}^{''})>10$ for  $m_{\eta^{0}_{R_{i}}}\sim85-100$ GeV are excluded. However, the LEP limit remains stringent compared to the LHC limits. }

\subsection{Lepton Flavor constraints}
The MEG II experiment, which searches for the decay $\mu^+ \to e^+ \gamma$, reports that no excess of events over the expected background has been observed, yielding an upper limit on the branching ratio \cite{MEG:2016leq, MEGII:2023ltw},
\begin{gather}
{\mathcal {B}} (\mu ^+ \rightarrow {\textrm{e}}^+ \gamma ) < 3.1 \times 10^{-13}~(90\%\rm~ C.L.)\,,
\end{gather}
\begin{figure}[htb!]
\centering
\begin{tikzpicture}[baseline={(current bounding box.center)},style={scale=0.9, transform shape}]
\begin{feynman}
\vertex (a){\(\color{black}{\ell_{\alpha}}\)};
\vertex [right=1.5cm of a] (b);
\vertex [above right=1cm and 1cm of b] (b1);
\vertex [below right=1cm and 1cm of b] (c1);
\vertex [above right=0.5cm and 1cm of b1] (b2);
\vertex [below right=0.5cm and 1cm of c1] (c2);
\diagram*{
(a)-- [line width=0.25mm,fermion, arrow size=1.2pt, style=bostonuniversityred,ultra thick] (b),
(b) -- [line width=0.25mm, plain,arrow size=1.2pt, style=black,ultra thick, edge label'={\(\color{black}{N_j}\)}] (c1),
(b) -- [line width=0.25mm, charged scalar, arrow size=1.2pt, style=black, ultra thick, edge label={\(\color{black}{\eta_j^- }\)}] (b1),
(b1) -- [line width=0.25mm, charged scalar,arrow size=1.2pt, style=black, ultra thick, edge label={\(\color{black}{\eta^-_j }\)}] (c1),
(b1) -- [line width=0.25mm, boson,arrow size=1.2pt, style=mediumtealblue,ultra thick] (b2),
(c1) -- [line width=0.25mm, fermion,  arrow size=1.2pt, style=mediumtealblue,ultra thick] (c2)};
\vertex[above right=0.5cm and 1cm of b1]{\(\color{black}{\gamma }\)};
\vertex[below right=0.5cm and 1cm of c1]{\(\color{black}{\ell_{\beta}}\)};
\node at (b)[circle,fill,style=black,inner sep=1pt]{};
\node at (b1)[circle,fill,style=black,inner sep=1pt]{};
\node at (c1)[circle,fill,style=black,inner sep=1pt]{};
\end{feynman}
\end{tikzpicture}
\begin{tikzpicture}[baseline={(current bounding box.center)},style={scale=0.9, transform shape}]
\begin{feynman}
\vertex (a){\(\color{black}{\ell_{\alpha}}\)};
\vertex[right=1.5cm of a] (a1);
\vertex[right=1cm of a1] (a2);
\vertex[right=1cm of a2] (b);
\vertex[above right=1.5cm and 1cm of b] (b1);
\vertex[below right=1.5cm and 1cm of b] (b2);
\diagram*{
(a)-- [line width=0.25mm,fermion, arrow size=1.2pt, style=bostonuniversityred,ultra thick] (a1),
(a1) -- [line width=0.25mm, charged scalar, half left,arrow size=1.2pt, style=black,ultra thick, edge label={\(\color{black}{\eta_j^-}\)}] (a2),
(a1) -- [line width=0.25mm, plain, half right, arrow size=1.2pt, style=black, ultra thick, edge label'={\(\color{black}{N_j}\)}] (a2),
(a2) -- [line width=0.25mm, fermion, arrow size=1.2pt, style=gray, ultra thick, , edge label={\(\color{black}{\ell_{\beta}}\)}] (b),
(b) -- [line width=0.25mm, fermion,arrow size=1.2pt, style=mediumtealblue, ultra thick] (b2),
(b) -- [line width=0.25mm, boson,arrow size=1.2pt, style=mediumtealblue,ultra thick] (b1)};
\vertex[above right=1.5cm and 1cm of b] {\(\color{black}{\gamma }\)};
\vertex[below right=1.5cm and 1cm of b] {\(\color{black}{\ell_{\beta}}\)};
\node at (b)[circle,fill,style=black,inner sep=1pt]{};
\node at (a1)[circle,fill,style=black,inner sep=1pt]{};
\node at (a2)[circle,fill,style=black,inner sep=1pt]{};
\end{feynman}
\end{tikzpicture}
\begin{tikzpicture}[baseline={(current bounding box.center)},style={scale=0.9, transform shape}]
\begin{feynman}
\vertex (a){\(\color{black}{\ell_{\alpha}}\)};
\vertex [right=1.5cm of a] (b);
\vertex [above right=1.5cm and 3cm of b] (b4);
\vertex [below right=0.5cm and 1cm of b] (b1);
\vertex [below right=1cm and 2cm of b] (b2);
\vertex [below right=0.5cm and 1cm of b2] (b3);
\diagram*{
(a)-- [line width=0.25mm,fermion, arrow size=1.2pt, style=bostonuniversityred,ultra thick] (b),
(b) -- [line width=0.25mm, fermion, arrow size=1.2pt, style=gray,ultra thick, edge label'={\(\color{black}{\ell_{\alpha}}\)}] (b1),
(b1) -- [line width=0.25mm, plain, half left, arrow size=1.2pt, style=black,ultra thick, edge label={\(\color{black}{N_j}\)}] (b2),
(b1) -- [line width=0.25mm, charged scalar, half right,arrow size=1.2pt, style=black,ultra thick, edge label'={\(\color{black}{\eta_j^-}\)}] (b2),
(b2) -- [line width=0.25mm, fermion, arrow size=1.2pt, style=mediumtealblue, ultra thick] (b3),
(b) -- [line width=0.25mm, boson,arrow size=1.2pt, style=mediumtealblue, ultra thick] (b4)};
\vertex[above right=1.5cm and 3cm of b] {\(\color{black}{\gamma }\)};
\vertex[below right=0.5cm and 1cm of b2] {\(\color{black}{\ell_{\beta}}\)};
\node at (b)[circle,fill,style=black,inner sep=1pt]{};
\node at (b1)[circle,fill,style=black,inner sep=1pt]{};
\node at (b2)[circle,fill,style=black,inner sep=1pt]{};
\end{feynman}
\end{tikzpicture}
\caption{1-loop Feynman diagrams related to $\ell_{\alpha}\to\ell_{\beta}\gamma$.}
\label{feyn:lfv}
\end{figure}
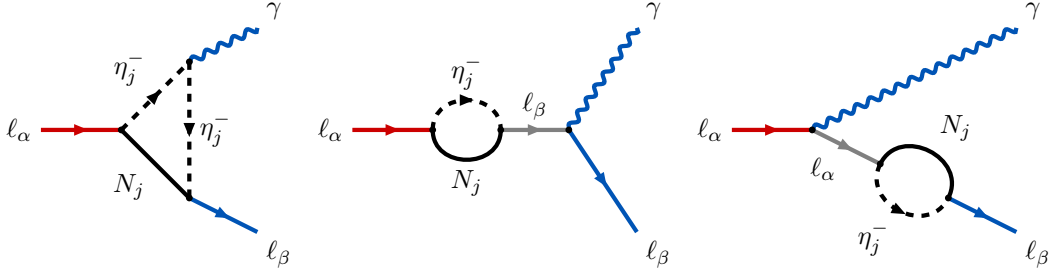
The branching fraction corresponds to $\ell_{\alpha}\to\ell_{\beta}\gamma$ is given by \cite{Toma:2013zsa,Vicente:2014wga,Valiente:2023blo,Hundi:2022iva},
\begin{align}
\mathcal{B}(\ell_{\alpha}\to\ell_{\beta}\gamma)=\dfrac{3(4\pi)^3\alpha_{\rm em}}{4 G_F^2}|F_D|^2~\mathcal{\rm BR}(\ell_{\alpha}\to\ell_{\beta}\nu_{\alpha}\overline{\nu_{\beta}})\,,
\end{align}
where $\alpha_{\rm em}$ and $G_F$ are the electromagnetic fine structure and Fermi constant, respectively. For SM leptonic decay branching, $\ell_{\alpha}\to\ell_{\beta}\nu_{\alpha}\overline{\nu_{\beta}}$, see \cite{ParticleDataGroup:2024cfk}. $F_D$ is the dipole form factor, given by,
\begin{align}
F_D=\sum_{i=1}^2\dfrac{h^*_{ii\beta}h_{ii\alpha}}{2(4\pi)^2}\dfrac{1}{m_{\eta^+_i}^2}G(x_i)\,,
\end{align}
where, $x_i=\dfrac{m_{N_i^{}}^2}{m_{\eta^+_i}^2}$ and $G(x)=\dfrac{1-6x+3x^2+2x^3-6x^2~\log~x}{6(1-x)^4}$.

During our analysis, we consistently account for this limit. Since the required Yukawa couplings are very small ($|h_{ii\alpha}|\sim 10^{-5}$), we don't need to be concerned about this limit. However, other limits, such as $\mu \to eee$, would be more suppressed, as the same couplings contribute to these processes.
\section{Thermal Leptogensis analysis}\label{sec:thermal-lepto}
\begin{figure}[htb!]
\centering
\includegraphics[width=0.45\linewidth]{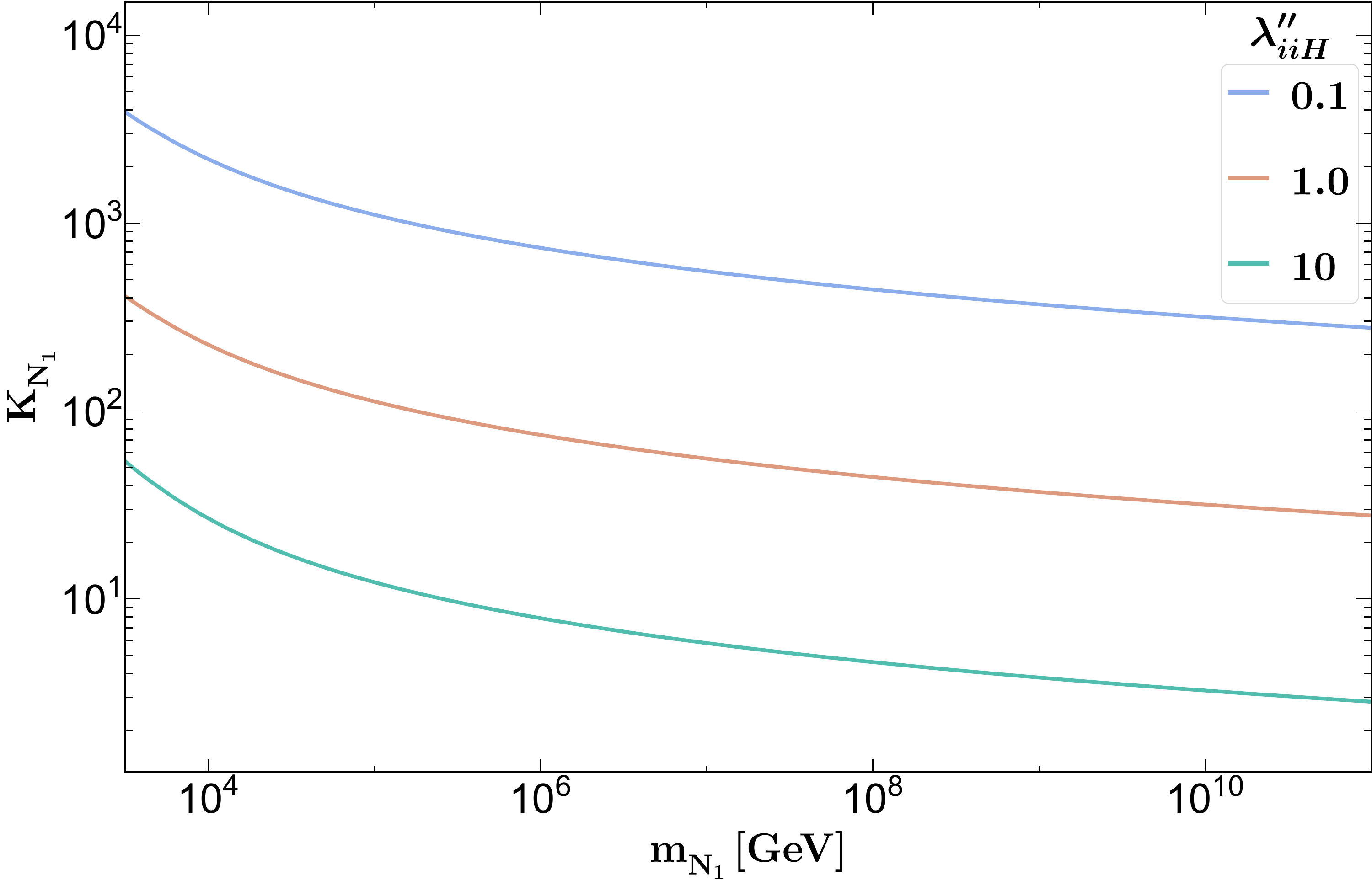}~~
\includegraphics[width=0.45\linewidth]{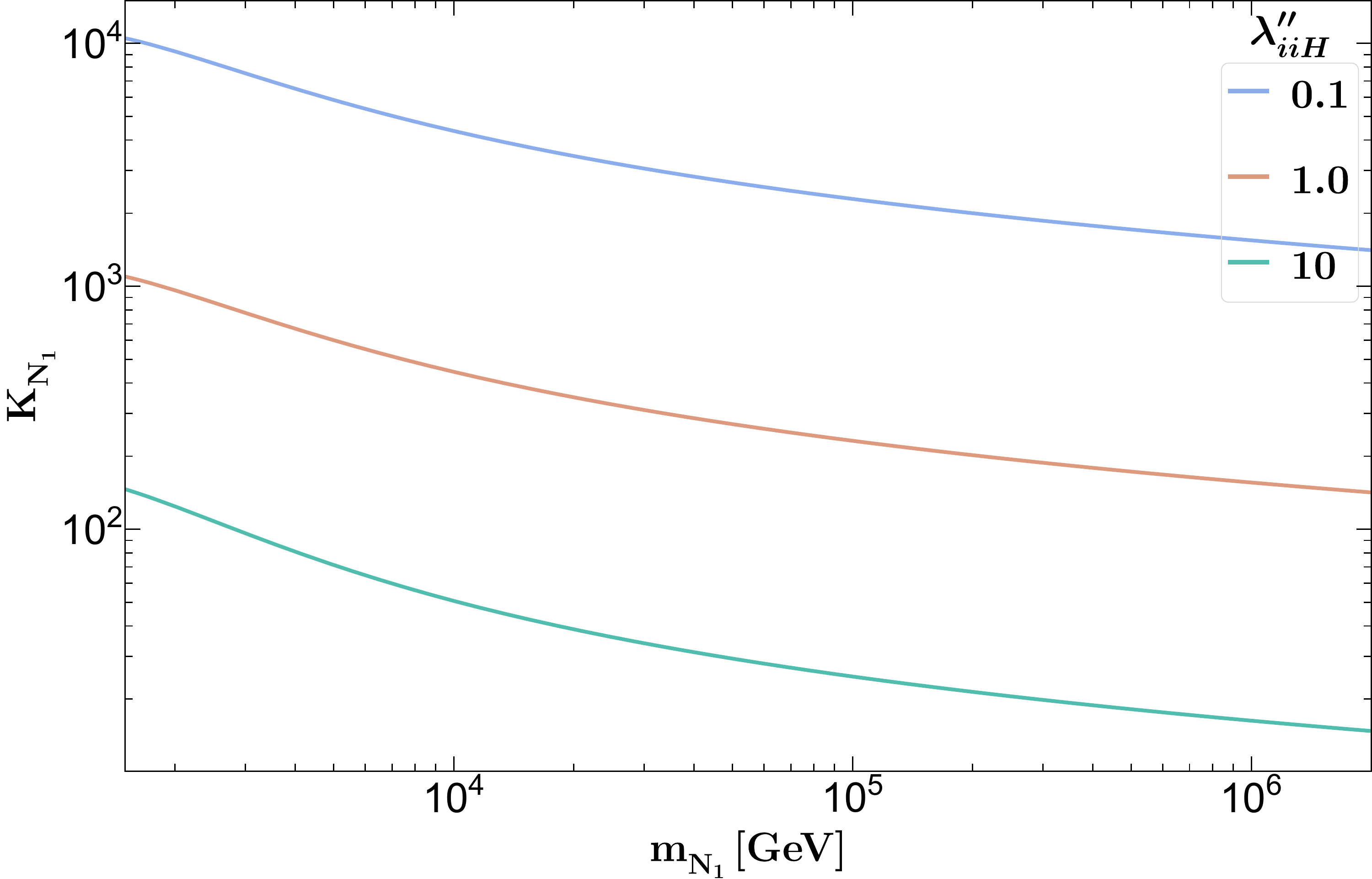}
\caption{We have used
$\{m_{N_1^{}}^{}>\mu_{\eta_2^{}}^{}+m_{\phi}^{},
~\lambda^{\prime}_{iiH}=0.01,
~m_{\eta_{I_1}^0}=0.5 ~{\rm TeV},~m_{\phi}^{}=0.4~{\rm TeV},
~m_{\eta_{I_2}^0}=m_{\eta_{I_1}^0}+m_{\phi}^{}+1{\rm ~GeV},
~\mu_{12\phi}=m_{\phi},~y_{12\phi}^{}=1,
~m_{N_2}^{}=3m_{N_1}^{},
~m_{\eta_i^+}=m_{\eta_{I_i}^0}+3{\rm ~GeV}$,
~$m_{\eta_{R_i}^0}=(m_{\eta_{I_i}^0}^2+\lambda^{\prime\prime}_{ii\rm H}v^2)^{1/2},
~a=0.1,~b=0.3\}$ to calculate the decay parameter, $K_{N_1^{}}^{}$.}
\label{fig:K1}
\end{figure}

In the minimal scotogenic model, a net lepton asymmetry can be generated from the decay of the lightest right-handed neutrino. The Yukawa coupling involved in Leptogenesis is subjected to satisfy the light neutrino data through the Casas-Ibarra (CI) parametrization \cite{Casas:2001sr, Alberico:2003kd}. The Yukawa couplings are determined by the scalar quartic coupling $\lambda_{iiH}^{''}$ and the active and right-handed neutrino masses \cite{Hugle:2018qbw}. While for three right-handed neutrinos, the Yukawa couplings of the lightest right-handed neutrino can be made small by fixing the lightest active neutrino mass, we don't have such a choice with two right-handed neutrinos \cite{Mahanta:2019gfe}. With two right-handed neutrinos, the Yukawa couplings of the lightest right-handed neutrino are always large, resulting in strong washouts of the lepton asymmetry. In fig\,.~\ref{fig:K1} we show the decay parameter $K_{N_1^{}}^{}=\Gamma_{N_1^{}}^{}/H(z=1)$ with mass of the lightest right handed neutrino ($m^{}_{N_1^{}}$). The left panel plot of fig\,.~\ref{fig:K1} show the decay parameter with two RHNs in the minimal scotogenic model is always greater than one, while the right panel corresponds to our model. This suggests that we are always in a strong washout region. Due to the strong washouts, the scale of Leptogenesis is pushed beyond $m^{}_{N_1^{}} \sim 10^{9}~\rm GeV$. This lower bound for a vanilla leptogenesis mechanism is known as the Davidson Ibarra (DI) bound \cite{Davidson:2002qv}. In the case of the minimal scotogenic model, the DI bound can be found in \cite{Hugle:2018qbw}. The singlet DM enters the one loop vertex correction diagram of $N_{1} \longrightarrow l_{\alpha} \eta_1^{}$ in this model. Although we still are in a strong washout region with only two RHNs (see right panel of fig\,.~\ref{fig:K1}), we now have a coupling in the leptogenesis loop free from the neutrino mass generation. One can sufficiently enhance the asymmetry parameter $\varepsilon_{N_{1}}^{}$ by fixing the $y_{ij\phi}^{}$. The DI bound for the minimal scotogenic model is no longer applicable in this case.  
\begin{figure}[htb!]
\centering
\includegraphics[width=0.475\linewidth]{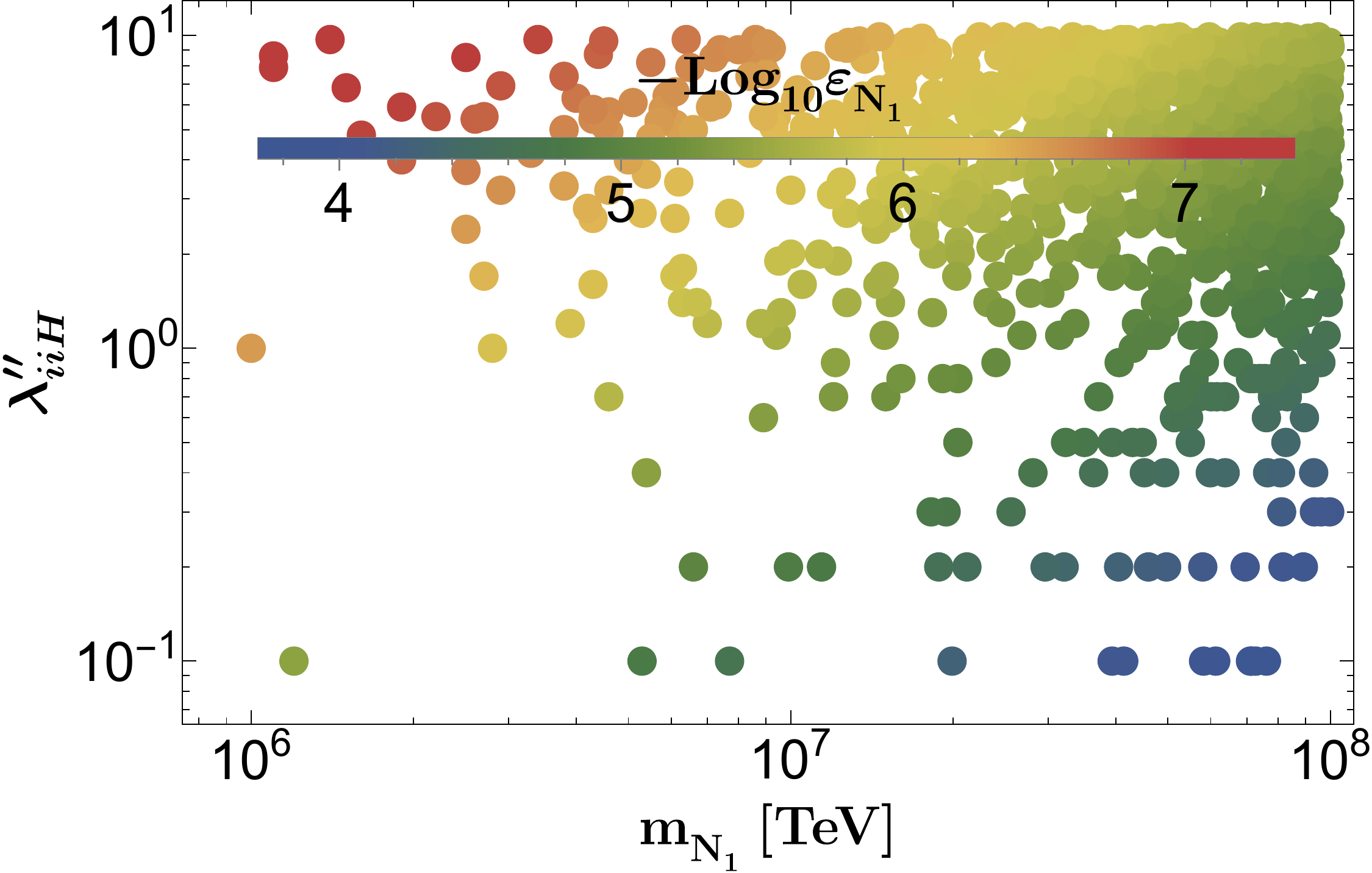}~~
\includegraphics[width=0.475\linewidth]{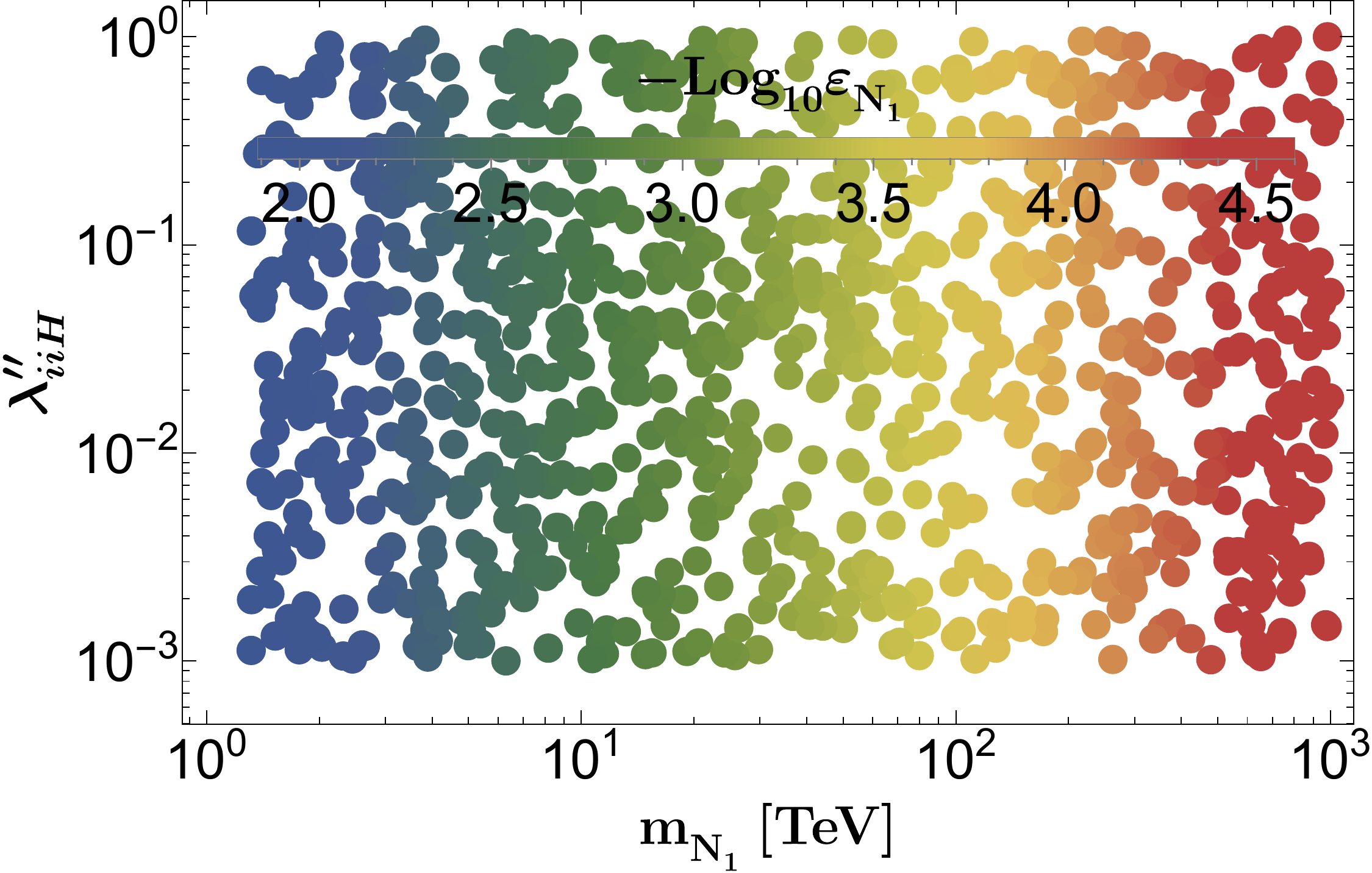}
\caption{Scan plot for the asymmetry parameter in $m_{N_1}^{}-\lambda_{iiH}^{\prime\prime}$ plane. For the left panel plot, the other important parameters are mentioned in the figure inset (for 2 RHN and one inert). The figure on the right corresponds to our model and uses the same parameters as those in fig\,.~\ref{fig:K1}.}
\label{fig:asymmetry}
\end{figure}

We calculate the CP asymmetry $(\varepsilon_{N_{1}}^{})$ parameter arising from the decay $N_{1} \longrightarrow l_{\alpha} \eta_1^{}$ in appendix\,.~\ref{app:asymmetry}.
In fig\,.~\ref{fig:asymmetry}, we show the variation of the asymmetry parameter through the dark rainbow color bar $\varepsilon^{}_{N_1}$ in $m_{N_{1}}^{}-\lambda^{\prime\prime}_{ii\rm H}$ plane. On the left panel plot of the fig\,.~\ref{fig:asymmetry}, asymmetry parameter $\varepsilon_{N_{1}}$ is plotted for the minimal scotogenic model while, on the right panel plot, it is shown for the extended model. In the left panel plot, it can be seen that there exists a correlation of the asymmetry parameter with the quartic coupling $\lambda^{\prime\prime}_{ii\rm H}$ as well as $m_{N_{1}}$. With the decrease in $\lambda_{ii\rm H}$ the Yukawa couplings $h_{11\alpha}$ increases increasing asymmetry parameter $\varepsilon_{N_{1}}^{}$. However, one can not decrease the $\lambda^{\prime\prime}_{ii\rm H}$ to arbitrarily small values, as it would increase the washouts leading to a strong washout of the generated asymmetry. With the increase in the masses of the RHNs, the corresponding Yukawa couplings increase, increasing the asymmetry parameter. On the right panel plot, it is seen that the correlation of the asymmetry parameter $\varepsilon_{N_{1}}^{}$ with $\lambda^{\prime\prime}_{ii\rm H}$ vanishes.  The dependence of $\varepsilon^{}_{N_{1}}$ on $\lambda_{iiH}^{''}$ is through the Yukawa coupling $h_{ii\alpha}$, by the Casas Ibarra parameterization given in appendix\,.~\ref{eq:CI}. Due to the presence of the Yukawa coupling $y_{12\phi}$ and $\mu_{12\phi}$ in the vertex correction diagram, the dependence of the asymmetry parameter on the Yukawa couplings $h_{11\alpha}$ vanishes~(see eq.~\ref{eq:epsilonvertex} and eq.~\ref{eq:epsilon}). 

In this model, we found that the asymmetry parameter $\varepsilon_{N_1}^{}$ is independent of the scalar couplings $\lambda_{11H}^{}$, unlike the minimal scotogenic model.
Required kinematics for Leptogenesis:
\begin{eqnarray}
m_{N_1}^{}>\mu_{\eta_2}^{}+m_{\phi}^{}\,,~\mu_{\eta_2}^{}>\mu_{\eta_1}^{}+m_{\phi}^{}\,,~m_{N_2}^{}>m_{N_1}^{}+m_{\phi}^{}\,.
\end{eqnarray}
The relevant coupled Boltzmann Equations for Leptogenesis is written as,
{\small
\begin{eqnarray}\begin{split}
\dfrac{dY^{}_{N_{1}}}{dz}  = &-D_{N_{1}}\left(Y_{N_{1}}-Y_{N_{1},0}^{}\right)-\dfrac{s}{H(z)z} \left(Y_{N_{1}}^{2}-Y_{N_{1},0}^{2}\right)\left[\langle \sigma v \rangle_{N_{1}N_{1}\longrightarrow \phi \phi}+\langle \sigma v \rangle_{N_{1}N_{1}\longrightarrow \eta_1 \eta_1^{\dagger}}+\langle \sigma v \rangle_{N_{1}N_{1}\longrightarrow \ell_{\alpha} \overline{\ell}_{\beta}}\right] \\
&-\dfrac{s}{H(z)z} \left(Y_{N_{1}}-Y_{N_{1},0}^{}\right)\left[Y_{\phi,0}^{}\langle \sigma v  \rangle_{N_{1}\phi \longrightarrow l_{\alpha} \eta_{2}^{\dagger}}+Y_{\eta_1,0}^{}\langle \sigma v \rangle_{N_{1}\eta_{1}\longrightarrow l_{\alpha} V_{\mu}}+Y_{N_2,0}^{}\langle \sigma v \rangle_{N_{1}N_2\longrightarrow \eta_1 \eta_2^{\dagger}}\right]\,,\\
\dfrac{dY_{B-L}^{}}{dz} = & -\varepsilon_{N_{1}}^{}D_{N_{1}}\left(Y_{N_{1}}-Y_{N_{1},0}^{}\right)-W_{\rm ID} Y_{B-L}-\dfrac{s}{H(z)z} Y_{B-L}\left[ 2\sum_{i=1,2}Y_{\eta_{i}^{},0}^{}  \langle \sigma v \rangle_{l_{\alpha} \eta^{\dagger}_{i}\longrightarrow \overline{l}_{\beta} \eta_{i}} \right.\\ &\left. +2 Y_{l,0}^{}\sum_{i=1,2}r^2_{\eta_{i}} \langle \sigma v  \rangle_{\eta_{i}^{\dagger}\eta_i^{\dagger} \longrightarrow l_{\alpha} l_{\beta}}+ Y_{l,0}^{}\sum_{i\neq j}r_{N_{i}} r_{\phi} \langle \sigma v  \rangle_{N_{i}\phi \longrightarrow l_{\alpha} \eta_{j}^{\dagger}} + Y_{l,0}^{}\sum_{i=1,2}r_{N_{i}}r_{\eta_{i}} \langle \sigma v \rangle_{\eta_{i}N_{i}\longrightarrow l_{\alpha} V_{\mu}} \right].
\label{eq:cbeq-lepto}
\end{split}\end{eqnarray}}
The Boltzmann equations are written in terms of the dimensionless variable $z=\rm m_{N_{1}}^{}/T$, $i^{th}$ particle co-moving number density $Y_{i}$ and equilibrium number density $Y_{i,0}^{}=Y_{i}^{\rm eq}$. Here $D_{N_{1}}^{}$ and $W_{\rm ID}^{}$ are the decay and inverse decay terms for $N_{1}$ defined as
\begin{eqnarray}
D_{N_{1}}^{}& = & K_{N_{1}}^{}z\dfrac{\kappa_{1}(z)}{\kappa_{2}(z)},\\
W_{\rm ID}^{}& = & \dfrac{1}{4}K_{N_{1}}^{}z^{3}\kappa_{1}(z)\,,
\end{eqnarray}
where $K_{N_{1}}^{}=\Gamma_{N_{1}}^{}/H(z=1)$ is known as the decay parameter and $\kappa_{i}$'s are the modified Bessel functions of second kind. $\langle \sigma v \rangle_{\rm AB\longrightarrow CD}$ represents the thermal averaged cross-section for a given process $\rm A+B\longrightarrow C+D$. Here $r_{j}^{}=Y_{j,0}^{}/Y_{l,0}^{}$.
The detailed derivation of neutrino mass and asymmetry parameters $(\varepsilon_{N_{1}}^{})$ is available in the appendix\,.~\ref{sec:Numass} and \ref{app:asymmetry}, respectively. The relevant Feynman diagrams corresponding to the leptogenesis scenarios are shown in fig\,.~\ref{fig:lepto-scattering}.
\begin{figure}[htb!]
\centering
\includegraphics[width=0.475\linewidth]{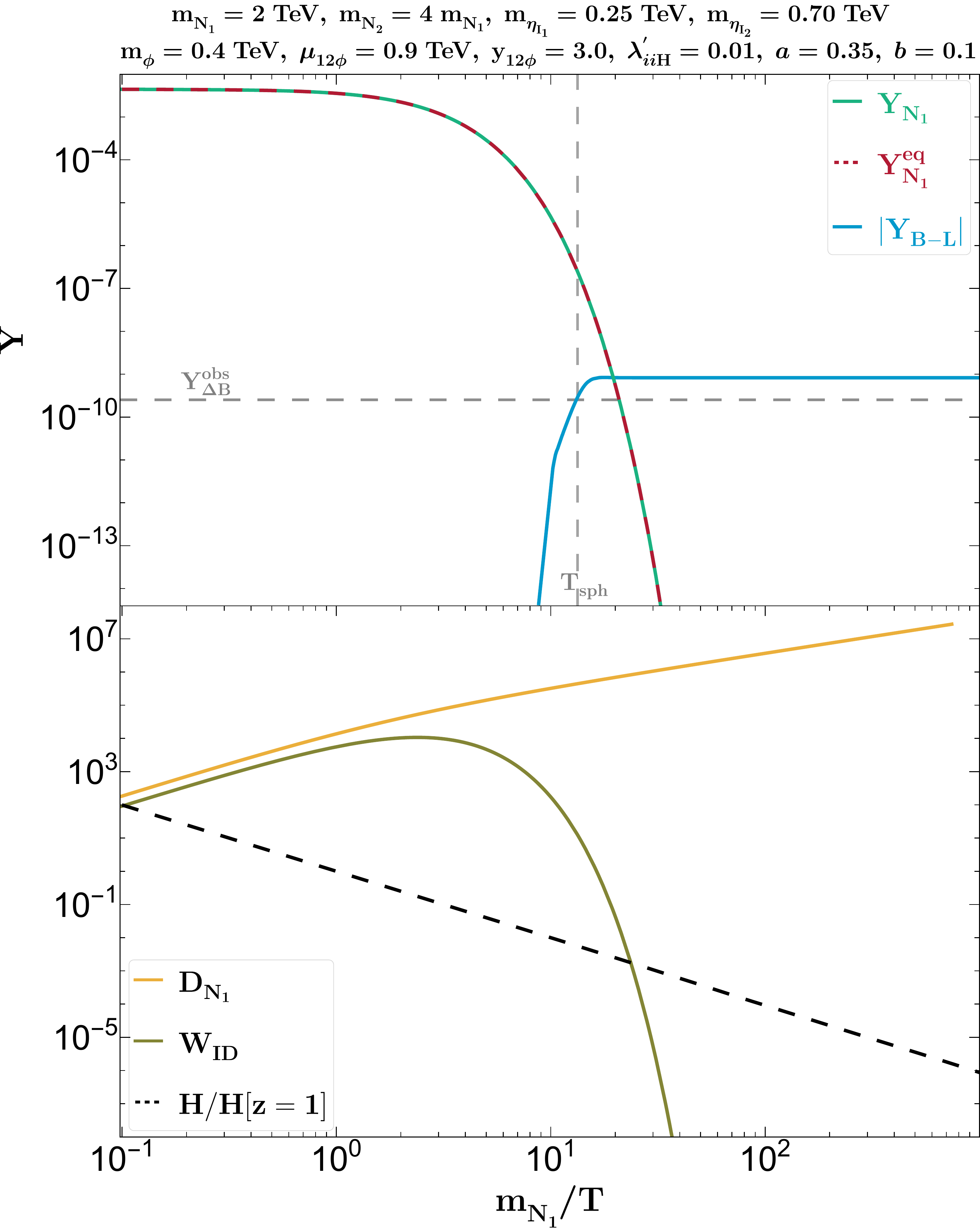}~~
\includegraphics[width=0.475\linewidth]{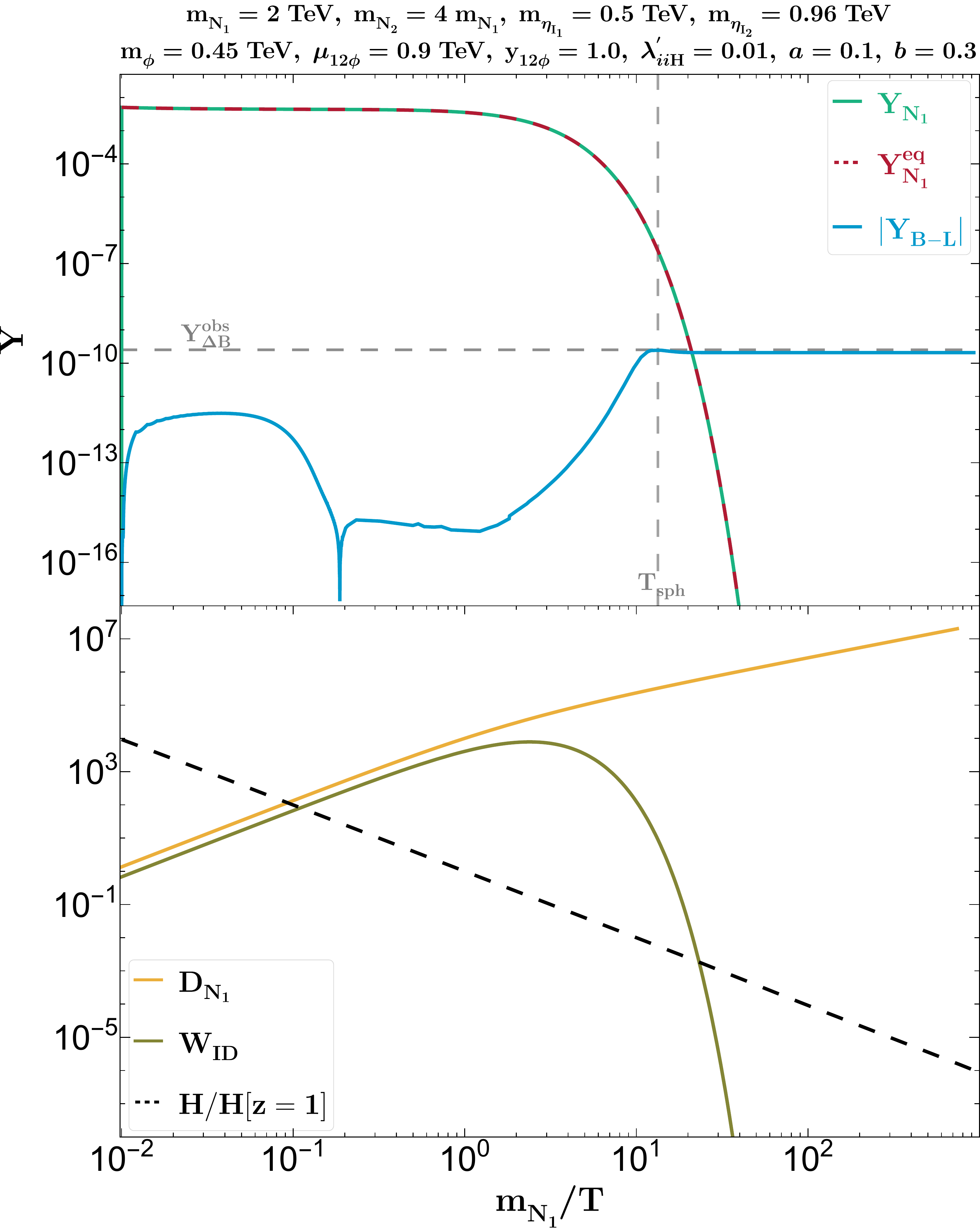}
\caption{The left figure corresponds to the initial condition $\rm  Y_{N_1} = Y_{N_1}^{eq}$, while for the right figure $\rm Y_{N_1^{}}^{} \sim 10^{-20}$. The horizontal and vertical gray dashed lines indicate $Y_{\Delta B}^{\rm obs}$ and $T_{\rm sph}^{}$, respectively. The other solid and dashed lines represent the variation of parameters with $z=\rm m_{N_1^{}}^{}/T$, as indicated in the figure's inset. The masses of the real and charged components of the inert doublet are defined as $m^{}_{\eta^{0}_{R_i^{}}} = m_{\eta_{I_i}^0}^{} + 2$ and $m_{\eta_i^+} = m^{}_{\eta^{0}_{I_i}} + 3$, respectively.}
\label{fig:lepto-yield}
\end{figure}

In fig\,.~\ref{fig:lepto-yield}, we have represented the solution of cBEQ (eq\,.~\ref{eq:cbeq-lepto}), yielding the $\rm Y_{B-L}^{}$ and $\rm Y_{N_1}^{}$ evaluation with $z$. In Leptogenesis, the observed baryon asymmetry is generated by the sphaleron process, where the sphaleron freeze-out occurs at $T_{\rm sph}$, transform $\rm Y_{B-L}$ asymmetry to the $\rm Y_{\Delta B}^{}$ asymmetry. We solve it for two benchmark points in the left and right panel of fig\,.~\ref{fig:lepto-yield}. The blue and green solid lines depict the evolution of $\rm |Y_{B-L}^{}|$ and $\rm Y_{N_1^{}}$, respectively, while the dark red and black dashed lines represent $\rm \rm Y_{N_1}^{eq}$ and ${\rm H/H}(z=1)$, respectively. The figures in the bottom panel illustrate how the decay and inverse decay rates vary with $z$. The intersection of $\rm D_{N_1}^{}$ and $\rm W_{ID}^{}$ with ${\rm H/H}(z=1)$, decide the asymmetry production and washout. At the first and second intersections, both are becoming in and out-equilibrium processes. Although the rates are equal at equilibrium, the production and washout are also influenced by $\rm Y^{}_{N_1}$ and $\rm Y_{B-L}^{}$, leading to fascinating dynamics. In the bottom panel of fig\,.~\ref{fig:lepto-yield}, we have not shown the scattering rate relevant for asymmetry evaluation, as they are less significant compared to the decay $(\rm D_{N_1^{}}^{})$ and inverse decay  $(\rm W_{ID}^{})$. However, some scattering processes significantly impact the number density of $N_{1}$ keeping it near its equilibrium abundance, reducing the asymmetry.

In fig\,.~\ref{fig:lepto-yield}, we show two plots representing two different scenarios: the left plot assumes that $\rm N_1^{}$ is initially in equilibrium, while the right plot considers a case where $\rm N_1^{}$ is not in equilibrium and starts from a very small initial abundance. In the left panel of fig\,.~\ref{fig:lepto-yield}, $\rm N_1^{}$ is initially in equilibrium so the decay ($N_{1}^{}\rightarrow l \eta_1^{}$) and the inverse decay ($l \eta_1^{}\rightarrow N_{1}^{}$) rates are almost close to each other at high temperature, $z\lesssim 9$. Because of this, at high temperatures, the produced asymmetry is washed out and the remaining asymmetry is negligible, $\rm |Y_{B-L}^{}|\lesssim 10^{-14}$. But, when the temperature falls below $m_{N_1}^{}$, the inverse decay rate would start to fall and finally go below the Hubble rate. During this time ($9<z<20$), the decay will become dominant and the produced asymmetry will survive due to the gradual decrease in strength of the inverse decay washout. The gradual increase in asymmetry will not continue indefinitely. It will stop when $\rm Y_{N_1^{}}^{eq}$ falls to a very small value 
 due to the decays and can no longer produce asymmetry at low temperatures ($z \gtrsim 20$) and it freezes in. A similar explanation applies to the right panel plots in fig\,.~\ref{fig:lepto-yield} when $ z\gtrsim 2$. However, below it $i.e\,.~z\lesssim 2$ the asymmetry is generated through the out-of-equilibrium decay of $N_1^{}$ and strongly washed out when it comes into thermal equilibrium. Once it does, it mimics the behavior seen in the left panel plot.

Since numerous parameters play a significant role in Leptogenesis, we are fixing some at specific values and expressing others in terms of $\{m^{}_{\eta_{I_1}^0},~m_{\phi}^{},~m_{N_1}^{},~y_{12\phi}^{},~\mu_{12\phi}^{}\}$:
\begin{eqnarray}
\begin{split}
m_{N_2}^{}=3m_{N_1}^{},~m_{\eta^{0}_{R_{i}}}=m_{\eta^{0}_{I_{i}}}^{}+2{\rm ~GeV},~m_{\eta^{+}_{i}}^{}=m_{\eta^{0}_{I_i}}^{}+3{\rm ~GeV},~a=1,~b=0.4\,,\qquad\\
m_{\phi}^{}<\mu_{12\phi}^{}<2m_{\phi}^{},~\lambda^{\prime}_{11H}=\lambda^{\prime}_{22H}=0.1,~m_{\eta^{0}_{I_2}}^{}=m_{\eta^{0}_{I_1}}^{}+m_{\phi}+1{\rm ~GeV},~m_{N_1}^{}> m_{\eta_{I_2}^0}^{}+m_{\phi}^{}\,.
\label{eq:lepto-const}
\end{split}
\end{eqnarray}
The choice of $a,~b$ is not unique, and the dependency is shown in fig\,.~\ref{fig:ab} for a sample benchmark.
\begin{figure}[htb!]
\centering
\includegraphics[width=0.6\linewidth]{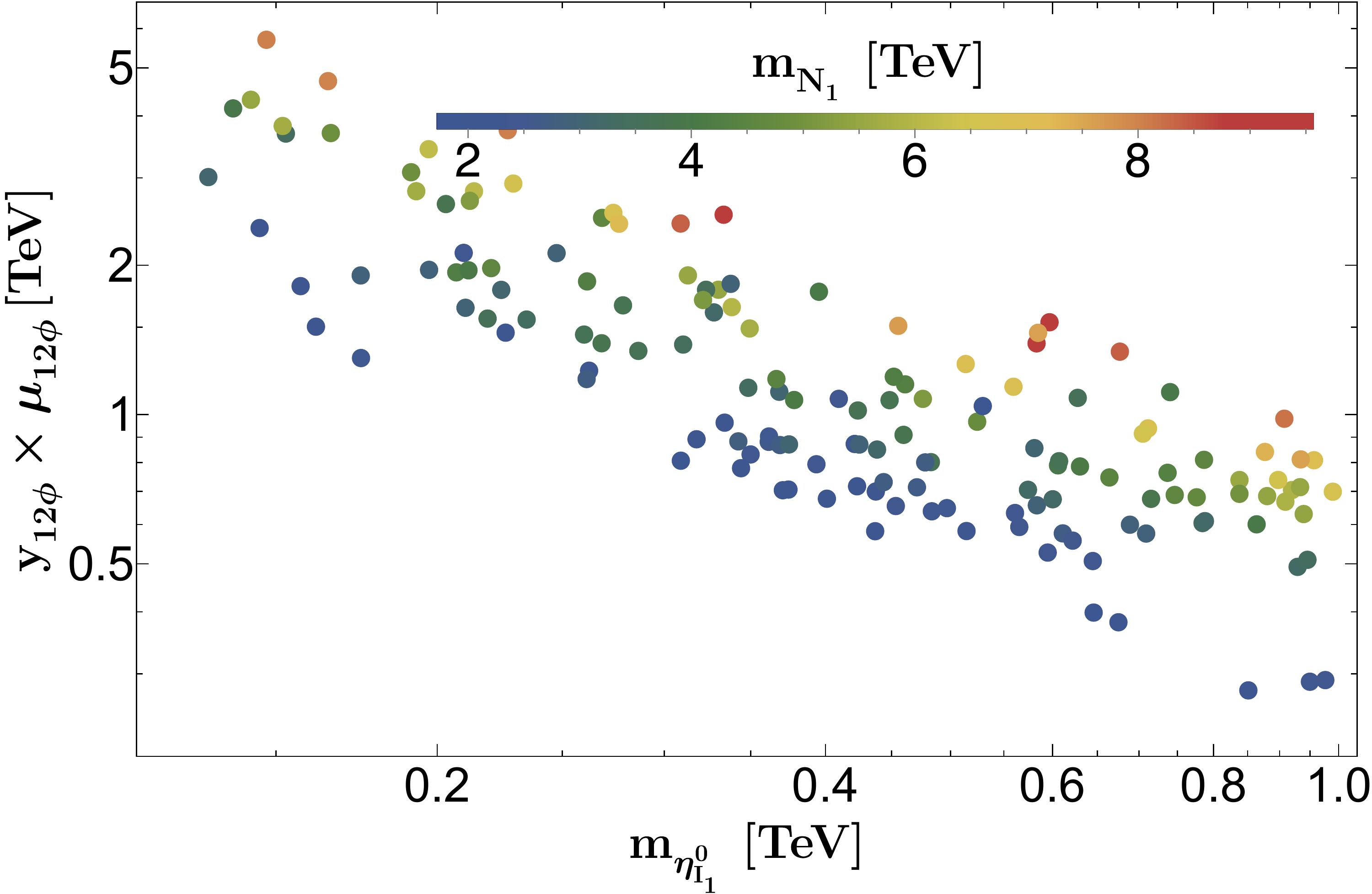}
\caption{In this figure, the dark rainbow-colored points represent the RHN mass $m_{N_1}^{}$. All of these points respect the observed baryon asymmetry ($Y_{\Delta B}^{\rm obs}\simeq 8.75^{+0.23}_{-0.23}\times 10^{-11}$) and account for the correct active neutrino masses.}
\label{fig:lepto-scan}
\end{figure}

In fig\,.~\ref{fig:lepto-scan} we show the points in $m_{\eta_{I_1}^{0}}^{}-y_{12\phi}^{}\mu_{12\phi}^{}$ plane that satisfies the neutrino mass and baryon asymmetry while respecting the current LEP constraints on the charged scalar. Since the free couplings $y_{12\phi}$ and $\mu_{12\phi}^{}$ both positively contribute to the asymmetry parameter $\varepsilon_{N_{1}}^{}$ we keep $y_{12\phi}^{}\mu_{12\phi}^{}$ in the y-axis.  The dimensionfull coupling $\mu_{12\phi}^{}$ is defined as a function of $m_{\phi}$ and is varied within the range, $1 < \mu^{}_{12\phi}/m_{\phi} < 2$. The different values of $m_{N_{1}}^{}$ are shown as a rainbow color bar. From fig\,.~\ref{fig:lepto-scan} it is seen that there exist a positive correlation between $m_{N_{1}}^{}$ and $y_{12\phi}^{}\mu_{12\phi}^{}$. In eq\,.~\ref{eq:asymmetry1}, it is clear that, the asymmetry parameter $\varepsilon^{}_{N_{1}}$ decreases with the increase in $m_{N_{1}}^{}$ and increases with the enhancement of $\eta_{1}$ $(\equiv m_{\eta_{I_1}^{0}}^{2}/m_{N_{1}}^{2})$. Therefore it is required to increase the free couplings $y_{12\phi}^{}\mu_{12\phi}^{}$ to generate sufficient asymmetry. Hence a larger value of $m_{N_{1}}^{}$ require larger $y_{12\phi}\mu_{12\phi}$ to satisfy the observed baryon asymmetry. Similarly, in fig\,.~\ref{fig:lepto-scan} it is seen that the points satisfying the observed asymmetry have a negative gradient with $m_{\eta_{I_1}^{0}}$. It is also due to the dependency of $\varepsilon_{N_1}^{}$ on $\eta_{1}^{}$. For a fixed $m^{}_{N_{1}}$, with the increase in $m_{\eta_{I_1}^{0}}^{}$ the asymmetry parameter $\varepsilon_{N_{1}}^{}$ increases. To compensate for this, we need smaller values of $y_{12\phi}^{}\mu_{12\phi}^{}$. 

\begin{table}[htb!]
\centering
\setlength{\tabcolsep}{4.5pt} 
\renewcommand{\arraystretch}{1.7} 
\begin{tabular}{|c |c c c c c |}\hline
\rowcolor{gray!25}{\rm BM}&$m_{\eta_{I_1}^0}^{}\left[\rm ~GeV\right]$&$m_{\phi}^{}\left[\rm ~GeV\right]$&$m_{N_1}^{}\left[\rm ~TeV\right]$&$\mu_{12\phi}^{}\left[\rm GeV\right]$&$y_{12\phi}^{}$\\\hline
\rowcolor{magenta!10}I & 232.37   & 656.80  & 6.388  & 1037.20  & 2.555 \\
\rowcolor{orange!10}II  & 786.91   & 124.31  & 3.670  & 222.43  & 2.603 \\
\rowcolor{lime!10}III & 355.11   & 365.58  & 2.205  & 384.67  & 2.608 \\
\rowcolor{green!10}IV & 340.88   & 467.62  & 5.932  & 598.88  & 2.688 \\
\rowcolor{cyan!10}V & 130.78   & 618.41  & 3.661  & 1220.52  & 3.001 \\\hline
\end{tabular}
\caption{The sample benchmark (BM) points are giving the observed baryon asymmetry after addressing the active neutrino masses, while other parameters are considered following the eq\,.~\ref{eq:lepto-const}.}
\label{tab:lepto-scan}
\end{table}
Before concluding this section, we have illustrated several points, in tab\,.~\ref{tab:lepto-scan}, that meet the constraints on active neutrino masses and the observed baryon asymmetry. However, the LFV constraints are not applicable in this context, as the asymmetry-respecting parameter $|h_{ii\alpha}|$ is effectively negligible.
\section{Dark Matter analysis}\label{sec:dm}
The lightest particle under a discrete symmetry is stable and becomes a DM. In tab\,.~\ref{tab:tab1}, we have define the particles charges under $\mathbb{Z}_2\otimes\mathbb{Z}_2^{\prime}$ symmetry. We have chosen the $\eta_{I_1}^0$ (CP-odd part of inert doublet $\eta_{1}^{}$) and $\phi$ (real scalar singlet) as our DMs by considering other parameter masses larger than these two. The required kinematics for the stabilisation of DMs ($\eta^0_{I_1},~\phi$):
\begin{eqnarray}
\begin{split}
m_{\eta^0_{I_2}}>m_{\eta^0_{I_1}}+m_{\phi}\,,~m_{N_2}^{}>m_{\eta^0_{I_2}}\,,
m_{\eta^0_{R_i}}>m_{\eta^0_{I_i}}\,,~m_{\eta^+_i}>m_{\eta^0_{I_i}}\,,~m_{N_1}^{}>m_{\eta^0_{I_1}}\,.
\end{split}
\label{eq:dm_stable}
\end{eqnarray}
The heavier dark sector particles contribute to relics through co-annihilation channels.
In this two-component real scalar DM scenario, both DMs interact with the visible sector via gauge and Higgs portal interactions.
In eq\,.~\ref{eq:poten}, we see that two inert doublets have gauge portal interactions, by which $\eta_{I_1}^0$ is in thermal equilibrium, and the lightest one acts as a viable thermal dark matter candidate.
In contrast, $\phi$ interacts with the visible sector only via the Higgs portal interactions. It's also coupled to the inert doublets through $\eta_1^{\dagger}\eta_2^{}\phi$ and $\eta_i^{\dagger}\eta_i^{}\phi^2$ terms.
Interestingly, the mass coupling associated with $\eta_1^{\dagger}\eta_2^{}\phi$ is also contributed in baryon asymmetry along with DM masses $m_{\eta_{I_1}^0}^{}$ and $m_{\phi}^{}$. 
However, Depending on the value of $\lambda_{\phi H}$, two scenarios arise: $(i)$ WIMP-WIMP, for $\lambda_{\phi H} \sim 0.1$, and $(ii)$ WIMP-pFIMP, for $\lambda_{\phi H} \sim 10^{-12}$, with dark matter masses in the $\rm GeV-TeV$ range. In this article, we exclusively focus on the WIMP-WIMP scenario; and we leave the alternate possibilities for future works. 

In a two-component DM scenario, the relic density of DM is calculated by solving the coupled Boltzmann Equation (cBEQ). For simplicity, we can neglect some of the processes that are ineffective in DM freezeout.
RHNs \( ({\rm N}_i) \) masses ($\sim \rm TeV$) are quite larger than DMs masses, necessary to account for both the correct baryon asymmetry and neutrino masses.
So, \( ({\rm N}_i) \)s remained out of thermal equilibrium and couldn't participate in DM freezeout, which occurs at \( \rm T_{\rm FO}\) is smaller than the sphaleron decoupling temperature, $ T_{\rm sph}$ where the $\rm Y_{B-L}$ is transferred into $\rm Y_{\Delta B}$. Finally, the coupled Boltzmann equations are written as,
\begin{eqnarray}
\begin{split}
\dfrac{dY_{\eta_{1}}}{dz} &\,=\, -\dfrac{s}{H\bigg(\dfrac{m_{\eta_{1}}}{m_{N_{1}}}z\bigg) \bigg(\dfrac{m_{\eta_{1}}}{m_{N_{1}}} \bigg)z} \bigg[ \big( Y_{\eta_{1}}^{2}-Y_{\eta_{1}^{},0}^{2} \big) \langle \sigma v \rangle^{\rm eff}_{\eta_{1}\eta_{1}\longrightarrow \rm SM~SM} + 
\left(Y_{\eta_{1}}^{2}-Y_{\phi}^2\dfrac{Y_{\eta_{1}^{},0}^{2}}{Y_{\phi,0}^{2}} \right) \langle \sigma v \rangle^{\rm eff}_{\eta_{1}\eta_{1}\longrightarrow \phi \phi}
\bigg]\,, \\
\dfrac{dY_{\phi}}{dz} &\,= \,-\dfrac{s}{H\bigg(\dfrac{m_{\phi}}{m_{N_{1}}}z\bigg)\bigg(\dfrac{m_{\phi}}{m_{N_{1}}} \bigg)z} \left[\left( Y_{\phi}^{2}- Y_{\phi,0}^{2} \right) \langle \sigma v \rangle_{\phi \phi\longrightarrow\rm SM~SM}+\left( Y_{\phi}^{2}-Y^2_{\eta_1} \dfrac{Y_{\phi,0}^{2}}{Y^2_{\eta_{1},0}} \right) \langle \sigma v \rangle^{\rm eff}_{ \phi \phi \longrightarrow\eta_{1}\eta_{1}}\right]\,.
\end{split}\end{eqnarray}
where $Y_{i,0}=Y_i^{\rm eq},~z={\rm m_{N_1}^{}/T},~\langle\sigma v\rangle^{\rm eff}$ denoted the effective annihilation cross section because of the co-annihilating particles, which eventually decay into the DM \cite{PhysRevD.43.3191, Edsjo:1997bg} and other parameters carried their usual meaning. We have solved the cBEQ using MicrOMEGAs-6.0 \cite{Alguero:2023zol}, and the results are presented in the subsequent sections. \textcolor{black}{ Due to the strong gauge annihilations, the inert doublet DM is under-abundant for $80 ~\rm GeV \lesssim m^{}_{_{DM}}\lesssim 500 ~\rm GeV $. This range could change depending on the mass difference between the charged and neutral CP-even scalar. In our model, in the presence of the second inert doublet, there will be additional co-annihilation making the DM even less abundant. However, it is possible to generate the correct relic in the presence of the other DM $(\phi)$ that can share the remaining relic.} 

In our analysis we fix the  parameters $\{m_{\eta_{I_1}^0}^{},~m_{\phi}^{},~\mu_{12\phi}^{},\lambda_{11\phi},~\lambda^{\prime}_{11\rm H}\}$ at the following values and express others in terms of them:
\begin{gather}\begin{split}
|h_{ii\ell}|=7.2\times 10^{-6} ~(a=0,~b=0.175),~\lambda_{ij}=1,~\lambda_{22\phi}=\lambda_{11\phi}\,,\quad\qquad\\
y_{12\phi}^{}=1,~0<\mu_{12\phi}^{}<2m_{\phi}^{},~\lambda^{\prime}_{22H}=1,~ m_{\eta^{0}_{I_2}}^{}=m_{\eta^{0}_{I_1}}^{}+m_{\phi}+1{\rm ~GeV},\quad~~\\ m_{N_1}=2m_{\eta^{0}_{I_2}}^{}, 
m_{N_2}=3m_{N_1},~m_{\eta^{0}_{R_{i}}}=m_{\eta^{0}_{I_{i}}}^{}+5{\rm ~GeV},~ m_{\eta^{+}_{i}}^{}=m_{\eta^{0}_{I_{i}}}^{}+5{\rm ~GeV}\,.
\label{eq:relic-para}
\end{split}
\end{gather}
Although different cosmological and astrophysical observations confirm the existence of DM, its detection still has not been confirmed. We will discuss the possible detection prospects via direct and indirect searches in the sections below. 
\subsection{Direct detection limits}
The most sensible method to detect a thermal DM is by observing the DM-nuclear/electron scattering rate through inelastic scattering with SM particles in direct detection experiments. The unobserved DM direct detection signal, from different underground experiments like XENON1T \cite{XENON:2018voc}, XENONnT \cite{XENON:2023cxc}, LUX-ZEPLIN \cite{LZ:2022lsv}, DARWIN/XLZD (projected) \cite{Baudis:2024jnk}, PandaX-xT (projected) \cite{PandaX:2024oxq}, puts an upper limit on the DM-nucleon scattering cross-section. There is a lower limit obtained from the coherent elastic neutrino-nucleus scatterings, known as neutrino-fog, where the discrimination of DM signal from neutrino background is challenging \cite{Tang:2023xub}.
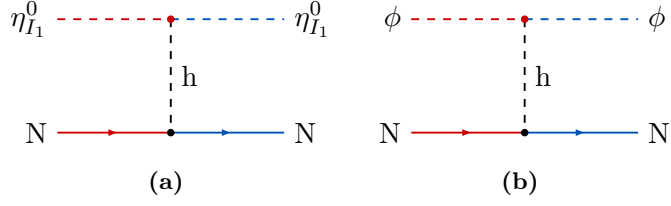
\begin{figure}[htb!]
\centering
\subfloat[]{\begin{tikzpicture}
\begin{feynman}
\vertex (a);
\vertex[left=1.5cm of a] (a1){\(\eta_{I_1}^{0}\)};
\vertex[right=1.5cm of a] (a2){\(\eta_{I_1}^{0}\)}; 
\vertex[below=1.5cm of a] (b); 
\vertex[below=1.5cm of a] (c); 
\vertex[left=1.5cm of b] (c1){\(\rm N\)};
\vertex[right=1.5cm of b] (c2){\(\rm N\)};
\diagram* {
(a1) -- [line width=0.25mm,scalar, arrow size=0.7pt, style=bostonuniversityred] (a),
(a) -- [line width=0.25mm,scalar, arrow size=0.7pt, style=mediumtealblue] (a2),
(a) -- [line width=0.25mm,scalar, edge label={\(\color{black}{\rm h}\)}, style=black] (b),
(c1) -- [line width=0.25mm,fermion, arrow size=0.7pt, style=bostonuniversityred] (b)  -- [line width=0.25mm,fermion, arrow size=0.7pt, style=mediumtealblue] (c2)};
\node at (a)[circle,fill,style=bostonuniversityred,inner sep=1pt]{};
\node at (b)[circle,fill,style=black,inner sep=1pt]{};
\end{feynman}
\end{tikzpicture}\label{feyn:dd-eta1-N}}\quad
\subfloat[]{\begin{tikzpicture}
\begin{feynman}
\vertex (a);
\vertex[left=1.5cm of a] (a1){\(\phi\)};
\vertex[right=1.5cm of a] (a2){\(\phi\)}; 
\vertex[below=1.5cm of a] (b); 
\vertex[below=1.5cm of a] (c); 
\vertex[left=1.5cm of b] (c1){\(\rm N\)};
\vertex[right=1.5cm of b] (c2){\(\rm N\)};
\diagram* {
(a1) -- [line width=0.25mm,scalar, arrow size=0.7pt, style=bostonuniversityred] (a),
(a) -- [line width=0.25mm,scalar, arrow size=0.7pt, style=mediumtealblue] (a2),
(a) -- [line width=0.25mm,scalar, edge label={\(\color{black}{\rm h}\)}, style=black] (b),
(c1) -- [line width=0.25mm,fermion, arrow size=0.7pt, style=bostonuniversityred] (b)  -- [line width=0.25mm,fermion, arrow size=0.7pt, style=mediumtealblue] (c2)};
\node at (a)[circle,fill,style=bostonuniversityred,inner sep=1pt]{};
\node at (b)[circle,fill,style=black,inner sep=1pt]{};
\end{feynman}
\end{tikzpicture}\label{feyn:dd-phi-N}}
\caption{Feynman diagrams correspond to the direct detection of $\eta_{I_1}^{0}$ (left) and $\phi$ (right).}
\label{feyn:wimp-dd}
\end{figure}

Here, the direct detection of $\eta_{I_1}^0$ and $\phi$ is possible via the Higgs mediated diagrams shown in fig\,.~\ref{feyn:wimp-dd}. In a two-component scalar DM setup, the effective spin-independent DM-nucleon scattering cross section is written as \cite{Bhattacharya:2016ysw},
\begin{align}
\sigma_{\rm N \eta^0_{I_1} }^{\rm eff}=\dfrac{\Omega_{\eta_{I_1}^0}}{\Omega_{\eta_{I_1}^0}+\Omega_{\phi}^{}}\sigma_{\rm N \eta^0_{I_1} }^{\rm SI}\,,\quad{\rm and}\quad
\sigma_{\rm N\phi }^{\rm eff}=\dfrac{\Omega_{\phi}^{}}{\Omega_{\eta_{I_1}^0}+\Omega_{\phi}^{}}\sigma^{\rm SI}_{\rm N\phi }\,.
\end{align}
\begin{figure}[htb!]
\centering
\subfloat[]{\includegraphics[width=0.475\linewidth]{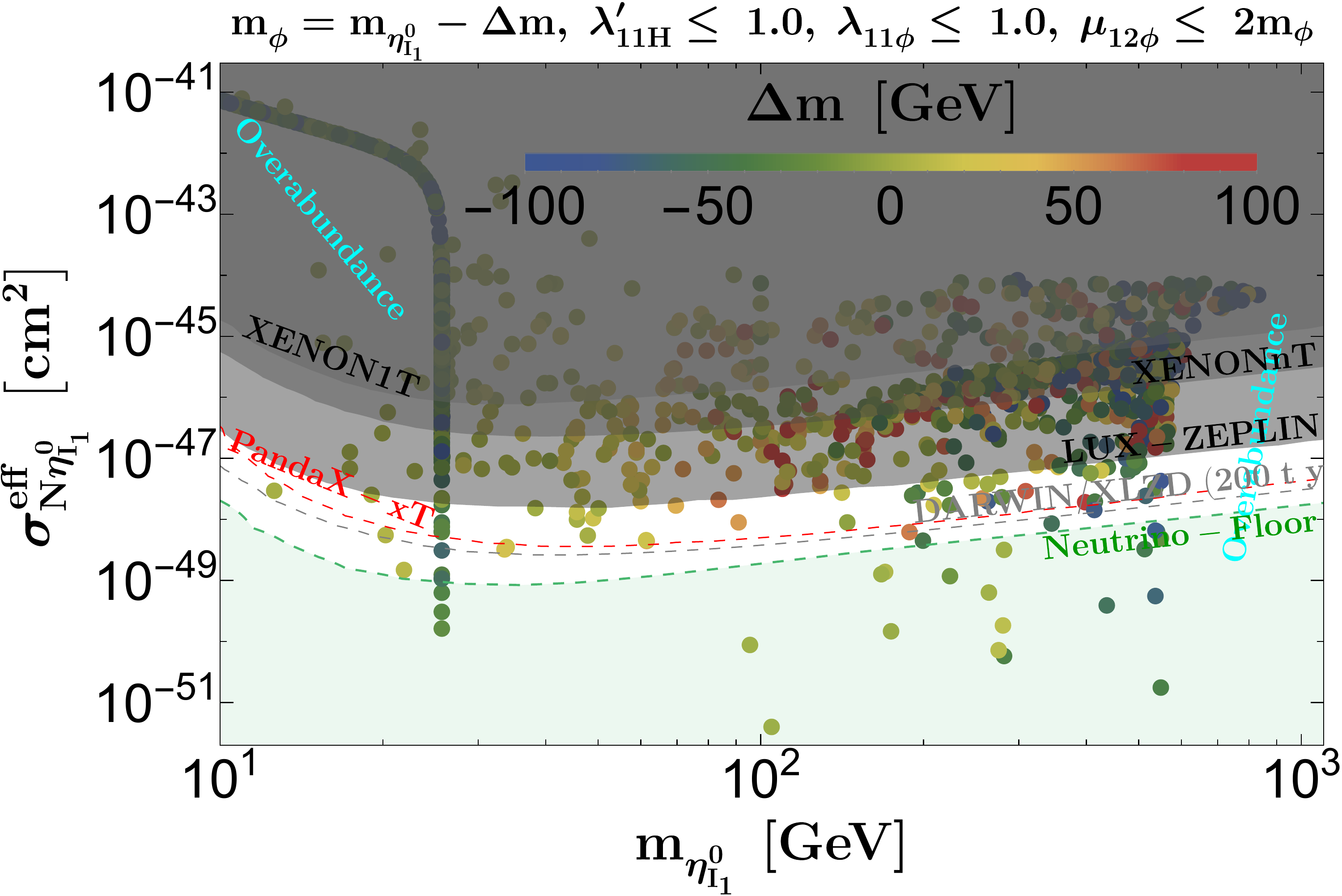}\label{fig:8a}}\quad
\subfloat[]{\includegraphics[width=0.475\linewidth]{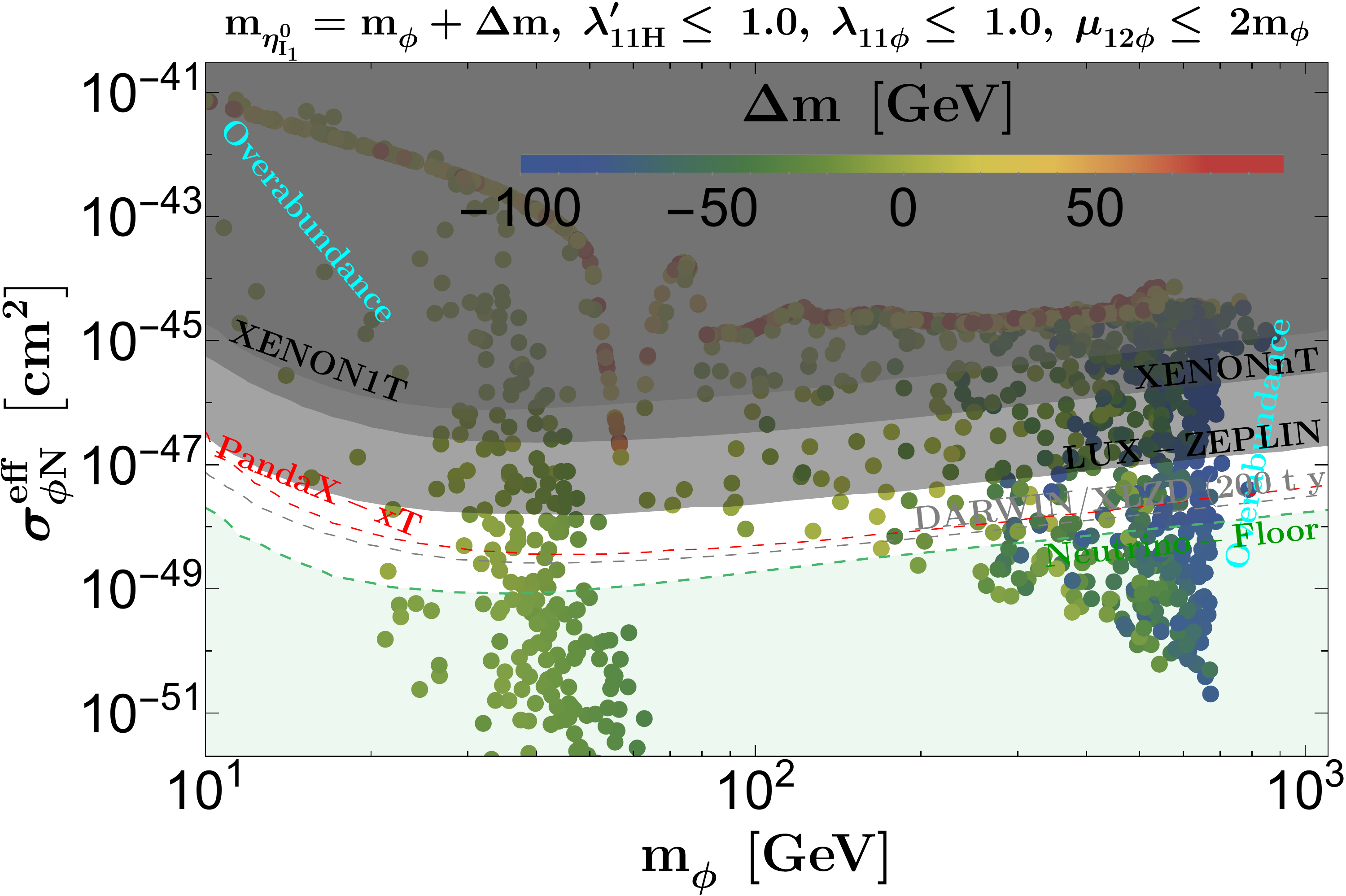}\label{fig:8b}}

\subfloat[]{\includegraphics[width=0.475\linewidth]{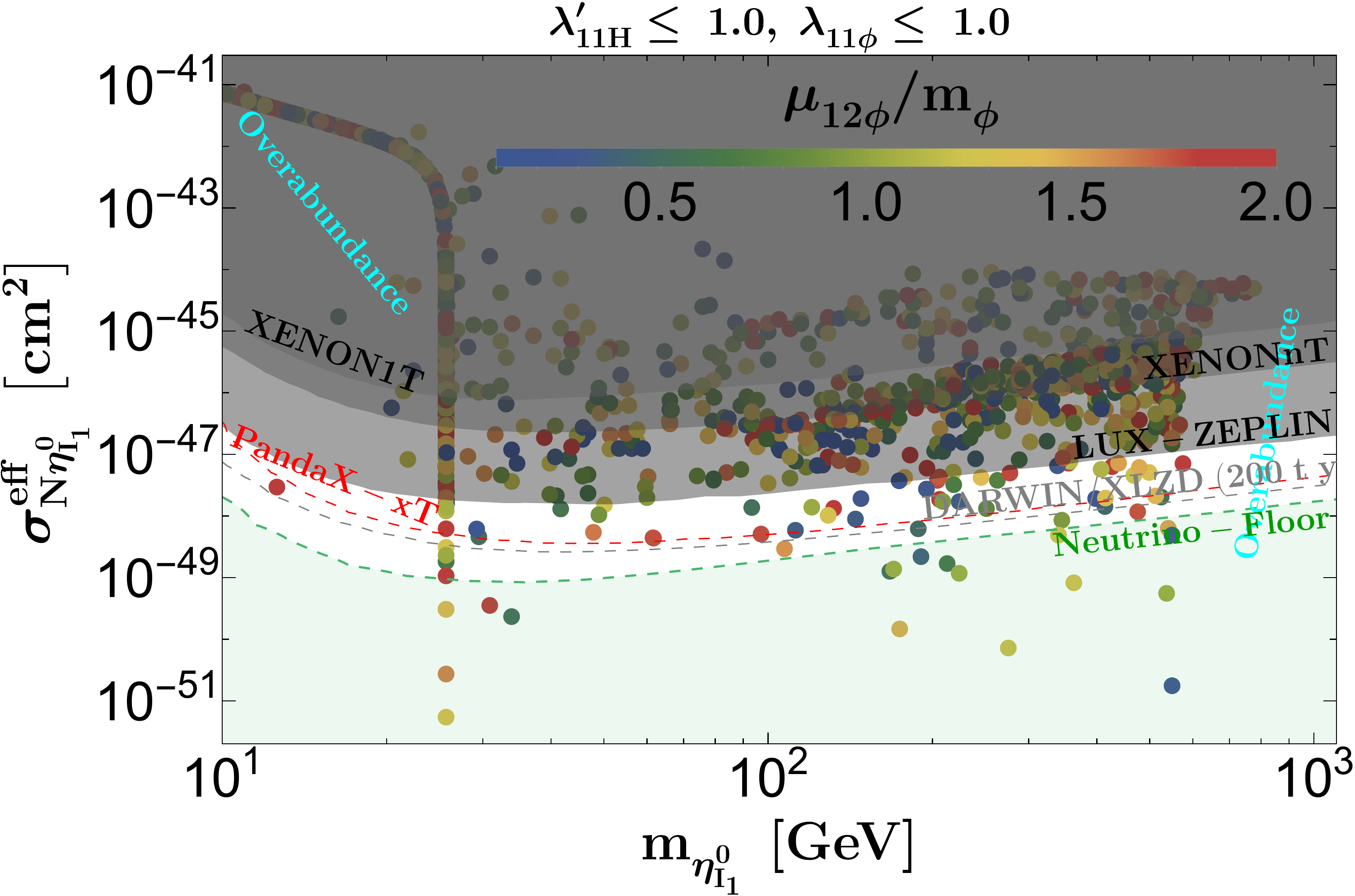}\label{fig:8c}}\quad
\subfloat[]{\includegraphics[width=0.475\linewidth]{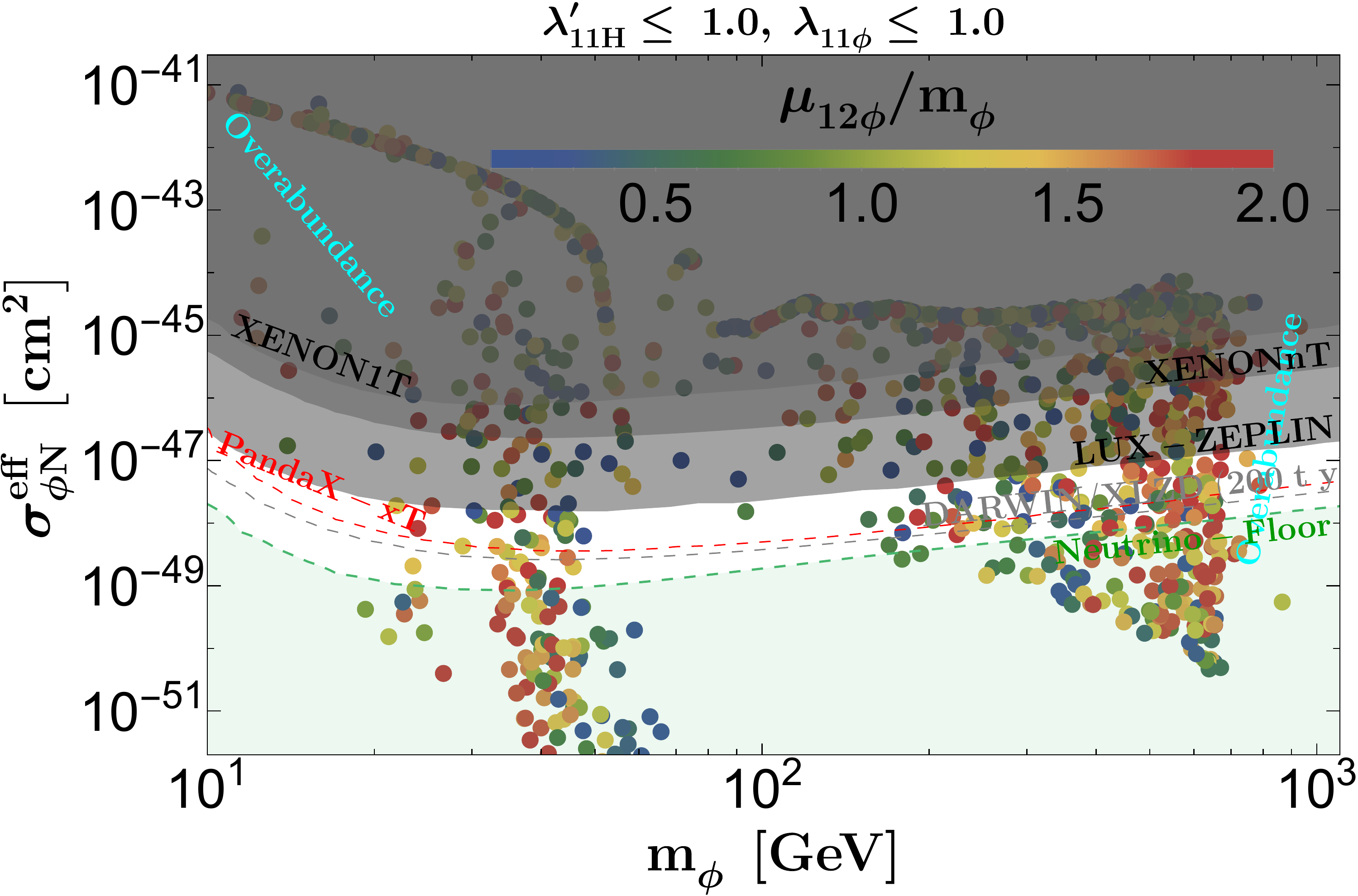}\label{fig:8d}}

\subfloat[]{\includegraphics[width=0.475\linewidth]{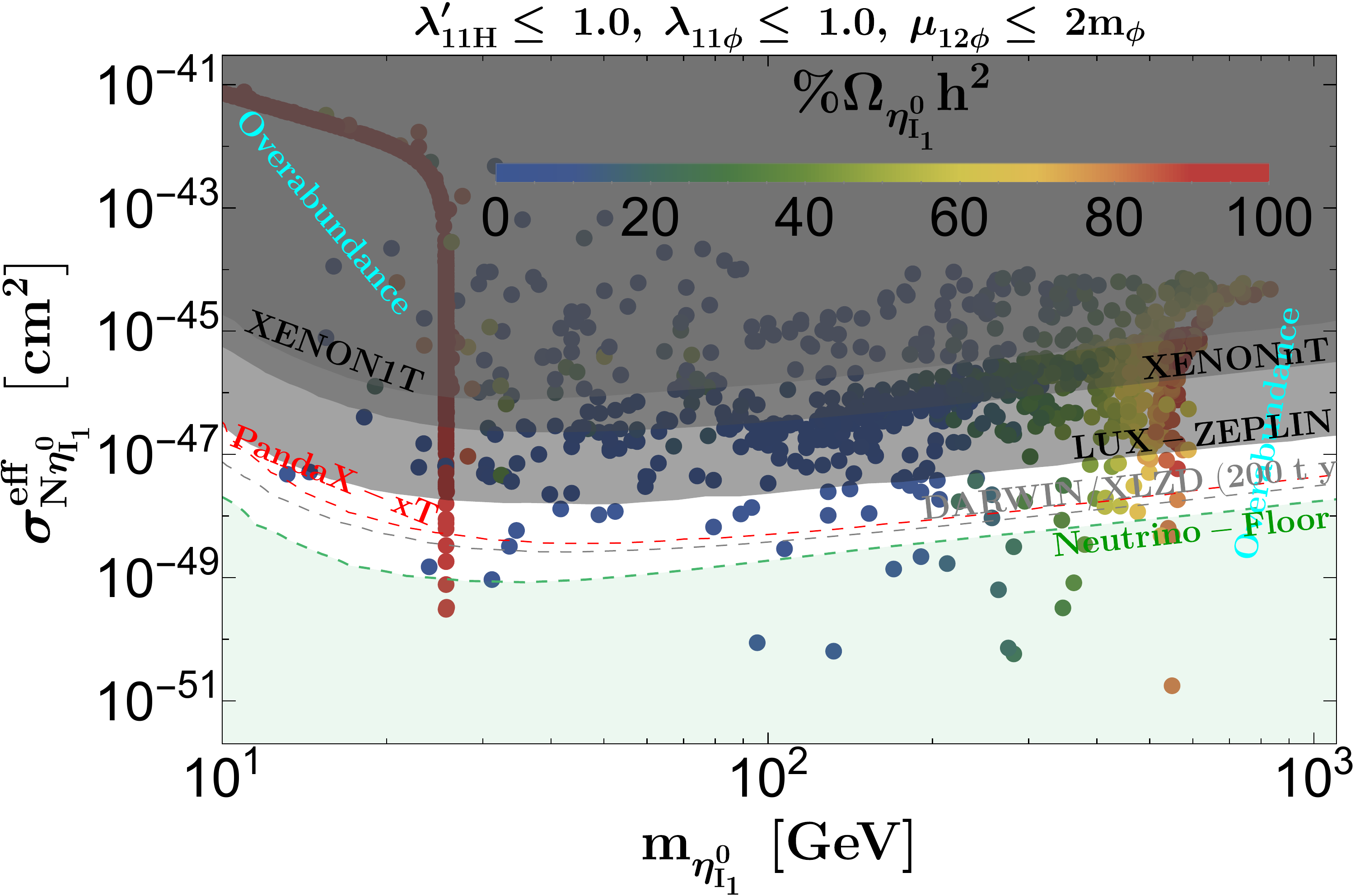}\label{fig:8e}}\quad
\subfloat[]{\includegraphics[width=0.475\linewidth]{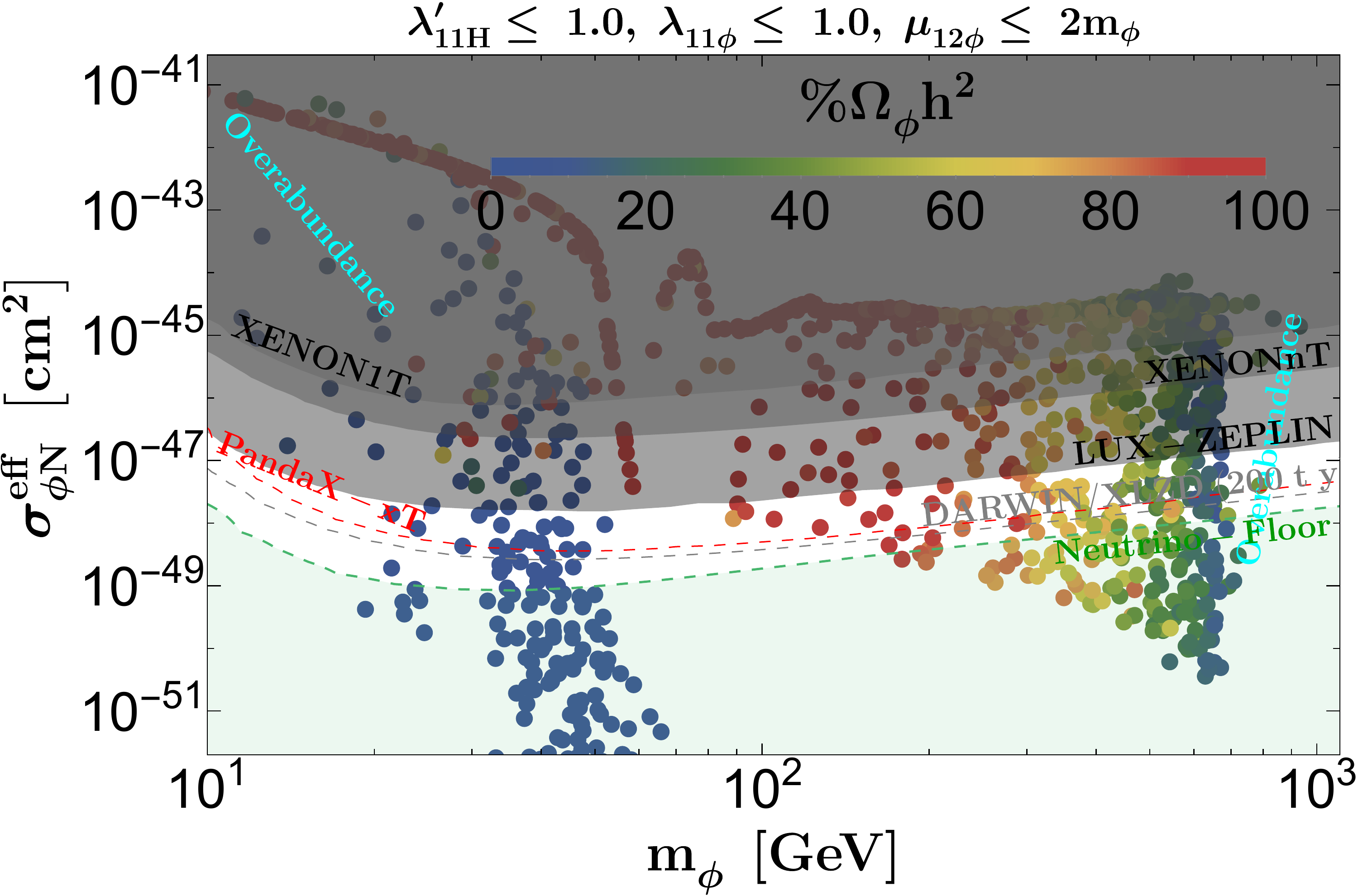}\label{fig:8f}}
\caption{Figs\,.~\ref{fig:8a}, \ref{fig:8c}, \ref{fig:8e} and \ref{fig:8b}, \ref{fig:8d}, \ref{fig:8f}, represents the relic allowed parameter space in $m_{\eta_{I_1}^0}-\sigma_{\rm N\eta_{I_1}^0}^{\rm eff}$ and $m_{\phi}-\sigma_{\rm N\phi}^{\rm eff}$ plane, respectively. The rainbow colorbar represents the variation of $\mu_{12\phi},~\Delta m=m_{\eta_{I_1}^0}^{}-m_{\phi}$, and $\%\Omega_{i}h^2=\Omega_{i}h^2\left(\sum\limits_{i}^{}\Omega_ih^2\right)^{-1}100$, where $i=\eta_{I_1}^0$ and $\phi$, has mentioned above the colorbar. The observed SI DD limit excludes the grey-shaded regions from XENON1T, XENONnT, and LUX-ZEPLIN, while the dashed lines correspond to projected limits from PandaX-xT and DARWIN/XLZD (200 t y). Light green represents the neutrino floor.}
\label{fig:relic-scan}
\end{figure}

\textcolor{black}{
In fig\,.~\ref{fig:relic-scan}, we show the DM relic density allowed parameter space in $m_{\eta_{I_1}^0}-\sigma_{\rm N\eta_{I_1}^0}^{\rm e ff}$ plane (left panel) and $m_{\phi}-\sigma_{\rm N\phi}^{\rm eff}$ plane (right panel). All plots in fig\,.~\ref{fig:relic-scan} are filled with color points having DM mass from $25\rm~GeV$ to $\rm 700~\rm GeV$. The rainbow color bar represents the variation of $\Delta m\equiv m_{\phi}-m_{\eta_{I_1}^0}$ (top), $\mu_{12\phi}$ (middle), and $\% \Omega_{i}h^2$ (bottom). The coupling associated with $\rm h\phi \phi$ and $\rm h\eta_{I_1}^0\eta_{I_1}^0$ are strongly constrained by the present SI DD constraint, thereby we keep them sufficiently small. In spite of that, a significant part of the scanned points are excluded by the grey-shaded regions from the stringent bound of the SI DD limit of the LUX-ZEPLIN experiment. The dashed lines correspond to the future projection limits set by PandaX-xT and XLZD experiments. In the large mass limit of both $m_{\phi}$ and $m_{\eta_{I_1}^0}$ is overabundant. The DM-DM conversion plays a crucial role in providing a larger direct search allowed parameter space for the two-component DM model as the conversion contributes significantly to the  relic density of the heavier DM component, however do not contribute to DD at the tree level.}


\textcolor{black}{ The (effective) conversion processes, $\phi\phi\to\eta_{i}^{\dagger}\eta_i^{}$ with $m_{\phi}>m_{\eta_{I_1}^0}$, help $\phi$ to achive underabundance with small $\lambda_{\phi H}$, as constrained by DD. The conversion rate depends on the ratio of the equilibrium number densities of both DMs, which requires a small mass splitting between the DM components $(\Delta m\lesssim 10 ~\rm GeV)$. This is clearly visible in figs\,.~\ref{fig:8a} and \ref{fig:8b}, where we have shown the role of $\Delta m$ in the relic-allowed parameter space. Another factor in conversion is the coupling $\mu_{12\phi}$, which enhances the effective annihilation cross-section. From figs\,.~\ref{fig:8c} and \ref{fig:8d} we see that there is no specific $\mu_{12}$ dependence on either of the DM masses as it can be appropriately tuned with $\Delta m$ to get correct conversion and therefore relic density. Finally, in figs\,.~\ref{fig:8e} and \ref{fig:8f}, we show the effective contribution of DMs in total relic, which shows all possible relative abundance combinations are accessible in this model with appropriate choices of DM masses and splittings. We note in particular, high $\eta_{I_1}^0$ mass region is $\eta_{I_1}^0$ dominant while the middle $m_{\eta_{I_1}^0}$ region is dominated by $\phi$, as expected.}

\subsection{Indirect detection limits}
The self-annihilation of DM particles at the galactic core can produce SM particles, like gamma rays, neutrinos, positrons, etc., which could be possible to detect by various telescopes and detectors. However, the non-observation of the excess in these signals, gamma rays (Fermi-LAT), cosmic rays (AMS-02), and neutrinos (Ice-Cube), etc., allows us to set an upper limit on the DM self-annihilation cross section guided us for the theoretical models and future searching for DM.
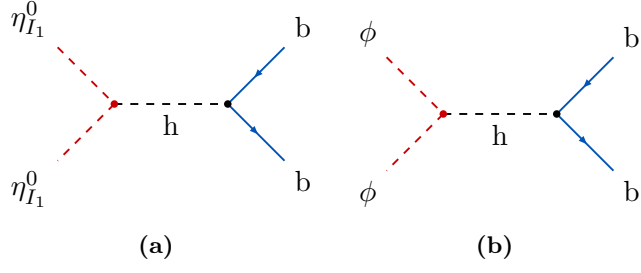
\begin{figure}[htb!]
\centering
\subfloat[]{\begin{tikzpicture}
\begin{feynman}
\vertex (a);
\vertex[above left=0.75cm and 0.75cm of a] (a1);
\vertex[below left=0.75cm and 0.75cm of a] (a2);
\vertex[right=1.5cm of a] (b); 
\vertex[above right=0.75cm and 0.75cm of b] (c1);
\vertex[below right=0.75cm and 0.75cm of b] (c2);
\diagram*{
(a1) -- [line width=0.25mm,scalar, arrow size=0.7pt, style=bostonuniversityred] (a),
(a) -- [line width=0.25mm,scalar, arrow size=0.7pt, style=bostonuniversityred] (a2),
(a) -- [line width=0.25mm,scalar, edge label'={\(\color{black}{\rm h}\)}, style=black] (b),
(c1) -- [line width=0.25mm,fermion, arrow size=0.7pt,style=mediumtealblue] (b)  -- [line width=0.25mm,fermion, arrow size=0.7pt,style=mediumtealblue] (c2)};
\node at (a)[circle,fill,style=bostonuniversityred,inner sep=1pt]{};
\node at (b)[circle,fill,style=black,inner sep=1pt]{};
\vertex[above left=0.75cm and 0.75cm of a] (a11){\(\eta_{I_1}^0\)};
\vertex[below left=0.75cm and 0.75cm of a] (a22){\(\eta_{I_1}^0\)}; 
\vertex[above right=0.75cm and 0.75cm of b] (c11){\(\rm b\)};
\vertex[below right=0.75cm and 0.75cm of b] (c22){\(\rm b\)};
\end{feynman}
\end{tikzpicture}
\label{fig:eta1I0eta1I0-h-bb}}~
\subfloat[]{\begin{tikzpicture}
\begin{feynman}
\vertex (a);
\vertex[above left=0.75cm and 0.75cm of a] (a1);
\vertex[below left=0.75cm and 0.75cm of a] (a2); 
\vertex[right=1.5cm of a] (b); 
\vertex[above right=0.75cm and 0.75cm of b] (c1);
\vertex[below right=0.75cm and 0.75cm of b] (c2);
\diagram* {
(a1) -- [line width=0.25mm,scalar, arrow size=0.7pt, style=bostonuniversityred] (a),
(a) -- [line width=0.25mm, scalar, arrow size=0.7pt, style=bostonuniversityred] (a2),
(a) -- [line width=0.25mm,scalar, edge label'={\(\color{black}{\rm h}\)}, style=black] (b),
(c1) -- [line width=0.25mm,fermion, arrow size=0.7pt,style=mediumtealblue] (b)  -- [line width=0.25mm,fermion, arrow size=0.7pt,style=mediumtealblue] (c2)};
\node at (a)[circle,fill,style=bostonuniversityred,inner sep=1pt]{};
\node at (b)[circle,fill,style=black,inner sep=1pt]{};
\vertex[above left=0.75cm and 0.75cm of a] (a11){\(\phi\)};
\vertex[below left=0.75cm and 0.75cm of a] (a22){\(\phi\)}; 
\vertex[above right=0.75cm and 0.75cm of b] (c11){\(\rm b\)};
\vertex[below right=0.75cm and 0.75cm of b] (c22){\(\rm b\)};
\end{feynman}
\end{tikzpicture}
\label{fig:phiphi-h-bb}}
\caption{The Feynman diagrams represents the self annihilation of \(\eta_{I_1}^0\) and \(\phi\) into $b~\overline{b}$ shown in figs\,.~\ref{fig:eta1I0eta1I0-h-bb}, and \ref{fig:phiphi-h-bb}, respectively.}
\label{feyn:2wimp-id}
\end{figure}
In this framework, we have estimated the self-annihilation of $\eta_{I_1}^0$ and $\phi$, is possible via the Higgs mediated diagrams shown in fig\,.~\ref{feyn:2wimp-id}, into the bottom pair as it gives the most stringent constraint for DM annihilation set by Fermi-LAT observation. More interestingly, In a two-component scalar DM setup, the effective DM self-annihilation cross-sections are written as \cite{Bhattacharya:2016ysw},
\begin{align}
\langle\sigma v\rangle_{\rm \eta_{I_1}^0\eta_{I_1}^0\to b~\overline{b}}^{\rm eff}=\dfrac{\Omega^2_{\eta_{I_1}^0}}{(\Omega_{\eta_{I_1}^0}+\Omega_{\phi}^{})^2}\langle\sigma v\rangle_{\rm \eta_{I_1}^0\eta_{I_1}^0\to b~\overline{b}}^{\rm }\,,\quad{\rm and}\quad 
\langle\sigma v\rangle_{\rm \phi~\phi\to b~\overline{b}}^{\rm eff}=\dfrac{\Omega^2_{\phi}}{(\Omega_{\eta_{I_1}^0}+\Omega_{\phi}^{})^2}\langle\sigma v\rangle_{\rm \phi~\phi\to b~\overline{b}}^{\rm }\,.
\end{align}
where the cross-section is calculated at the WIMP freeze-out temperature, $ T^{}_{\rm FO}\sim m^{}_{\rm DM}/25$.
\begin{figure}[htb!]
\centering
\subfloat[]{\includegraphics[width=0.475\linewidth]{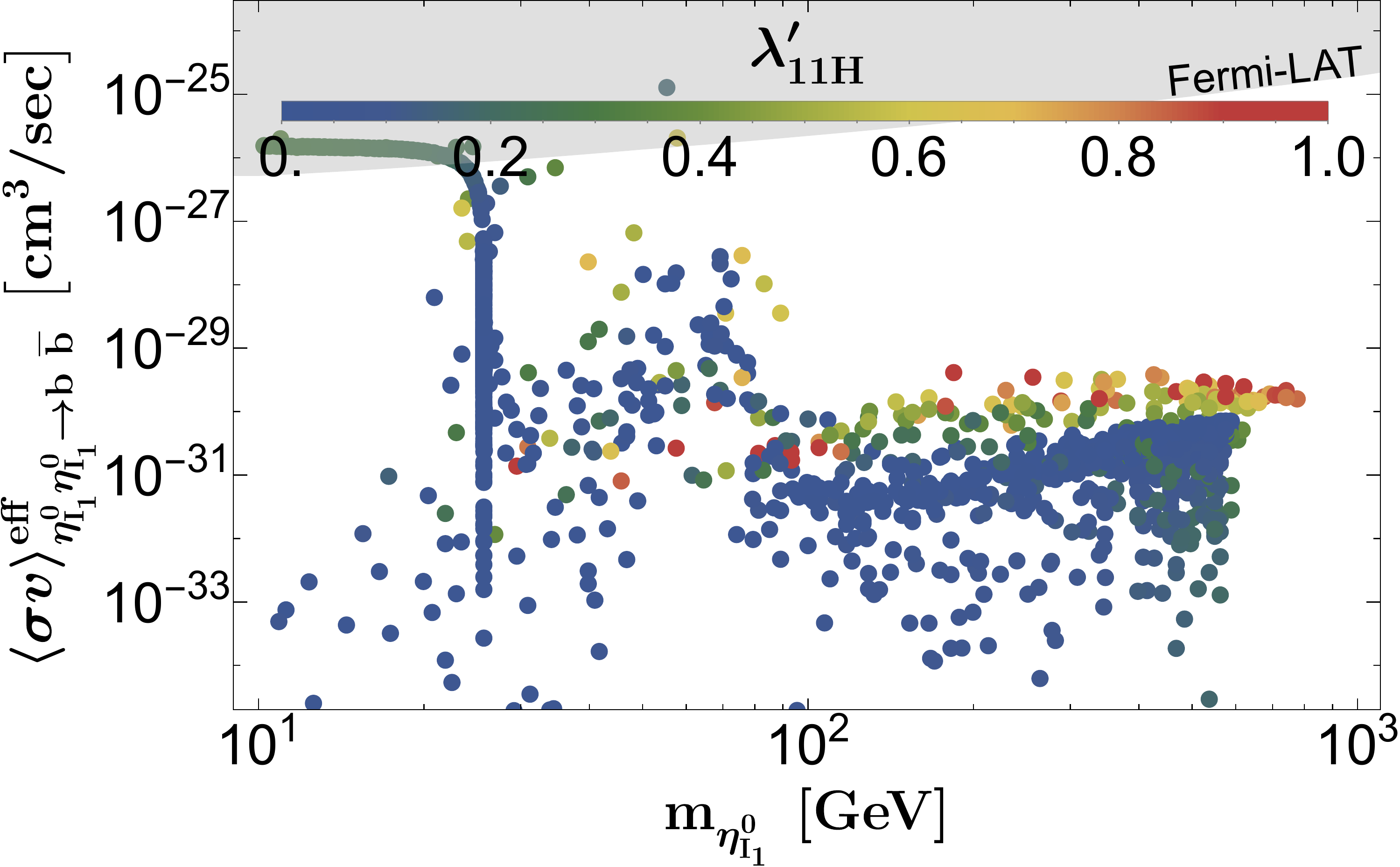}\label{fig:10a}}\quad
\subfloat[]{\includegraphics[width=0.475\linewidth]{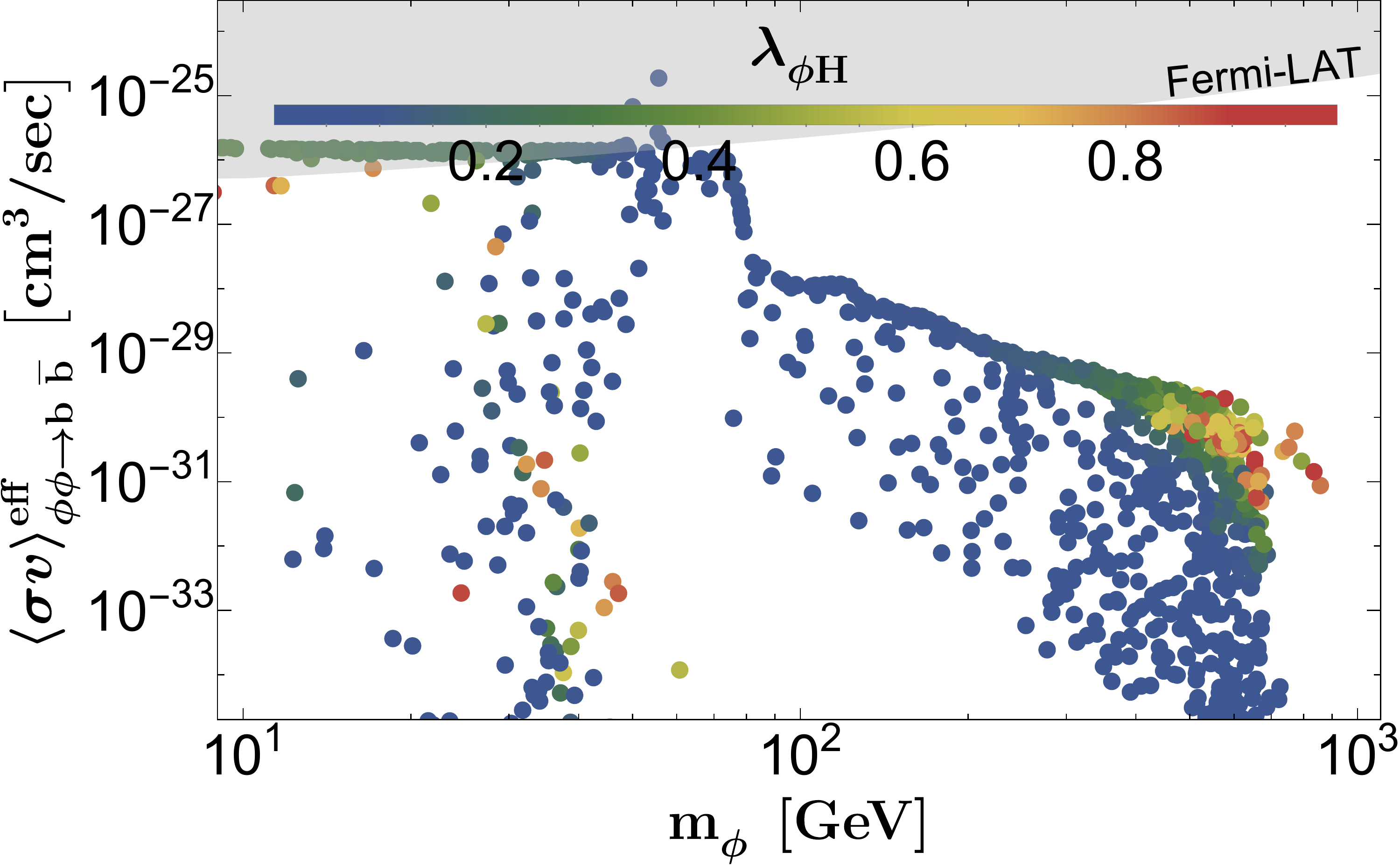}\label{fig:10b}}

\subfloat[]{\includegraphics[width=0.475\linewidth]{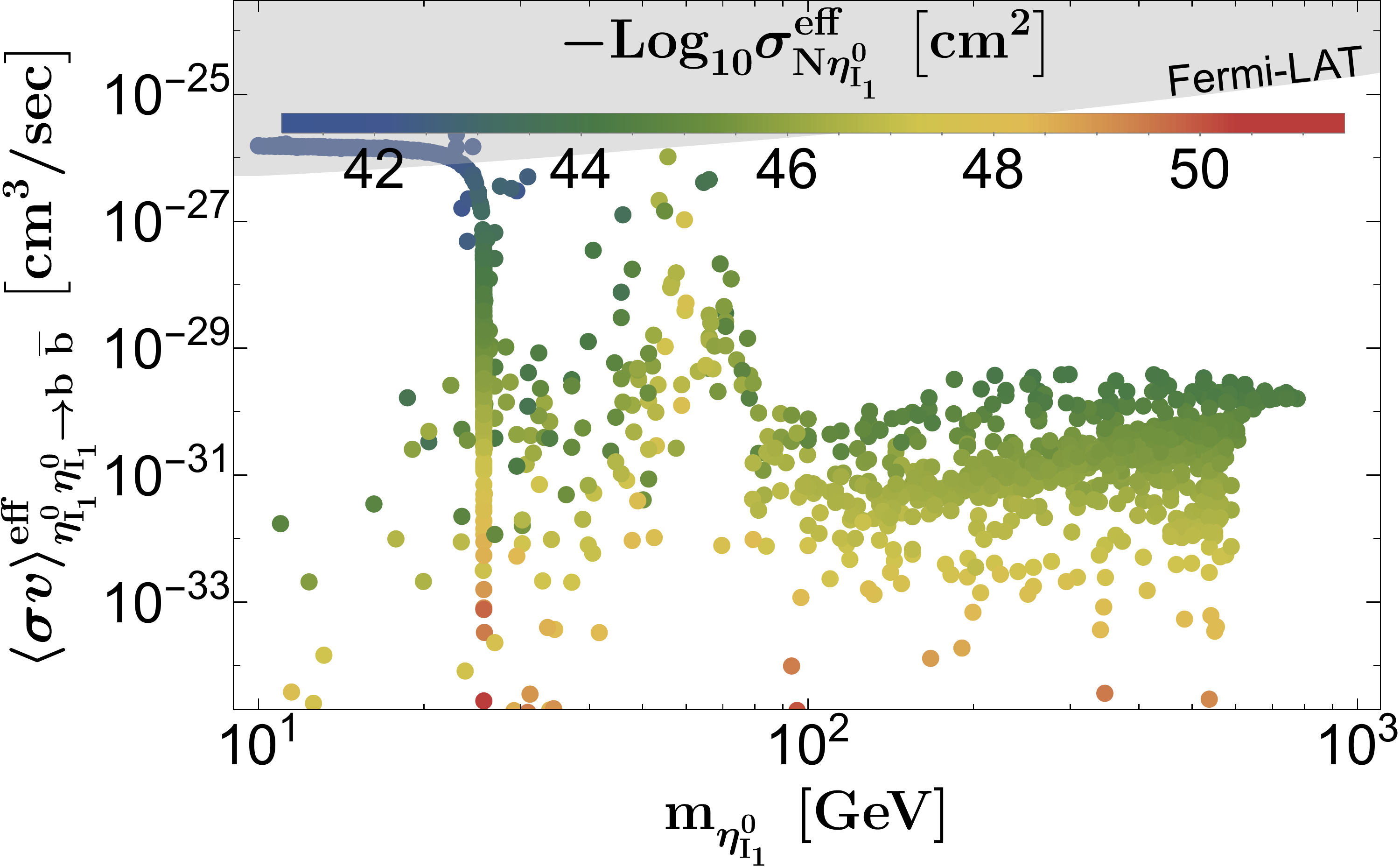}\label{fig:10c}}\quad
\subfloat[]{\includegraphics[width=0.475\linewidth]{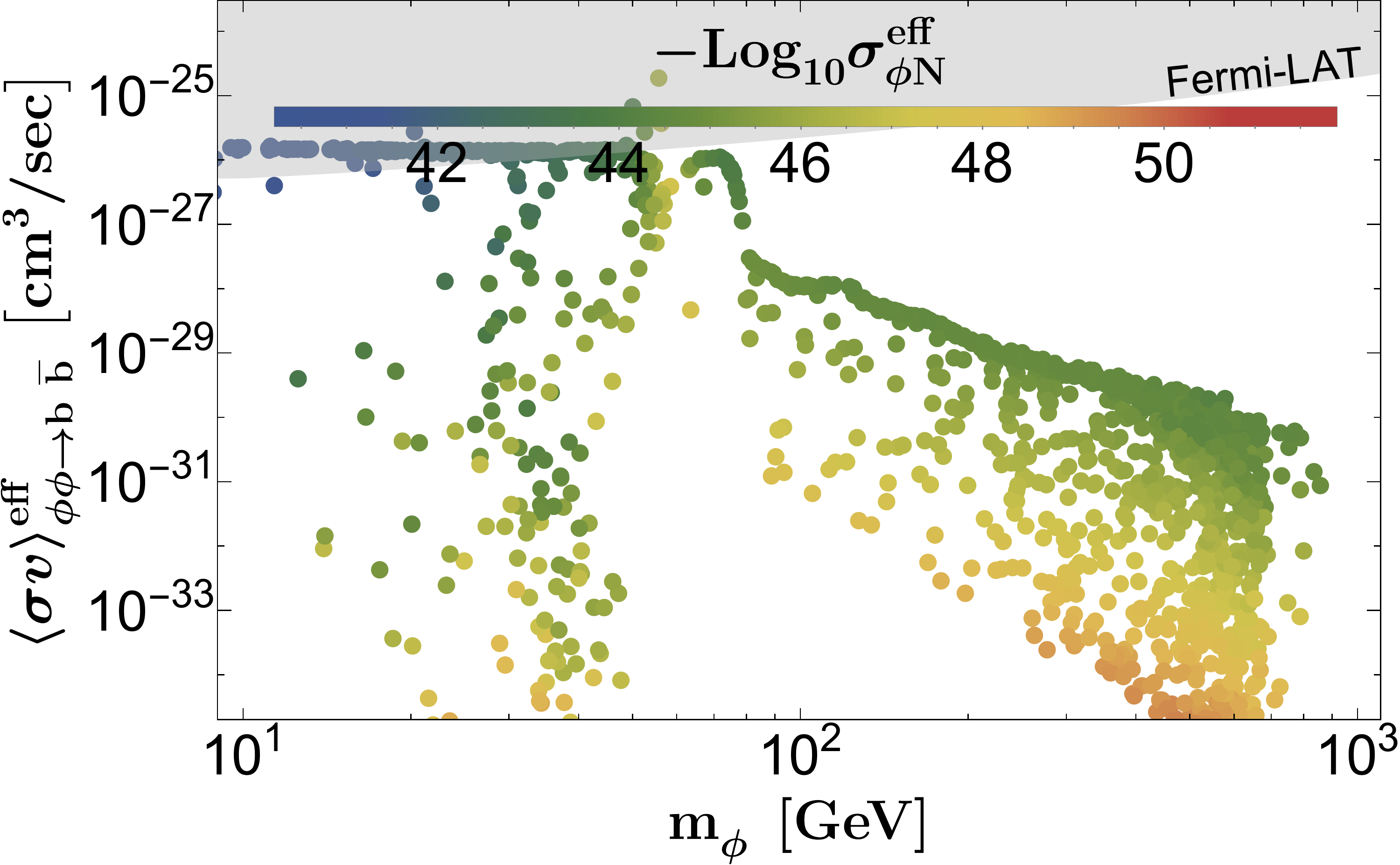}\label{fig:10d}}
\caption{Figures are shown the Indirect detection limit on relic allowed parameter space in $m_{i}-\langle\sigma v\rangle_{i~i\to b~\overline{b}}^{\rm eff}$ plane, where $i=\eta_{I_1}^0,~{\rm and}~\phi$. The grey-shaded regions are excluded from the recent Fermi-LAT limit on the DM annihilation bottom pair. In this scan, we have taken: $\mu_{12\phi}\leq 2m_{\phi}$ and $\lambda_{11\phi}\leq 1.0$. The rainbow color shows the variation of the respective parameters mentioned above in the color bar.}
\label{fig:scan-id}
\end{figure}
The annihilation cross-section, $\langle\sigma v\rangle_{\rm \eta_{I_1}^0\eta_{I_1}^0\to b~\overline{b}}^{\rm }$, proportionally depends on the parameters $m_{\eta_{I_1}^0}$ and $\lambda_{h\eta_{I_1}^0\eta_{I_1}^0}=(\lambda_{11H}+\lambda^{\prime}_{11H}-\lambda^{\prime\prime}_{11H})v=(2m_{\eta_{I_1}^0}^2-2m_{\eta_{1}^+}^2)/v+\lambda^{\prime}_{11H}v$, on the contrary $\langle\sigma v\rangle_{\rm \phi\phi\to b~\overline{b}}^{\rm }$ depends on the $m_{\phi}$ and $\lambda_{h\phi\phi}=\lambda_{\phi H}v$. In fig\,.~\ref{fig:scan-id}, we have shown the variation of these parameters through the rainbow colorbar. In all plots, the gray regions are excluded by the upper limit on the DM annihilation to the bottom pair from Fermi-LAT, and this is stringent compared to other indirect observations. In case of $\eta_{I_1}^0$, with the enhancement of $\lambda^{\prime}_{1H}$ the $\langle\sigma v\rangle_{\rm \eta_{I_1}^0\eta_{I_1}^0\to b~\overline{b}}^{\rm }$ is increased which depicted in fig\,.~\ref{fig:10a}. If the DM mass $m_{\eta_{I_1}^0}\sim m_h/2$, some points are excluded by Fermi-LAT due to the Higgs resonance enhancement in the cross-section, and the same explanation is also valid for the second DM $\phi$ which has been illustrated in fig\,.~\ref{fig:10b}. Below $m_{h}/2$, the cross-section gradually decreases, and some points go inside the Fermi-LAT exclusion regime. As these couplings are also involved in SI direct detection of $\eta_{I_1}^0$ and $\phi$, so its decrement also decreases the DD cross-sections, as we see in figs\,.~\ref{fig:10c} and \ref{fig:10d}. Here, the gradual decrement of both DD and ID cross-section with the vertex factor $\lambda_{h\eta_{I_1}^0\eta_{I_1}^0}$ and $\lambda_{h\phi\phi}$, is portrayed by the transition of color gradient from blue to red.
\section*{Summary}

\textcolor{black}{In this section, we connect to the three phenomena which motivates us to search for the physics beyond the Standard Model, namely, $(i)$ Neutrino mass governed by $\{m_{\eta_{R_k}^0},~m_{\eta^0_{I_k}}^{},~m_{N_k}^{},~h_{kk\alpha}\}$, $(ii)$ Baryon asymmetry governed by $\{m_{\eta^0_{R_k}}^{},~m_{\eta^0_{I_k}}^{},~m_{\eta_k^+}^{},~m_{N_k}^{},m_{\phi},~y_{12\phi},~\mu_{12\phi},~h_{kk\alpha}\}$, and $(iii)$ DM relic density and its detection possibilities guided by $\{m_{\eta^0_{R_k}}^{},~m_{\eta^0_{I_k}}^{},~m_{\eta_k^+},~m_{\phi}$, $~\lambda_{\phi H},~\lambda^{\prime}_{kkH},~\mu_{12\phi},~\lambda_{mn},~\lambda_{kk\phi}\}$ parameters in a one plane. The common parameters which connects the neutrino mass, leptogenesis and DM phenomenology are the neutral component of the inert doublet masses $\{m_{\eta^0_{R_k}}^{},m_{\eta^0_{I_k}}^{}\}$. The RHN being much heavier than DMs, they do not play any role in the DM relic density. The Casas-Ibarra (CI) parametrization provides  the Yukawa couplings $h_{kk\alpha}$ consistent with the light neutrino masses and mixing where $\{m_{\eta^0_{R_k}}^{},m_{\eta^0_{I_k}}^{},m_{N_k}^{}\}$ parameters are taken as input. The observed BAU adjusts the remaining free parameters, $\{m_{\eta_k^+},m_{\phi},y_{12\phi},\mu_{12\phi}\}$. Finally, the correct DM relic density is achieved by fixing the parameters $\{\lambda_{\phi H},\lambda^{\prime}_{kkH},\lambda_{mn},\lambda_{kk\phi}\}$. The quartic coupling $\lambda_{\phi H}$ is directly connected with DD and ID of DM. The Yukawa coupling $h_{kk\alpha}$ can have upper limits from the current bounds on lepton flavor-violating decays $(\ell_{\alpha}\to\ell_{\beta}\gamma)$. In this work the values of the Yukawa couplings $h_{kk\alpha}$ are small $\mathcal{O}$($10^{-6}$) and the lepton flavor violating decays are suppressed within the limit.}

\begin{figure}[htb!]
\centering
\includegraphics[width=0.475\linewidth]{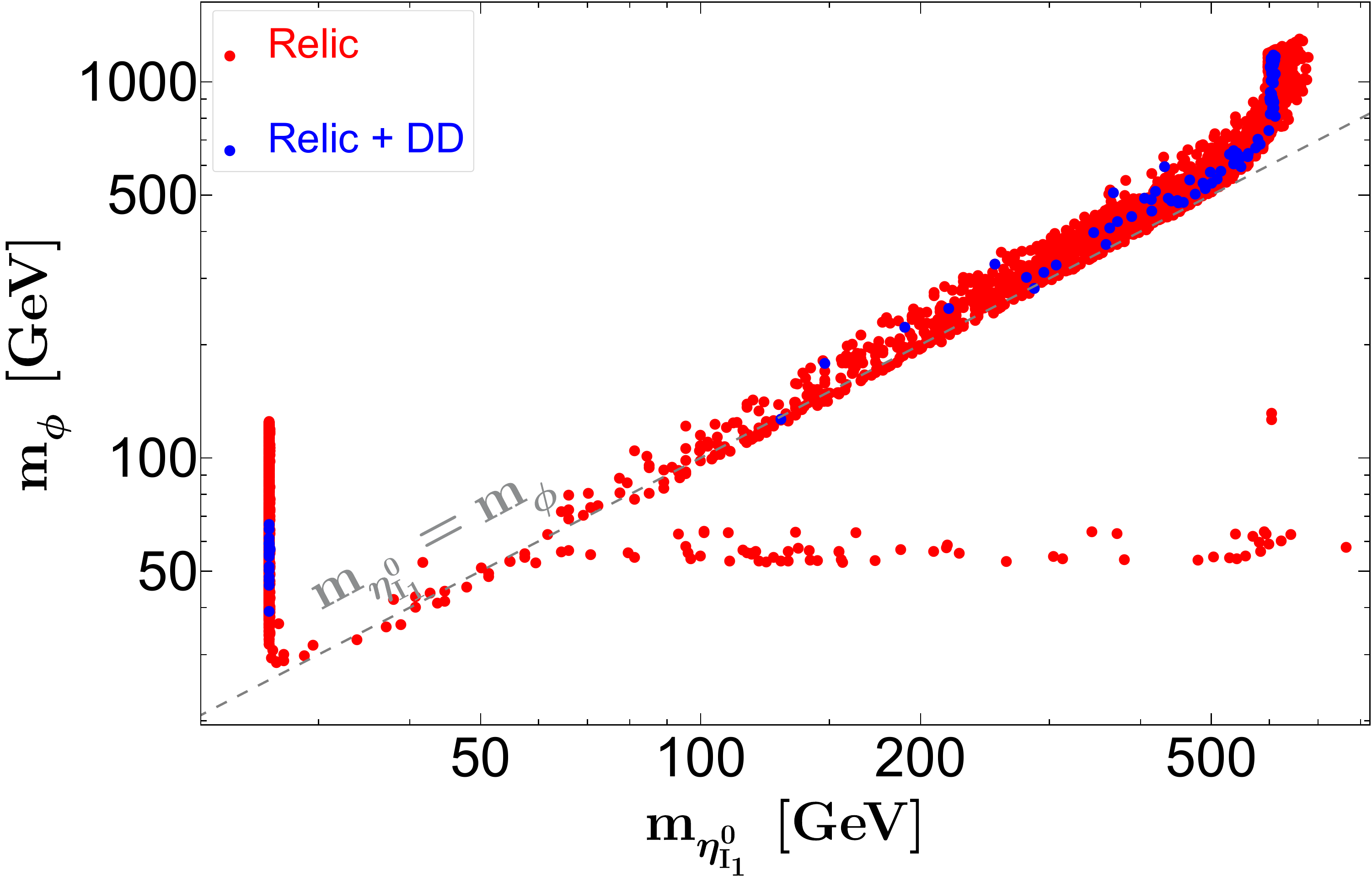}~~
\includegraphics[width=0.475\linewidth]{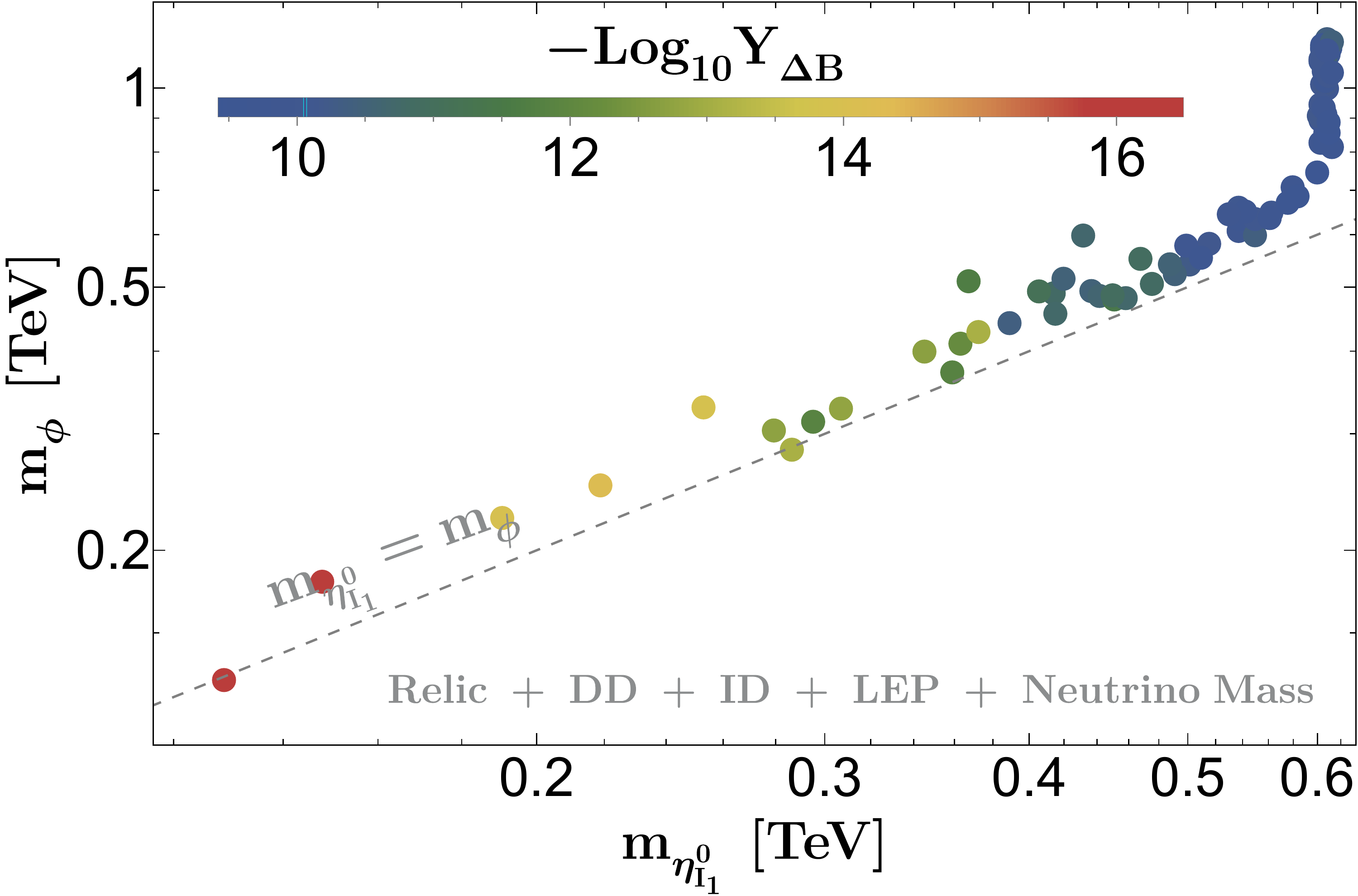}
\label{fig:summary}
\caption{The red points represent the total relic density satisfied by both DM candidates. The blue points correspond to the total relic density satisfied and allowed by the DD constraint from the LZ-2022 experiment for both DMs. Lastly, the rainbow color points indicate the regions where the relic density, DD, ID provided by Fermi-LAT, and LEP limits on charged scalars mass are satisfied by both DM candidates. The color bar shows the baryon asymmetry, $Y_{\Delta B}(z^{}_{\rm sph})$, corresponding to each point, while $Y_{\Delta B}^{\rm obs}$ indicated by cyan line on the legendbar.}
\label{fig:sum}
\end{figure}

\textcolor{black}{In fig\,.~\ref{fig:sum}, we show the relic density, neutrino mass, BAU, DD, ID, and flavor violating decay-allowed parameter space in $m_{\eta_{I_1}^0}-m_{\phi}$ plane, obeying eq\,.~\ref{eq:relic-para}.
The left panel of fig\,.~\ref{fig:sum} represents the parameter space allowed by the DM relic density shown by the red dotted points, while the blue points correspond to regions consistent with both the relic density and DD constraints. Here, we have used the stringent exclusion limit from SI DM-nucleon scattering cross-sections provided by the LUX-ZEPLIN data. Also, we restrict the DM mass ratio ($m_{\phi}/m_{\eta_{I_1}^0}$) upto three for simplicity. We see that the full DM mass range is allowed with a small $\Delta m$. We may remind that the inert doublet DM results in an underabundance in the mass range $\rm m_h/2 < m^{}_{_{DM}} < 500~\mathrm{GeV}$, which is compensated here by the second DM component $\phi$. The self-annihilation of $\phi$ into SM particles is possible only via the Higgs portal interactions, and its freeze-out point depends on the $\lambda_{\phi\rm H}$, which DD also constrains.}

In the right panel of fig\,.~\ref{fig:sum}, we show the variation of the baryon asymmetry in $m_{\eta_{I_1}^0}-m_{\phi}$ plane, while all points respect the DM relic density, neutrino mass, DD, ID and LEP-II constraints on charged scalar mass. Recent observation of Fermi-LAT puts an upper limit on the DM annihilation to bottom pairs, which are less effective compared to DD constraint. Due to this, the allowed parameter space remains unaltered after imposing ID constraints.
The parameters $h_{kk\alpha},~y_{12\phi}^{},~m^{}_{N_{1}},~m^{}_{N_{2}}$ are responsible for Leptogenesis and related to active neutrino mass generation, but does not have any role in DM phenomenology.
Using this freedom, we calculated the yield of baryon asymmetry $(\rm Y_{\Delta B}^{})$. The rainbow color bar shows the variation of $\rm Y^{}_{\Delta B}$ while the \textcolor{black}{cyan} line on the color bar represents the observed baryon asymmetry $\rm(Y_{\Delta B}^{obs})$. But, still, we observe that a dependence on DM mass is present via the decay of $N_1$ ($\to \ell_{\alpha}~\eta_1$) in the asymmetry parameter, which is reflected in the plot.
With the increase in the mass of $\eta_{I_1}^0$, the $m_{N_1}^{}$ (as described by eq\,.~\ref{eq:relic-para}) also increases, leading to an increase in $\Gamma_{N_1}^{}$, which in turn increases the asymmetry.
\textcolor{black}{Finally, we have implemented the conservative LEP-II bound on the charged scalar mass ($m_{\eta^+_1}\geq70$), and also respect the $13$ TeV LHC data on chargino search as discussed in page 5 of Sec.\ref{sec:constrnt}}.
\section{Summary and Conclusions}\label{sec:summary}
The simplest scenario that addresses neutrino mass generation and BAU is type-I seesaw model, where the right-handed neutrinos (RHNs) need to be heavy,  
$\sim 10^{10}$ \textcolor{black}{GeV} to satisfy the active neutrino masses and baryon asymmetry within the observed limit. On the contrary, the WIMP (mass $\sim$ GeV) freeze out 
occurs much later. So, having a separate DM candidate added to such a framework won't correlate all the phenomena together. 
The simplest way to connect them is to consider the 
scotogenic model, where DM freeze-out and leptogenesis could happen nearly at the same scale, and the masses of RHNs get down to $\sim $TeV scale. 
However, the BAU and DM phenomena are not directly coupled in such a framework. Our primary focus was to find out a model setup with 
minimal particle content that not only addresses all of them together, but also provides an inter dependence, so that the prediction in one sector affects the other.

We find out that an extension of scotogenic model that includes two RHNs, two inert doublets, and one real scalar, stabilised appropriately under 
$\mathbb{Z}_2^{}\otimes\mathbb{Z}_2^{\prime}$ symmetry with two real scalar DM components, does the job, establishing a novel connection between the 
DM phenomenology and the matter-antimatter asymmetry. The asymmetry generation in leptogenesis strongly depends on the DM masses and coupling with the 
RHNs through the vertex correction of $N_1^{}$ decays into $\ell\eta_1^{}$. On the other hand, neutrino mass generation strongly depends on the inert doublet 
masses and $h_{kk\alpha}$ couplings, which also play a significant role in leptogenesis. Due to the presence of $N\ell\eta$ vertex, the radiative lepton flavor 
violating decay becomes possible, but the limit from MEG-II is less sensitive for the parameter space relevant for us. As the LEP-II experiments already put a stringent lower limit on the singly charged scalar $\sim 70$ GeV, excludes some of our parameter space. 

Operationally, we perform all the detailed calculations to arrive at our final results. We have solved the cBEQ of $\rm Y_{N_1^{}}^{}$ and $\rm Y_{B-L}^{}$ numerically for 
some chosen benchmark points that address the correct active neutrino mass limits.
We also provide a scan plot in $m_{\eta_{I_1}^{0}}-(y_{12\phi}\times\mu_{12\phi})$ plane, 
where the points are explaining the observed baryon asymmetry and neutrino masses simultaneously.  We do the DM calculation via solving coupled BEQs as well as scanning it via micrOmegas. The mass hierarchy between DM components plays a crucial role in leptogenesis, as both the DMs are considered as on-shell particles in the 1-loop vertex correction calculation. The parameters $\{m_{\eta_{I_1}^{0}},~m_{\phi},~\mu_{12\phi}^{}\}$ relevant for Leptogenesis, also turns crucial for DM phenomenology. 
In the minimal scotogenic model, with one inert doublet, the relic density 
allowed parameter space is very restrictive, depends on the mass splitting between the charged and neutral components of the inert doublet, and most of the parameter space 
accessible to collider observation is under-abundant due to the presence of gauge portal interactions. Here, on the contrary, in the presence of two DM components, the 
parameter space available for both leptogenesis and DM constraints, is enhanced, allowing a future detection of the charged component of the inert doublet in the collider, 
indicating to a specific parameter space relevant for DM, leptogenesis and neutrino mass generation. A summary plot also indicates the future detectability of DM in the Direct and Indirect searches after addressing the relevant constraints.
\appendix
\section{Parameters before EWSB}
\begin{align}
\eta_i=\begin{pmatrix} \eta^{+}_i\\\eta_i^0\end{pmatrix}\,; \quad H=\begin{pmatrix} H^{+}\\H^0\end{pmatrix}\,;\quad   \mathbb{L}=\begin{pmatrix} \nu_{\ell}\\\ell\end{pmatrix} \,.
\end{align}

\begin{align}\begin{split}
\mathcal{D}_{\mu}=\partial_{\mu}+ i g \dfrac{\tau^a}{2}W^a_{\mu}+ig^{\prime}YB_{\mu}=\partial_{\mu}+\dfrac{i}{2}\begin{pmatrix} g W_{\mu}^3+g^{\prime}B_{\mu}&\sqrt{2}gW_{\mu}^+\\\sqrt{2}gW_{\mu}^- & -g W_{\mu}^3+g^{\prime}B_{\mu} \end{pmatrix}\,.
\end{split}\end{align}

{\small
\begin{eqnarray}
\begin{split}
(\mathcal{D}^{\mu}\eta_i)^{\dagger}(\mathcal{D}_{\mu}\eta_i)=& \partial^{\mu} \eta_i^{\dagger}\partial_{\mu} \eta_i+\dfrac{1}{4}
\begin{pmatrix} \eta^{-}_i&\eta_i^{0^*}\end{pmatrix}
\begin{pmatrix} g W_{\mu}^3+g^{\prime}B_{\mu}&\sqrt{2}gW_{\mu}^+\\\sqrt{2}gW_{\mu}^- & -g W_{\mu}^3+g^{\prime}B_{\mu} 
\end{pmatrix}
\begin{pmatrix} g W_{\mu}^3+g^{\prime}B_{\mu}&\sqrt{2}gW_{\mu}^+\\\sqrt{2}gW_{\mu}^- & -g W_{\mu}^3+g^{\prime}B_{\mu} 
\end{pmatrix}
\begin{pmatrix} 
\eta^{+}_i\\\eta_i^0
\end{pmatrix}
\\&\hspace{1.4cm} -\dfrac{i}{2}\left[
\begin{pmatrix} 
\eta^{-}_i&\eta_i^{0^*}
\end{pmatrix}
\begin{pmatrix} g W_{\mu}^3+g^{\prime}B_{\mu}&\sqrt{2}gW_{\mu}^+\\\sqrt{2}gW_{\mu}^- & -g W_{\mu}^3+g^{\prime}B_{\mu} 
\end{pmatrix}
\begin{pmatrix} 
\partial^{\mu}\eta^{+}_i\\\partial^{\mu}\eta_i^0
\end{pmatrix}
-h.c.\right]
\\=&\partial^{\mu} \eta_i^{\dagger}\partial_{\mu} \eta_i+\dfrac{1}{4}\left[2g^2W^+_{\mu}W^{\mu^-}(\eta_i^+\eta_i^-+\eta_i^0\eta_i^{0^*})+2\sqrt{2}gg^{\prime}(B_{\mu}W^{\mu^-}\eta^+_i\eta^{0^*}_i+h.c.)\right]\\&
-\dfrac{i}{2}\left[\left(g W_{\mu}^3+g^{\prime}B_{\mu}\right)\left(\partial^{\mu}\eta^+_i\right)\eta^-_i +\sqrt{2}gW_{\mu}^+\left(\partial^{\mu}\eta^0_i\right)\eta_i^-+\sqrt{2}gW_{\mu}^-\left(\partial^{\mu}\eta^+_i\right)\eta_i^{0^{*}}\right.\\&\left. +\left(g^{\prime}B_{\mu}-g W_{\mu}^3\right)\left(\partial^{\mu}\eta^0_i\right)\eta^{0^*}_i -h.c.\right]+\dfrac{1}{4}\left[|g^{\prime}B_{\mu}+gW_{\mu}^3|^2\eta^+_i\eta^-_i+|g^{\prime}B_{\mu}-gW_{\mu}^3|^2\eta^0_i\eta^{0^*}_i\right]\,.
\end{split}\end{eqnarray}}

\begin{eqnarray}
\begin{split}
&\overline{\mathbb{L}}_Li\gamma^{\mu}\mathcal{D}_{\mu}\mathbb{L}_L+\overline{\ell}_Ri\gamma^{\mu}\mathcal{D}_{\mu}\ell_R\\\,=\,&\overline{\mathbb{L}}_{L}i\gamma^{\mu}\left(\partial_{\mu}+i g \dfrac{\tau^a}{2}W^a_{\mu}-i\dfrac{g^{\prime}}{2}B_{\mu}\right)\mathbb{L}_{L}+\overline{\ell}_Ri\gamma^{\mu}\left(\partial_{\mu}-ig^{\prime}B_{\mu}\right)\ell_R\\\,=\,&
\begin{pmatrix} 
\overline{\nu}_{\ell_L} & \overline{\ell}_L
\end{pmatrix}
\left(i\slashed{\partial}-\dfrac{1}{2}\gamma^{\mu}
\begin{pmatrix} 
g W_{\mu}^3-g^{\prime}B_{\mu}&\sqrt{2}gW_{\mu}^+\\\sqrt{2}gW_{\mu}^- & -g W_{\mu}^3-g^{\prime}B_{\mu} 
\end{pmatrix}\right)
\begin{pmatrix} 
\nu_{\ell_L}\\\ell_L
\end{pmatrix}
+\overline{\ell}_R\left(i\slashed{\partial}+\gamma^{\mu}g^{\prime}B_{\mu}\right)\ell_R\\
\,=\,&\overline{\ell}i\slashed{\partial}\ell-(gW_{\mu}^3-g^{\prime}B_{\mu})\overline{\nu_{\ell}}\dfrac{1}{2}\gamma^{\mu}\mathbb{P}_L\nu_{\ell} +\overline{\ell}\gamma^{\mu}\left[\dfrac{1}{2}(g W^3_{\mu}+g^{\prime}B_{\mu})\mathbb{P}_L+g^{\prime}B_{\mu}\mathbb{P}_R\right]\ell-\dfrac{g}{\sqrt{2}}\left(W_{\mu}^+\overline{\nu}_{\ell}\gamma^{\mu}\mathbb{P}_L\ell+h.c.\right)\,.
\end{split}
\end{eqnarray}

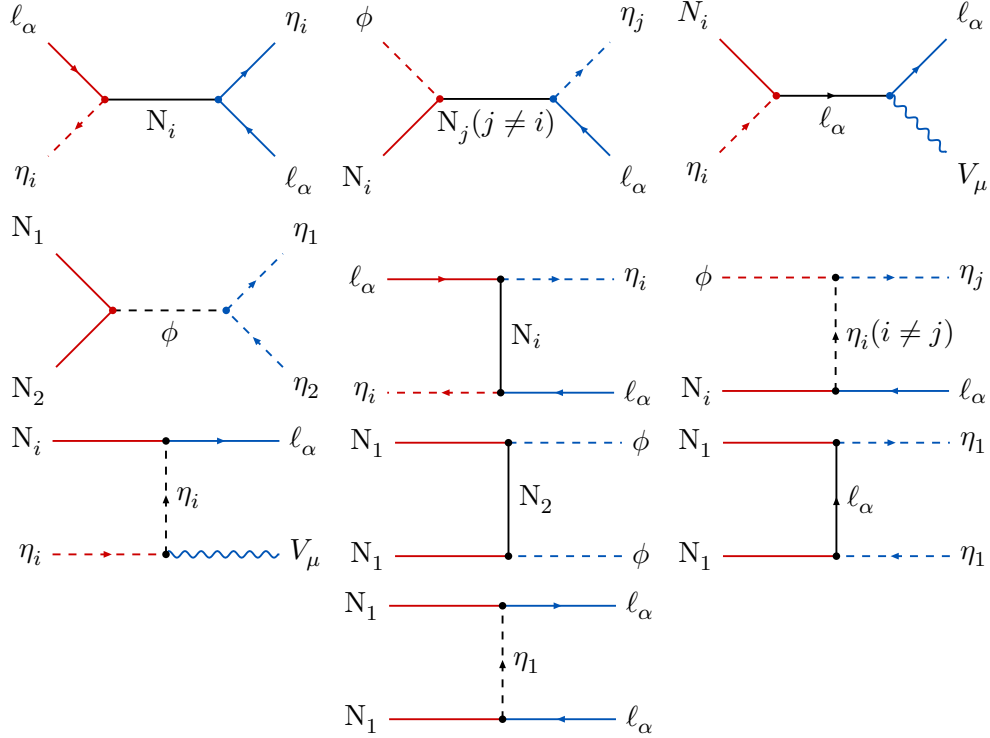
\begin{figure}[htb!]
\centering
\begin{tikzpicture}
\begin{feynman}
\vertex (a);
\vertex[above left=0.75cm and 0.75cm of a] (a1);
\vertex[below left=0.75cm and 0.75cm of a] (a2); 
\vertex[right=1.5cm of a] (b); 
\vertex[above right=0.75cm and 0.75cm of b] (c1);
\vertex[below right=0.75cm and 0.75cm of b] (c2);
\diagram* {
(a1) -- [line width=0.25mm,fermion, arrow size=0.7pt, style=bostonuniversityred] (a),
(a) -- [line width=0.25mm, charged scalar, arrow size=0.7pt, style=bostonuniversityred] (a2),
(a) -- [line width=0.25mm,plain, arrow size=0.7pt, edge label'={\(\color{black}{{\rm N}_i^{}}\)}, style=black] (b),
(c2) -- [line width=0.25mm,fermion, arrow size=0.7pt,style=mediumtealblue] (b)  -- [line width=0.25mm,fermion, arrow size=0.7pt,style=mediumtealblue] (c1)};
\node at (a)[circle,fill,style=bostonuniversityred,inner sep=1pt]{};
\node at (b)[circle,fill,style=mediumtealblue,inner sep=1pt]{};
\vertex[above left=0.75cm and 0.75cm of a] (a11){\(\ell_{\alpha}^{}\)};
\vertex[below left=0.75cm and 0.75cm of a] (a22){\(\eta_i^{}\)}; 
\vertex[above right=0.75cm and 0.75cm of b] (c11){\(\eta_i^{}\)};
\vertex[below right=0.75cm and 0.75cm of b] (c22){\(\ell_{\alpha}^{}\)};
\end{feynman}
\end{tikzpicture}
\begin{tikzpicture}
\begin{feynman}
\vertex (a);
\vertex[above left=0.75cm and 0.75cm of a] (a1);
\vertex[below left=0.75cm and 0.75cm of a] (a2); 
\vertex[right=1.5cm of a] (b); 
\vertex[above right=0.75cm and 0.75cm of b] (c1);
\vertex[below right=0.75cm and 0.75cm of b] (c2);
\diagram* {
(a1) -- [line width=0.25mm,scalar, arrow size=0.7pt, style=bostonuniversityred] (a),
(a) -- [line width=0.25mm, plain, arrow size=0.7pt, style=bostonuniversityred] (a2),
(a) -- [line width=0.25mm,plain, arrow size=0.7pt, edge label'={\(\color{black}{{\rm N}_j^{}(j\neq i)}\)}, style=black] (b),
(c2) -- [line width=0.25mm,fermion, arrow size=0.7pt,style=mediumtealblue] (b)  -- [line width=0.25mm,charged scalar, arrow size=0.7pt,style=mediumtealblue] (c1)};
\node at (a)[circle,fill,style=bostonuniversityred,inner sep=1pt]{};
\node at (b)[circle,fill,style=mediumtealblue,inner sep=1pt]{};
\vertex[above left=0.75cm and 0.75cm of a] (a11){\(\phi\)};
\vertex[below left=0.75cm and 0.75cm of a] (a22){\({\rm N}_i^{}\)}; 
\vertex[above right=0.75cm and 0.75cm of b] (c11){\(\eta_j^{}\)};
\vertex[below right=0.75cm and 0.75cm of b] (c22){\(\ell_{\alpha}^{}\)};
\end{feynman}
\end{tikzpicture}
\begin{tikzpicture}
\begin{feynman}
\vertex (a);
\vertex[above left=0.75cm and 0.75cm of a] (a1);
\vertex[below left=0.75cm and 0.75cm of a] (a2); 
\vertex[right=1.5cm of a] (b); 
\vertex[above right=0.75cm and 0.75cm of b] (c1);
\vertex[below right=0.75cm and 0.75cm of b] (c2);
\diagram* {
(a1) -- [line width=0.25mm,plain, arrow size=0.7pt, style=bostonuniversityred] (a),
(a2) -- [line width=0.25mm, charged scalar, arrow size=0.7pt, style=bostonuniversityred] (a),
(a) -- [line width=0.25mm,fermion, arrow size=0.7pt, edge label'={\(\color{black}{\ell_{\alpha}^{}}\)}, style=black] (b),
(c2) -- [line width=0.25mm,boson, arrow size=0.7pt,style=mediumtealblue] (b)  -- [line width=0.25mm,fermion, arrow size=0.7pt,style=mediumtealblue] (c1)};
\node at (a)[circle,fill,style=bostonuniversityred,inner sep=1pt]{};
\node at (b)[circle,fill,style=mediumtealblue,inner sep=1pt]{};
\vertex[above left=0.75cm and 0.75cm of a] (a11){\({N}_i^{}\)};
\vertex[below left=0.75cm and 0.75cm of a] (a22){\(\eta_i^{}\)}; 
\vertex[above right=0.75cm and 0.75cm of b] (c11){\(\ell_{\alpha}^{}\)};
\vertex[below right=0.75cm and 0.75cm of b] (c22){\(V_{\mu}^{}\)};
\end{feynman}
\end{tikzpicture}
\begin{tikzpicture}
\begin{feynman}
\vertex (a);
\vertex[above left=0.75cm and 0.75cm of a] (a1);
\vertex[below left=0.75cm and 0.75cm of a] (a2); 
\vertex[right=1.5cm of a] (b); 
\vertex[above right=0.75cm and 0.75cm of b] (c1);
\vertex[below right=0.75cm and 0.75cm of b] (c2);
\diagram* {
(a1) -- [line width=0.25mm,plain, arrow size=0.7pt, style=bostonuniversityred] (a),
(a2) -- [line width=0.25mm, plain, arrow size=0.7pt, style=bostonuniversityred] (a),
(a) -- [line width=0.25mm,scalar, arrow size=0.7pt, edge label'={\(\phi\)}, style=black] (b),
(c2) -- [line width=0.25mm, charged scalar, arrow size=0.7pt,style=mediumtealblue] (b)  -- [line width=0.25mm, charged scalar, arrow size=0.7pt,style=mediumtealblue] (c1)};
\node at (a)[circle,fill,style=bostonuniversityred,inner sep=1pt]{};
\node at (b)[circle,fill,style=mediumtealblue,inner sep=1pt]{};
\vertex[above left=0.75cm and 0.75cm of a] (a11){\({\rm N}_1^{}\)};
\vertex[below left=0.75cm and 0.75cm of a] (a22){\({\rm N}_2^{}\)}; 
\vertex[above right=0.75cm and 0.75cm of b] (c11){\(\eta_1^{}\)};
\vertex[below right=0.75cm and 0.75cm of b] (c22){\(\eta_2^{}\)};
\end{feynman}
\end{tikzpicture}
\begin{tikzpicture}
\begin{feynman}
\vertex (a);
\vertex[left=1.5cm of a] (a1);
\vertex[right=1.5cm of a] (a2); 
\vertex[below=1.5cm of a] (b); 
\vertex[below=1.5cm of a] (c); 
\vertex[left=1.5cm of b] (c1);
\vertex[right=1.5cm of b] (c2);
\diagram* {
(a1) -- [line width=0.25mm, fermion, arrow size=0.7pt, style=bostonuniversityred] (a),
(a) -- [line width=0.25mm, charged scalar, arrow size=0.7pt, style=mediumtealblue] (a2),
(a) -- [line width=0.25mm, plain, arrow size=0.7pt, edge label={\({\rm N}_i\)}, style=black] (b),
(c2) -- [line width=0.25mm, fermion, arrow size=0.7pt, style=mediumtealblue] (b),
(b)-- [line width=0.25mm, charged scalar, arrow size=0.7pt, style=bostonuniversityred ] (c1)};
\node at (a)[circle,fill,style=black,inner sep=1pt]{};
\node at (b)[circle,fill,style=black,inner sep=1pt]{};
\vertex[left=1.5cm of a] (a11){\(\ell_{\alpha}^{}\)};
\vertex[right=1.5cm of a] (a22){\(\eta_i^{}\)}; 
\vertex[left=1.5cm of b] (c11){\(\eta_i^{}\)};
\vertex[right=1.5cm of b] (c22){\(\ell_{\alpha}\)};
\end{feynman}
\end{tikzpicture}
\begin{tikzpicture}
\begin{feynman}
\vertex (a);
\vertex[left=1.5cm of a] (a1);
\vertex[right=1.5cm of a] (a2); 
\vertex[below=1.5cm of a] (b); 
\vertex[below=1.5cm of a] (c); 
\vertex[left=1.5cm of b] (c1);
\vertex[right=1.5cm of b] (c2);
\diagram* {
(a1) -- [line width=0.25mm, scalar, arrow size=0.7pt, style=bostonuniversityred] (a),
(a) -- [line width=0.25mm, charged scalar, arrow size=0.7pt, style=mediumtealblue] (a2),
(b) -- [line width=0.25mm, charged scalar, arrow size=0.7pt, edge label'={\( \eta_i^{}(i\neq j)\)}, style=black] (a),
(c2) -- [line width=0.25mm, fermion, arrow size=0.7pt, style=mediumtealblue] (b),
(b)-- [line width=0.25mm, plain, arrow size=0.7pt, style=bostonuniversityred ] (c1)};
\node at (a)[circle,fill,style=black,inner sep=1pt]{};
\node at (b)[circle,fill,style=black,inner sep=1pt]{};
\vertex[left=1.5cm of a] (a11){\(\phi\)};
\vertex[right=1.5cm of a] (a22){\(\eta_j^{}\)}; 
\vertex[left=1.5cm of b] (c11){\({\rm N}_i^{}\)};
\vertex[right=1.5cm of b] (c22){\(\ell_{\alpha}\)};
\end{feynman}
\end{tikzpicture}
\begin{tikzpicture}
\begin{feynman}
\vertex (a);
\vertex[left=1.5cm of a] (a1);
\vertex[right=1.5cm of a] (a2); 
\vertex[below=1.5cm of a] (b); 
\vertex[below=1.5cm of a] (c); 
\vertex[left=1.5cm of b] (c1);
\vertex[right=1.5cm of b] (c2);
\diagram* {
(a1) -- [line width=0.25mm, plain, arrow size=0.7pt, style=bostonuniversityred] (a),
(a) -- [line width=0.25mm, fermion, arrow size=0.7pt, style=mediumtealblue] (a2),
(b) -- [line width=0.25mm, charged scalar, arrow size=0.7pt, edge label'={\( \eta_i^{}\)}, style=black] (a),
(c2) -- [line width=0.25mm, boson, arrow size=0.7pt, style=mediumtealblue] (b),
(c1)-- [line width=0.25mm, charged scalar, arrow size=0.7pt, style=bostonuniversityred ] (b)};
\node at (a)[circle,fill,style=black,inner sep=1pt]{};
\node at (b)[circle,fill,style=black,inner sep=1pt]{};
\vertex[left=1.5cm of a] (a11){\({\rm N}_i^{}\)};
\vertex[right=1.5cm of a] (a22){\(\ell_{\alpha}^{}\)}; 
\vertex[left=1.5cm of b] (c11){\({\rm \eta}_i^{}\)};
\vertex[right=1.5cm of b] (c22){\(V_{\mu}^{}\)};
\end{feynman}
\end{tikzpicture}
\begin{tikzpicture}
\begin{feynman}
\vertex (a);
\vertex[left=1.5cm of a] (a1);
\vertex[right=1.5cm of a] (a2); 
\vertex[below=1.5cm of a] (b); 
\vertex[below=1.5cm of a] (c); 
\vertex[left=1.5cm of b] (c1);
\vertex[right=1.5cm of b] (c2);
\diagram* {
(a1) -- [line width=0.25mm, plain, arrow size=0.7pt, style=bostonuniversityred] (a),
(a) -- [line width=0.25mm, scalar, arrow size=0.7pt, style=mediumtealblue] (a2),
(b) -- [line width=0.25mm, plain, arrow size=0.7pt, edge label'={\( {\rm N}_2^{}\)}, style=black] (a),
(c2) -- [line width=0.25mm, scalar, arrow size=0.7pt, style=mediumtealblue] (b),
(c1)-- [line width=0.25mm, plain, arrow size=0.7pt, style=bostonuniversityred ] (b)};
\node at (a)[circle,fill,style=black,inner sep=1pt]{};
\node at (b)[circle,fill,style=black,inner sep=1pt]{};
\vertex[left=1.5cm of a] (a11){\({\rm N}_1^{}\)};
\vertex[right=1.5cm of a] (a22){\(\phi\)}; 
\vertex[left=1.5cm of b] (c11){\({\rm N}_1^{}\)};
\vertex[right=1.5cm of b] (c22){\(\phi\)};
\end{feynman}
\end{tikzpicture}
\begin{tikzpicture}
\begin{feynman}
\vertex (a);
\vertex[left=1.5cm of a] (a1);
\vertex[right=1.5cm of a] (a2); 
\vertex[below=1.5cm of a] (b); 
\vertex[below=1.5cm of a] (c); 
\vertex[left=1.5cm of b] (c1);
\vertex[right=1.5cm of b] (c2);
\diagram* {
(a1) -- [line width=0.25mm, plain, arrow size=0.7pt, style=bostonuniversityred] (a),
(a) -- [line width=0.25mm, charged scalar, arrow size=0.7pt, style=mediumtealblue] (a2),
(b) -- [line width=0.25mm, fermion, arrow size=0.7pt, edge label'={\( {\rm \ell}_{\alpha}^{}\)}, style=black] (a),
(c2) -- [line width=0.25mm, charged scalar, arrow size=0.7pt, style=mediumtealblue] (b),
(c1)-- [line width=0.25mm, plain, arrow size=0.7pt, style=bostonuniversityred ] (b)};
\node at (a)[circle,fill,style=black,inner sep=1pt]{};
\node at (b)[circle,fill,style=black,inner sep=1pt]{};
\vertex[left=1.5cm of a] (a11){\({\rm N}_1^{}\)};
\vertex[right=1.5cm of a] (a22){\(\eta_1^{}\)}; 
\vertex[left=1.5cm of b] (c11){\({\rm N}_1^{}\)};
\vertex[right=1.5cm of b] (c22){\(\eta_1^{}\)};
\end{feynman}
\end{tikzpicture}
\begin{tikzpicture}
\begin{feynman}
\vertex (a);
\vertex[left=1.5cm of a] (a1);
\vertex[right=1.5cm of a] (a2); 
\vertex[below=1.5cm of a] (b); 
\vertex[below=1.5cm of a] (c); 
\vertex[left=1.5cm of b] (c1);
\vertex[right=1.5cm of b] (c2);
\diagram* {
(a1) -- [line width=0.25mm, plain, arrow size=0.7pt, style=bostonuniversityred] (a),
(a) -- [line width=0.25mm, fermion, arrow size=0.7pt, style=mediumtealblue] (a2),
(b) -- [line width=0.25mm, charged scalar, arrow size=0.7pt, edge label'={\( {\rm \eta}_1^{}\)}, style=black] (a),
(c2) -- [line width=0.25mm, fermion, arrow size=0.7pt, style=mediumtealblue] (b),
(c1)-- [line width=0.25mm, plain, arrow size=0.7pt, style=bostonuniversityred ] (b)};
\node at (a)[circle,fill,style=black,inner sep=1pt]{};
\node at (b)[circle,fill,style=black,inner sep=1pt]{};
\vertex[left=1.5cm of a] (a11){\({\rm N}_1^{}\)};
\vertex[right=1.5cm of a] (a22){\(\ell_{\alpha}^{}\)}; 
\vertex[left=1.5cm of b] (c11){\({\rm N}_1^{}\)};
\vertex[right=1.5cm of b] (c22){\(\ell_{\alpha}^{}\)};
\end{feynman}
\end{tikzpicture}
\caption{The relevant Feynman diagrams for Leptogenesis where $i=1,2~;~V_{\mu}=B_{\mu},~W^3_{\mu},~W_{\mu}^{\pm}$ \cite{Pilaftsis:2003gt, Giudice:2003jh} and $\alpha$ define the lepton generation.}
\label{fig:lepto-scattering}
\end{figure}
\section{Neutrino mass generation}
\label{sec:Numass}
\begin{figure}[htb!]
\centering
\begin{tikzpicture}
\begin{feynman}
\vertex (a){\color{black}{$\nu_{\alpha}$}};
\vertex [right = 1.5cm of a] (b);
\vertex [right = 3cm of b] (c);
\vertex [right = 1.5cm of c] (d){\color{black}{$\nu_{\beta}^c$}};
\vertex [ above right = 1.5cm and 1.5cm of b] (e);
\vertex [ above left = 1.5cm and 1.5cm of e] (f){\color{black}{H}};
\vertex [ above right = 1.5cm and 1.5cm of e] (g){\color{black}{H}};
\diagram*{ 
(a) -- [line width=0.25mm, fermion, arrow size=1.2pt, style=bostonuniversityred, ultra thick] (b),
(b) -- [line width=0.25mm, plain,arrow size=1.2pt, style=black, ultra thick,edge label={$\color{black}{{\rm N}_k }$}] (c),
(d) -- [line width=0.25mm, fermion, arrow size=1.2pt, style=mediumtealblue, ultra thick] (c), 
(e) -- [line width=0.25mm, charged scalar, quarter right, style=black, ultra thick, arrow size=1.2pt,edge label'={$\color{black}{\eta_k^0 }$}] (b),
(e) -- [line width=0.25mm, charged scalar, quarter left, style=black, ultra thick, arrow size=1.2pt,edge label={$\color{black}{\eta_k^0 }$}] (c),
(f) -- [line width=0.25mm, charged scalar, style=gray, ultra thick, arrow size=1.2pt] (e),
(g) -- [line width=0.25mm, charged scalar,  style=gray, ultra thick, arrow size=1.2pt] (e)};
\node at (b)[circle,fill,style=black, inner sep=1pt]{};
\node at (c)[circle,fill,style=black, inner sep=1pt]{};
\node at (e)[circle,fill,style=gray, inner sep=1pt]{};
\end{feynman}
\end{tikzpicture}
\begin{tikzpicture}
\begin{feynman}
\vertex (a){\color{black}{$\nu_{\alpha}$}};
\vertex [right = 1.5cm of a] (b);
\vertex [right = 3cm of b] (c);
\vertex [right = 1.5cm of c] (d){\color{black}{$\nu_{\beta}^c$}};
\vertex [ above right = 1.5cm and 1.5cm of b] (e);
\vertex [ above right = 1.5cm and 0.7cm of b] (o){$\color{black}{\eta_{kR}^0(\eta_{kI}^0) }$};
\vertex [ above left = 1.5cm and 1.5cm of e] (f){\color{white}{H}};
\vertex [ above right = 1.5cm and 1.5cm of e] (g){\color{white}{H}};
\diagram*{
(a) -- [line width=0.25mm,fermion, arrow size=1.2pt, style=bostonuniversityred, ultra thick] (b),
(b) -- [line width=0.25mm, plain,style=black, ultra thick,edge label={$\color{black}{{\rm N}_k^{} }$}] (c),
(d) -- [line width=0.25mm,fermion, arrow size=1.2pt, style=mediumtealblue, ultra thick] (c), 
(e) -- [line width=0.25mm,scalar, quarter right, style=black, ultra thick] (b),
(e) -- [line width=0.25mm,scalar, quarter left, style=black, ultra thick, arrow size=1.2pt] (c)};
\node at (b)[circle,fill,style=black, inner sep=1pt]{};
\node at (c)[circle,fill,style=black, inner sep=1pt]{};
\end{feynman}
\end{tikzpicture}
\caption{Radiative neutrino majorana mass generation where $i,j$ are generation indices and $k=1,2$. The left and right figures correspond to before and after EWSB, respectively.}
\label{fig:NuMass}
\end{figure}
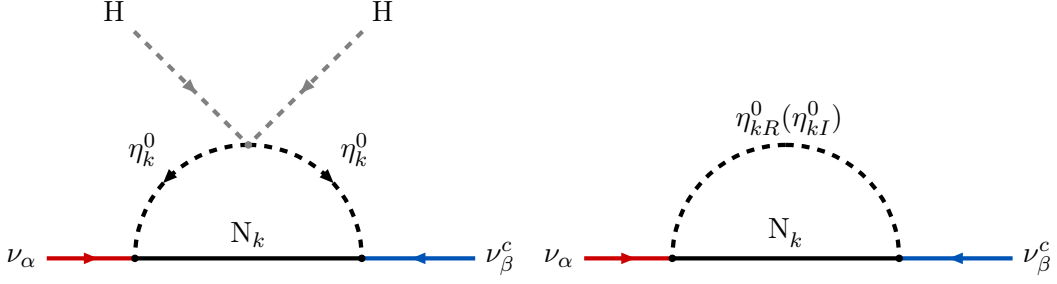

In this scenario, neutrino mass is generated via a one-loop radiative diagram, as shown in fig\,.~\ref{fig:NuMass}.
and the 1-loop correction due to this self-energy graph is written as,
\begin{align}
\overline{u}_{L_{\nu_{\alpha}}}(p)\Sigma_{\alpha\beta}(p)u^c_{L_{\nu_{\beta}}}(p)=\overline{u}_{L_{\nu_{\alpha}}}(p)\sum_{k=1,2}\left(\Sigma_{\alpha\beta}(p,\eta_{kR}^0)+\Sigma_{\alpha\beta}(p,\eta_{kI}^0)\right)u^c_{L_{\nu_{\beta}}}(p)\,.
\end{align}

\begin{align}
-i\Sigma_{\alpha\beta}\big(p,\eta^0_{kR(I)}\big)&=-C^{\alpha k}_{R(I)}\int \dfrac{d^4l}{(2\pi)^4}\dfrac{i}{(p-l)^2-m_{\eta^0_{kR(I)}}^2}\dfrac{i(\slashed{l}+M_{k})}{l^2-M_{k}^2}C^{\beta k}_{R(I)}\,.
\label{eq:eq7}
\end{align}
Here $C^{\alpha k}_{R(I)}$ and $C^{\beta k}_{R(I)}$ are the couplings, given by
\begin{align}
C^{\alpha k}_{R(I)}=\frac{f_{R(I)}}{\sqrt{2}}h^*_{kk\alpha }\quad \text{and} \quad C^{\beta k}_{R(I)}=\frac{f_{R(I)}}{\sqrt{2}}h^*_{kk\beta }\,,   
\end{align}
with $f_{R}=1$ and $f_{I}=i$.
\begin{align}
\mathbb{M}^k_{\alpha\beta}&=\Sigma_{\alpha\beta}(0,\eta_{kR}^0)+\Sigma_{\alpha\beta}(0,\eta_{kI}^0)\,.
\end{align}
Eq\,.~\ref{eq:eq7} is also valid for $p=0$, so we set this limit during mass matrix calculation. The odd power of $l$ should vanish after integration.
\begin{align}
\nonumber \Sigma_{\alpha\beta}(0,\eta_{kR(I)}^0)&=i C^{\alpha k}_{R(I)}\int \dfrac{d^4l}{(2\pi)^4}\dfrac{1}{l^2-m_{\eta^0_{kR(I)}}^2}\dfrac{(\slashed{l}+M_{k})}{l^2-M_{k}^2}C^{\beta k}_{R(I)}\\ &\equiv 
i C^{\alpha k}_{R(I)}C^{\beta k}_{R(I)}\int \dfrac{d^4l}{(2\pi)^4}\dfrac{1}{l^2-m_{\eta^0_{kR(I)}}^2}\dfrac{M_{k}}{l^2-M_{k}^2}\,;
\end{align}
\begin{align}
\mathbb{M}_{\alpha\beta}&=\sum_{k=1,2}\mathbb{M}^k_{\alpha\beta}=\sum_{k=1,2}\Sigma_{\alpha\beta}(0,\eta_{kR}^0)+\Sigma_{\alpha\beta}(0,\eta_{kI}^0)\,;
\end{align}
\begin{align}
\nonumber \mathbb{M}^k_{\alpha\beta}&=i\dfrac{h^*_{kk\alpha}h^*_{kk\beta}M_{k}}{2}\int \dfrac{d^4l}{(2\pi)^4}\Bigg(\dfrac{1}{\big(l^2-m_{\eta^0_{kR}}^2\big)(l^2-M_{k}^2)}-\dfrac{1}{\big(l^2-m_{\eta^0_{kI}}^2\big)(l^2-M_{k}^2)}\Bigg)\\\nonumber&
=h^*_{kk\alpha}\Bigg[\dfrac{M_{k}}{32\pi^2}\Bigg(\dfrac{m_{\eta^0_{kR}}^2}{m_{\eta^0_{kR}}^2-M^2_{k}}\ln{\dfrac{m_{\eta^0_{kR}}^2}{M^2_{k}}}-\dfrac{m_{\eta^0_{kI}}^2}{m_{\eta^0_{kI}}^2-M^2_{k}}\ln{\dfrac{m_{\eta^0_{kI}}^2}{M^2_{k}}}\Bigg)\Bigg]h^*_{kk\beta}\,.
\end{align}
\noindent The neutrino mass generated from the diagrams shown in fig\,.~\ref{fig:NuMass} is given by 

\begin{eqnarray}
\left(  \mathcal{M}  \right)_{\alpha \beta} & = & \sum_{i=1}^{2} \dfrac{h^{*}_{\alpha i} h^{*}_{\beta i} M_{i}}{32 \pi^{2}}  \left[  L(m_{\eta_{iR}}^{2} )  -L(m_{\eta_{iI}}^{2})\right],
\label{eq:Numass}
\end{eqnarray}

\noindent where the function $L$ has the following form 
\begin{eqnarray}
L(m^{2}) & = & \dfrac{m^{2}}{m^{2}-M_{i}^{2}} ln \left( \dfrac{m^{2}}{M_{i}^{2}} \right). 
\end{eqnarray}

\begin{align}
\nonumber\mathbb{M}_{\alpha\beta}&=\sum_{k}h_{kk\alpha }^*(\Lambda_{kk})^{-1} h^*_{kk\beta}
\\\nonumber&=h_{11\alpha}^*(\Lambda_{11})^{-1} h^*_{11\beta}+h_{22\alpha}^*(\Lambda_{22})^{-1} h^*_{22\beta}
\\&=
\begin{pmatrix}
h^*_{11\alpha} & h^*_{22\alpha}
\end{pmatrix} 
\begin{pmatrix}
\dfrac{1}{\Lambda_{11}} & 0 \\
0 & \dfrac{1}{\Lambda_{22}}
\end{pmatrix}
\begin{pmatrix}
h^*_{11\beta} \\ h^*_{22\beta}
\end{pmatrix} \,;
\end{align}

\begin{align}
\mathbb{M}=\begin{pmatrix}
h^*_{111} & h^*_{221}\\
h^*_{112} & h^*_{222}\\
h^*_{113} & h^*_{223}
\end{pmatrix} 
\begin{pmatrix}
\dfrac{1}{\Lambda_{11}} & 0 \\
0 & \dfrac{1}{\Lambda_{22}}
\end{pmatrix}
\begin{pmatrix}
h^*_{111} & h^*_{112}& h^*_{113}\\ h^*_{221}& h^*_{222}& h^*_{223}
\end{pmatrix} \equiv h^* \Lambda^{-1} (h^*)^T=h^* \Lambda^{-1} h^{\dagger}\,,
\end{align}
where 
\begin{eqnarray}
h= 
\begin{pmatrix}
h_{111} & h_{221} \\
h_{112} & h_{222} \\
h_{113} & h_{223} 
\end{pmatrix} ~{\rm and} ~
\Lambda^{-1}=
\begin{pmatrix}
\dfrac{1}{\Lambda_{11}} & 0 \\
0 & \dfrac{1}{\Lambda_{22}}
\end{pmatrix}\,,
\end{eqnarray}
\noindent and 

\begin{eqnarray}
\Lambda_{ii} & = & \dfrac{4\pi^{2}}{m_{\eta_{Ri}}^{2}-m_{\eta_{Ii}}^{2}}  \xi_{i} M_{i} = \dfrac{4\pi^{2}}{\lambda^{\prime\prime}_{iiH}}  \xi_{i} \dfrac{M_{i}}{v^2}.
\end{eqnarray}

\noindent The loop functions $\xi_{i}$ are given by

\begin{eqnarray}
\xi_{i} & = & \bigg( \dfrac{1}{8} \dfrac{M_{i}^{2}}{m_{\eta_{Ri}}^{2}-m_{\eta_{Ii}}^{2}} \left[L(m_{\eta_{Ri}}^{2}) -L(m_{\eta_{Ii}}^{2})\right]  \bigg)^{-1}\,.
\end{eqnarray}
Now, the light neutrino mass is diagonalized using the usual PMNS matrix $\rm U $, with Majorana and Dirac phases, which is determined from neutrino oscillation data and
$$\rm \mathbb{M}_{\nu}=diag(m_1,m_2,m_3)=U^{\dagger} \mathbb{M} U^*=U^{\dagger}h^* \sqrt{\Lambda^{-1}}\sqrt{\Lambda^{-1}} h^{\dagger}  U^*\,,$$
$$\rm \mathbb{M}_{\sqrt{\nu}}\mathbb{M}_{\sqrt{\nu}}=U^{\dagger}h^* \sqrt{\Lambda^{-1}}\sqrt{\Lambda^{-1}} h^{\dagger}  U^*\,,$$
$$\rm \mathbb{I}=\mathbb{M}_{\sqrt{\nu^{-1}}}U^{\dagger}h^* \sqrt{\Lambda^{-1}}\sqrt{\Lambda^{-1}} h^{\dagger}  U^*\mathbb{M}_{\sqrt{\nu^{-1}}}=\left[\sqrt{\Lambda^{-1}} h^{\dagger}  U^*\mathbb{M}_{\sqrt{\nu^{-1}}}\right]^T\left[\sqrt{\Lambda^{-1}} h^{\dagger}  U^*\mathbb{M}_{\sqrt{\nu^{-1}}}\right]\equiv \mathbb{R}^T\mathbb{R}\,.$$
where $\mathbb{R}$ is any $2 \times 3$ orthogonal matrix. Then, the Yuakawa coupling matrix satisfying the neutrino data can be written as,

\begin{equation}
h=U^*\mathbb{M}_{\sqrt{\nu}}\mathbb{R}^{\dagger}\sqrt{\Lambda}
\label{eq:CI}
\end{equation}

where \cite{Antusch:2011nz},
\begin{eqnarray}
\mathbb{M}_{\sqrt{\nu}}= 
\begin{pmatrix}
\sqrt{m_1} & 0 & 0 \\
0 & \sqrt{m_2} & 0 \\
0 & 0 & \sqrt{m_3} \\
\end{pmatrix} ~{\rm and} ~
\sqrt{\Lambda}=
\begin{pmatrix}
\sqrt{\Lambda_{11}} & 0 \\
0 & \sqrt{\Lambda_{22}} 
\end{pmatrix}\\
U=\begin{pmatrix}
c_{12}c_{13} & s_{12}c_{13}  & s_{13}e^{-i\delta} \\
-s_{12}c_{23}-c_{12}s_{23}s_{13}e^{i\delta} & c_{12}c_{23}-s_{12}s_{23}s_{13}e^{i\delta} & s_{23}c_{13} \\
s_{12}s_{23}-c_{12}c_{23}s_{13}e^{i\delta} & -c_{12}s_{23}-s_{12}c_{23}s_{13}e^{i\delta} & c_{23}c_{13} \\
\end{pmatrix}
\end{eqnarray}

Neutrino oscillation experiments only measure two neutrino mass squared differences \cite{Esteban:2018azc, Esteban:2020cvm}.
$$
\begin{rcases}
m_1=0 \\
m_2=\sqrt{\Delta m_{21}^2} \\
m_3=\sqrt{\Delta m_{31}^2}
\end{rcases}
\quad
\begin{tabular}{l}
NO
\end{tabular}\quad
\left\{\begin{array}{lr}
\theta_{12}/^{\circ}=33.44,~\theta_{23}/^{\circ}=49.2,~\theta_{13}/^{\circ}=8.57,~\delta_{CP}=197\\
\Delta m_{21}^2=7.42\times\rm 10^{-5}~eV^2\\
\Delta m_{31}^2=2.517\times\rm 10^{-3}~eV^2
\end{array}\right.
$$

$$
\begin{rcases}
m_1=\sqrt{-\Delta m_{32}^2-\Delta m_{21}^2} \\
m_2=\sqrt{-\Delta m_{32}^2} \\
m_3=0
\end{rcases}
\quad
\begin{tabular}{l}
IO
\end{tabular}\quad
\left\{\begin{array}{lr}
\theta_{12}/^{\circ}=33.45,~\theta_{23}/^{\circ}=49.3,~\theta_{13}/^{\circ}=8.60,~\delta_{CP}=282\\
\Delta m_{21}^2=7.42\times\rm 10^{-5}~eV^2\\
\Delta m_{32}^2=-2.498\times\rm 10^{-3}~eV^2
\end{array}\right.
$$

$$\mathbb{R}^{\rm NO}= 
\begin{pmatrix}
0 & \cos {\rm z} & \sin {\rm z} \\
0 & -\sin {\rm z} & \cos {\rm z} \\
\end{pmatrix} ~{\rm and}~ \mathbb{R}^{\rm IO}= 
\begin{pmatrix}
\cos {\rm z} & \sin {\rm z} & 0 \\
-\sin {\rm z} & \cos {\rm z} & 0 \\
\end{pmatrix}$$
where ${\rm z}=a+i~b$ and \{$a,~b$\} are our free parameters. Also, one can find the relations 
\begin{eqnarray}
\sum\limits_{\alpha=1}^3 h_{11\alpha}h^*_{11\alpha}=Tr(h^{\prime}h^{\dagger})\quad\text{and} \quad \sum\limits_{\alpha=1}^3 h_{22\alpha}h^*_{11\alpha}=Tr(h^{\prime\prime}h^{\dagger}), 
\end{eqnarray}
where 
\begin{eqnarray}
h^{\prime}= 
\begin{pmatrix}
h_{111} & h_{221} \\
h_{112} & h_{222} \\
h_{113} & h_{223} 
\end{pmatrix}
\begin{pmatrix}
1 & 0\\
0 & 0\\
\end{pmatrix}=
\begin{pmatrix}
h_{111} & 0 \\
h_{112} & 0 \\
h_{113} & 0 
\end{pmatrix}
~{\rm and} ~
h^{''}= 
\begin{pmatrix}
h_{111} & h_{221} \\
h_{112} & h_{222} \\
h_{113} & h_{223} 
\end{pmatrix}
\begin{pmatrix}
0 & 0\\
1 & 0\\
\end{pmatrix}=
\begin{pmatrix}
h_{221} & 0 \\
h_{222} & 0\\
h_{223} & 0 
\end{pmatrix}
\end{eqnarray}
\section{CP-asymmetry calculation}
\label{app:asymmetry}
\begin{figure}[htb!]
\centering
\subfloat[]{\begin{tikzpicture}
\begin{feynman}
\vertex (a){\(\rm\bf\color{black}{N_1 }\)};
\vertex [right=3cm of a] (b);
\vertex [above right=2cm and 2cm of b] (c){\(\rm\bf\color{black}{\ell_{\alpha} }\)};
\vertex [below right=2cm and 2cm of b] (d){\(\rm\bf\color{black}{\eta_1 }\)};
\diagram* {
(a) -- [line width=0.25mm,plain, arrow size=1.2pt, style=bostonuniversityred,ultra thick,momentum={[arrow style=cyan,label style=black]\(p\)}] (b),
(b)-- [line width=0.25mm,fermion, arrow size=1.2pt, style=mediumtealblue,ultra thick,momentum={[arrow style=cyan,label style=black]\(p_1\)}] (c),
(b) -- [line width=0.25mm,charged scalar,  arrow size=1.2pt, style=mediumtealblue,ultra thick,momentum'={[arrow style=cyan,label style=black]\(p_2\)}] (d)};
\node at (b)[circle,fill,style=mediumtealblue,inner sep=1pt]{};
\end{feynman}
\end{tikzpicture}}\quad
\subfloat[]{\begin{tikzpicture}
\begin{feynman}
\vertex (a){\(\rm\bf\color{black}{N_1 }\)};
\vertex [right=3cm of a] (b);
\vertex [above right=2cm and 2cm of b] (c);
\vertex [below right=2cm and 2cm of b] (d);
\vertex [right=2cm of c] (e){\color{black}{\(\ell_{\alpha}\)}};
\vertex [right=2cm of d] (f){\color{black}{\(\eta_1\)}};
\diagram* {(a) -- [line width=0.25mm,plain,arrow size=1.2pt, style=bostonuniversityred,ultra thick,momentum={[arrow style=cyan,label style=black]\(p\)}] (b),
(b)-- [line width=0.25mm, plain, arrow size=1.2pt, style=black,ultra thick, edge label'={\(\rm\bf\color{black}{N_{2} }\)},momentum={[arrow style=cyan,label style=black]\(q_2\)}] (c),
(d) -- [line width=0.25mm,scalar, style=black, ultra thick,arrow size=1.2pt,edge label'={\(\rm\color{black}{\phi }\)},momentum={[arrow style=cyan,label style=black]\(q_1\)} ] (b),
(c) -- [line width=0.25mm,charged scalar, style=black, ultra thick, arrow size=1.2pt, edge label'={\(\rm\bf\color{black}{\eta_2}\)},momentum={[arrow style=cyan,label style=black]\(q_3\)}] (d),
(c) -- [line width=0.25mm,fermion,  arrow size=1.2pt, style=mediumtealblue,ultra thick,momentum={[arrow style=cyan,label style=black]\(p_1\)}] (e),
(d) -- [line width=0.25mm,charged scalar, style=mediumtealblue, ultra thick,arrow size=1.2pt,momentum'={[arrow style=cyan,label style=black]\(p_2\)} ] (f)};
\node at (b)[circle,fill,style=black,inner sep=1pt]{};
\node at (c)[circle,fill,style=black,inner sep=1pt]{};
\node at (d)[circle,fill,style=black,inner sep=1pt]{};
\end{feynman}
\end{tikzpicture}}
\caption{Feynman diagrams corresponds to $N_1\to \ell_{\alpha} {\eta}_1$ is relevant for Leptogenesis.}
\label{fig:feynman-lepto}
\end{figure}
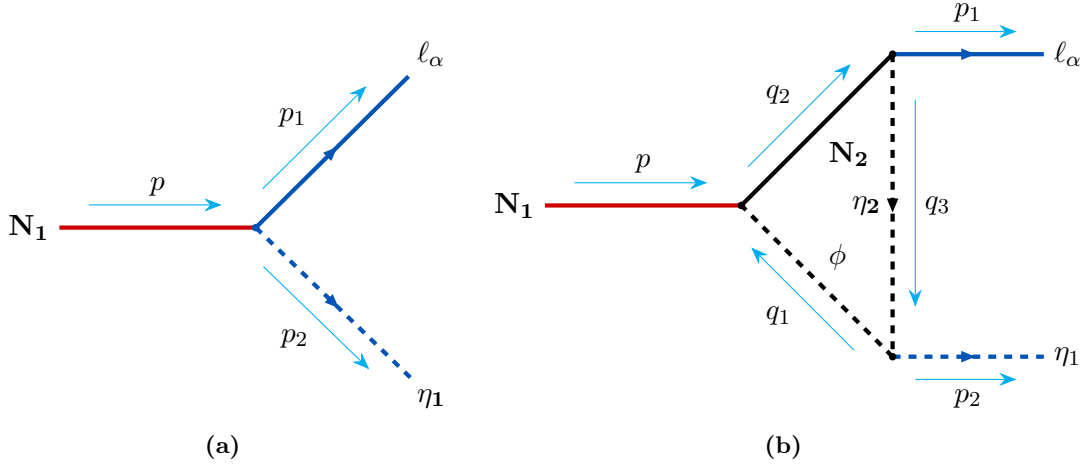
The vertex factors for the processes in fig\,.~\ref{fig:feynman-lepto} are,

\begin{gather}
N_j\to \ell_{\alpha} {\eta}_j:~-ih_{jj\alpha}P_R;\\
N_2\to N_1\phi:~iy_{12\phi}C^{\dagger};\quad\eta_2\to\eta_1\phi:~-i\mu_{12\phi}\,.
\end{gather}

\begin{align}
\Gamma_{N_1\to \ell_{\alpha}\overline{\eta}_1}=\Gamma_{N_1\to \overline{\ell}_{\alpha}\eta_1}={\color{black}2}\dfrac{h_{11\alpha}h^*_{11\alpha}}{32\pi m_{N_1}^3}(m_{N_1}^2+m_{\alpha}^2-m_{\eta_1}^2)\sqrt{(m_{N_1}^2-m_{\alpha}^2+m_{\eta_1}^2)^2-4m_{\eta_1}^2m_{N_1}^2}\,.
\end{align}
In the above equation, the factor {\color{black}2} arises due to the presence of two potential decay channels, $N_1\rightarrow \nu\eta^0_1$ and $N_1\rightarrow l^-\eta^+_1$.\\
The cross term which related to CP asymmetry parameter $\varepsilon=\dfrac{2 \int\limits_{\rm phase~space} Im[\mathcal{M}_{tree}\mathcal{M}_{loop}^{\dagger}]}{\Gamma_{N_1}}$
\begin{align}
\nonumber    {\rm \mathcal{I}_{ vertex}}&=\tiny{\int \dfrac{d^4q_2}{(2\pi)^4}\overline{u}_{\alpha}(-ih_{22\alpha}P_R)\dfrac{-i(\slashed{q_2}+m_{N_2})C}{q_2^2-m_{N_2}^2+i\epsilon}(iy_{12\phi}C^{\dagger})\dfrac{u_{N_1}^c(-i\mu_{12\phi })}{q_3^2-m_{\eta_2}^2+i\epsilon}\dfrac{1}{q_1^2-m_{\phi}^2+i\epsilon}\underbrace{\left[\overline{u}_{\alpha}(-ih_{11\alpha}P_R)u_{N_1}^c\right]^{\dagger}}_{\mathcal{M}_{tree}^{\dagger}}}\\
&=(-ih_{22\alpha}y_{12\phi}\mu_{12\phi}h_{11\alpha}^*)\int \dfrac{d^4q_2}{(2\pi)^4}\dfrac{\overline{u}_{\alpha}P_R(\slashed{q_2}+m_{N_2})}{q_2^2-m_{N_2}^2+i\epsilon}\dfrac{u_{N_1}^c}{q_3^2-m_{\eta_2}^2+i\epsilon}\dfrac{(u_{N_1}^TC^{\dagger}P_Lu_{\alpha})}{q_1^2-m_{\phi}^2+i\epsilon}
\end{align}
After summing over final spins and averaging over initial spins, we get,
\begin{align}
\nonumber  {\rm \mathcal{I}_{ vertex}}&=\tiny{(-ih_{22\alpha}y_{12\phi}\mu_{12\phi}h_{11\alpha}^*)\dfrac{1}{2}\int \dfrac{d^4q_2}{(2\pi)^4}\dfrac{Tr\left[P_R(\slashed{q_2}+m_{N_2})(-\slashed{p}+m_{N_1})P_L(\slashed{p}_1+m_{\alpha})\right]}{(q_2^2-m_{N_2}^2+i\epsilon)(q_3^2-m_{\eta_2}^2+i\epsilon)(q_1^2-m_{\phi}^2+i\epsilon)}}\\
&=(ih_{22\alpha}y_{12\phi}\mu_{12\phi}h_{11\alpha}^*)\int \dfrac{d^4q_2}{(2\pi)^4}\dfrac{m_{N_2}p.p_1-m_{N_1}p_1.q_2}{(q_2^2-m_{N_2}^2+i\epsilon)(q_3^2-m_{\eta_2}^2+i\epsilon)(q_1^2-m_{\phi}^2+i\epsilon)}
\end{align}

To calculate the imaginary part of the amplitude, which is related to the discontinuity of the amplitude, we use the cutting rule to calculate this. The proper cutting through the propagators associated with momenta $q_1$ and $q_3$. Thus, we make the replacement,

$$\dfrac{1}{q_1^2-m_{\phi}^2+i\epsilon}\xrightarrow{} -2\pi i \delta(q_1^2-m_{\phi}^2)\Theta(E_{\phi})=-2\pi i \delta((p- q_2)^2-m_{\phi}^2)\Theta(m_{N_1}-E_{N_2})\,,$$
$$\dfrac{1}{q_3^2-m_{\eta_2}^2+i\epsilon}\xrightarrow{} -2\pi i \delta((p_1-q_2)^2-m_{\eta_2}^2)\Theta(E_{\eta_2})\,,$$
where, 
$$2p.p_1=m_{N_1}^2+m_{\alpha}^2-m_{\eta_1}^2\,.$$
$$p_1.q_2=E_{\alpha}E_{N_2}-|{\bf p}_1||{\bf q}_2| \cos\theta\,~{\rm and~}\,|{\bf p}_1|=|{\bf p}_2|=\dfrac{1}{2m_{N_1}}\sqrt{(m_{N_1}^2+m_{\eta_1}^2-m_{\alpha}^2)^2-4m_{N_1}^2m_{\eta_1}^2}\,.$$
and $\theta$ is the angle between ${\bf p}_1$ and ${\bf q}_2$. Putting all of these together we obtain $(\epsilon\to 0)$,

\begin{gather}
\nonumber{\rm Disc(\mathcal{I}_{vertex}^{\prime})}=\dfrac{-i}{4\pi^2}\int dE_{N_2}d^3q_2\dfrac{m_{N_2}p.p_1-m_{N_1}(E_{\alpha}E_{N_2}-|{\bf p}_1||{\bf q}_2| \cos\theta))}{q_2^2-m_{N_2}^2}\\\times\delta[(p-q_2)^2-m_{\phi}^2]\delta[(p_1-q_2)^2-m_{\eta_2}^2]\Theta(m_{N_1}-E_{N_2})\Theta(E_{\alpha}-E_{N_2})
\end{gather}

$$\delta[(p-q_2)^2-m_{\phi}^2]=\delta[E_{N_2}^2-2m_{N_1}E_{N_2}+m_{N_1}^2-|{\bf q}_2|^2-m_{\phi}^2]=-\dfrac{\delta[E_{N_2}-(m_{N_1}-\sqrt{|{\bf q}_2|^2+m_{\phi}^2})]}{2\sqrt{|{\bf q}_2|^2+m_{\phi}^2}}$$

\begin{gather}
\nonumber{\rm Disc(\mathcal{I}_{vertex}^{\prime})}=\dfrac{i}{4\pi^2}\int \dfrac{|{\bf q_2}|^2}{2\sqrt{|{\bf q}_2|^2+m_{\phi}^2}}d|{\bf q_2}|d\Omega\dfrac{m_{N_2}p.p_1-m_{N_1}\big[E_{\alpha}(m_{N_1}-\sqrt{|{\bf q}_2|^2+m_{\phi}^2})-|{\bf p}_1||{\bf q}_2| \cos\theta\big]}{m_{N_1}^2-m_{N_2}^2+m_{\phi}^2-2m_{N_1}\sqrt{|{\bf q_2}|^2+m_{\phi}^2}}
\\\nonumber\times {\color{black}\delta[m_{\alpha}^2+m_{N_1}^2+m_{\phi}^2-m_{\eta_2}^2-2E_{\alpha}(m_{N_1}-\sqrt{|{\bf q}_2|^2+m_{\phi}^2})-2m_{N_1}\sqrt{|{\bf q_2}|^2+m_{\phi}^2}+2|{\bf p}_1||{\bf q}_2| \cos\theta]}
\\\nonumber\times\Theta[(m_{N_1}-\sqrt{|{\bf q}_2|^2+m_{\phi}^2})-E_{\alpha}]
\end{gather}
{\small\begin{align}
{\color{black}\delta\Bigg[\cos\theta-\frac{1}{2|{\bf p}_1||{\bf q}_2|}(-m_{N_{1}^{}}^{2}+m_{\eta_2^{}}^2-m_{\phi}^2-m_{\alpha}^2+\sqrt{|{\bf p_1}|^2+m_{\alpha}^2}(m_{N_{1}^{}}^{}-\sqrt{|{\bf q_2}|^2+m_{\phi}^2})+2m_{N_{1}^{}}^{}\sqrt{|{\bf q_2}|^2+m_{\phi}^2})\Bigg]\frac{1}{2|{\bf p}_1||{\bf q}_2|}}
\end{align}}
\begin{gather}
\nonumber{\rm Disc(\mathcal{I}_{vertex}^{\prime})}=\tiny{\dfrac{i}{8\pi}\int \dfrac{|{\bf q_2}|^2}{\sqrt{|{\bf q}_2|^2+m_{\phi}^2}}d|{\bf q_2}|\dfrac{m_{N_2}m_{N_1}E_{\alpha}-E_{\alpha}m_{N_1}^2+E_{\alpha}m_{N_1}\sqrt{|{\bf q_2}|^2+m_{\phi}^2}+m_{N_1}|{\bf p}_1||{\bf q}_2| \cos\theta}{m_{N_1}^2-m_{N_2}^2+m_{\phi}^2-2m_{N_1}\sqrt{|{\bf q_2}|^2+m_{\phi}^2}}}\\\nonumber~~~~~~~\times\frac{\delta\Bigg[\cos\theta-\frac{(-m_{N_{1}^{}}^{2}+m_{\eta_2^{}}^2-m_{\phi}^2-m_{\alpha}^2+\sqrt{|{\bf p_1}|^2+m_{\alpha}^2}(m_{N_{1}^{}}^{}-\sqrt{|{\bf q_2}|^2+m_{\phi}^2})+2m_{N_{1}^{}}^{}\sqrt{|{\bf q_2}|^2+m_{\phi}^2})}{2|{\bf p}_1||{\bf q}_2|}\Bigg]~d\cos\theta}{|{\bf p}_1||{\bf q}_2|}\\\nonumber\times\Theta[
(m_{N_1}-\sqrt{|{\bf q}_2|^2+m_{\phi}^2})-E_{\alpha}]
\end{gather}
\begin{gather}
\nonumber{\rm Disc(\mathcal{I}_{vertex}^{\prime})}=\dfrac{-i}{8\pi}\int_{0}^{A}\dfrac{m_{N_{1}^{}}^{2}|{\bf q}_0|~d|{\bf q}_0|}{\sqrt{|{\bf q}_0|^2+m_{\phi}^2}\times\sqrt{(m_{N_1}^2-m_{\alpha}^2+m_{\eta_1}^2)^2-4m_{\eta_1}^2m_{N_1}^2}} \\\times\dfrac{\Big(m_{N_{1}^{}}^{}-m_{\eta_{1}^{}}^{}+m_{\alpha}^{}-\dfrac{m_{N_{1}^{}}^{}}{m_{N_{2}^{}}^{}}(m_{N_1^{}}^2+m_{\alpha}^2-m_{\eta_1}^2)+m_{\phi}^{2}-2m_{N_1}\sqrt{|{\bf q_0}|^2+m_{\phi}^2}\Big)}{m_{N_1}^2-m_{N_2}^2+m_{\phi}^2-2m_{N_1}\sqrt{|{\bf q_0}|^2+m_{\phi}^2}},
\end{gather}
where the upper limit of the integration
\begin{gather}
\nonumber A=\sqrt{\Bigg(m_{N_{1}^{}}^{}-\dfrac{(m_{N_1^{}}^2+m_{\alpha}^2-m_{\eta_1}^2)}{2m_{N_{1}^{}}^{}}\Bigg)^{2}-m_{\phi}^{2}}
\end{gather}
comes from the theta function $\Theta
(m_{N_1}-\sqrt{|{\bf q}_2|^2+m_{\phi}^2})-E_{\alpha}).$
\begin{gather}
{\rm Disc(\mathcal{I}_{vertex}^{\prime})}=-\dfrac{m_{N_{1}^{}}^{}\Bigg((\sqrt{5-2C+C^2-4G}-2\sqrt{\sigma})+(D-r+\sqrt{r}~G){\rm ln}\Big[\dfrac{1-r+\sigma-(2-G)}{1-r+\sigma-2\sqrt{\sigma}}\Big]\Bigg)}{2 \sqrt{(2-G)^{2}-4\eta_{1}^{}}},
\end{gather}
where 
\begin{gather}
\nonumber C=(\eta_{1}^{}-l'),~D=(\eta_{2}^{}-l'),~G=(1-\eta_{1}^{}+l'),\\\nonumber~\eta_{1}^{}=\dfrac{m_{\eta_{1}^{}}^{2}}{m_{N_{1}^{}}^{2}},~\eta_{2}^{}=\dfrac{m_{\eta_{2}^{}}^{2}}{m_{N_{1}^{}}^{2}},~\sigma=\dfrac{m_{\phi}^{2}}{m_{N_{1}^{}}^{2}},~r=\dfrac{m_{N_{2}^{}}^{2}}{m_{N_{1}^{}}^{2}},~l'=\dfrac{m_{\alpha}^{2}}{m_{N_{1}^{}}^{2}}
\end{gather}
Now the imaginary part of ${\rm Im( \mathcal{I}_{ vertex}^{\prime})}=\dfrac{1}{2i}{\rm Disc(\mathcal{I}_{vertex}^{\prime})}$

$$\Gamma_{\rm total}=\Gamma+\overline{\Gamma}=2\Gamma_{N_1\to \ell_{\alpha}{\eta}_1}$$
2-body phase space factor $V_{N_{1} \longrightarrow \eta_{1} l}={\color{black}2}\dfrac{|{\bf p_1}|}{8\pi E_{cm}^2}={\color{black}2}\dfrac{|{\bf p_1}|}{8\pi m_{N_1}^2}$

\begin{align}
\varepsilon_{N_1\to \ell_{\alpha}{\eta}_1}=-\dfrac{4}{\Gamma_{\rm tot}}{\rm Im( \mathcal{F}_{ vertex})}{\rm Im( \mathcal{I}_{ vertex}^{\prime})}V_{N_1\to \ell{\eta}_1},
\end{align}
where ${\rm Im( \mathcal{F}_{ vertex})}$ is the vertex factor of the loop integration written as
\begin{equation}\label{eq:epsilonvertex}
{\rm Im( \mathcal{F}_{ vertex})}={\rm Im}(h_{22\alpha}y_{12\phi}\mu_{12\phi}h_{11\alpha}^*).
\end{equation}
Finally the CP asymmetry  from $N_1$ decay can be written as $\varepsilon_{N_1}=\sum\limits_{\alpha=1}^3\varepsilon_{N_1\to \ell_{\alpha}{\eta}_1}$
\begin{gather}  \varepsilon_{N_1}=\sum\limits_{\alpha=1}^3\frac{{\rm Im( \mathcal{F}_{vertex})}}{8\pi M_{N_{1}}^{}}\dfrac{\Bigg((\sqrt{(1-C)^2+4(1-G)}-2\sqrt{\sigma})+(D-r+\sqrt{r}G){\rm ln}\Big[\dfrac{1-r+\sigma-(2-G)}{1-r+\sigma-2\sqrt{\sigma}}\Big]\Bigg)}{ (h_{11\alpha}h^*_{11\alpha})~G~\sqrt{(2-G)^{2}-4\eta_{1}^{}}}.
\end{gather}
With the approximation $l'\rightarrow0$ this expression will look like
\begin{gather}
\label{eq:asymmetry1}
\varepsilon_{N_1^{}}^{}=\tiny{\sum\limits_{\alpha=1}^3\frac{{\rm Im( \mathcal{F}_{vertex})}}{8\pi M_{1}^{}}\frac{\Bigg[(1+\eta_{1}^{})-2\sqrt{\sigma}+\Bigg(\eta_{2}^{}-r+(1-\eta_{1}^{})\sqrt{r}\Bigg)\Bigg({\rm ln}\bigg[\sigma-r-\eta_{1}^{}\bigg]-{\rm ln}\bigg[1-r+\sigma-2\sqrt{\sigma}\bigg]\Bigg)\Bigg]}{(h_{11\alpha}h^*_{11\alpha})(1-\eta_{1}^{})^{2}}}. 
\end{gather}
Further with additional approximations $\eta_{1}^{}\rightarrow0,~\eta_{2}^{}\rightarrow0$ and $\sigma\rightarrow0$, this expression will transform to
\begin{gather}\label{eq:epsilon}
\varepsilon_{N_1^{}}^{}=\sum\limits_{\alpha=1}^3\frac{{\rm Im( \mathcal{F}_{vertex})}}{8\pi(h_{11\alpha}h^*_{11\alpha}) M_{1}^{}}\Bigg(1+r{\rm ln}\Big[1-\dfrac{1}{r}\Big]-\sqrt{r}{\rm ln}\Big[1-\dfrac{1}{r}\Big]\Bigg).  
\end{gather}
We have verified our asymmetry parameter in the limit $ m_{\eta_{1,2}} \to 0$ and $m_{\phi} \to 0$, comparing it with the results from \cite{LeDall:2014too}, except for a negative sign.

\begin{figure}[htb!]
\centering
\includegraphics[width=0.5\linewidth]{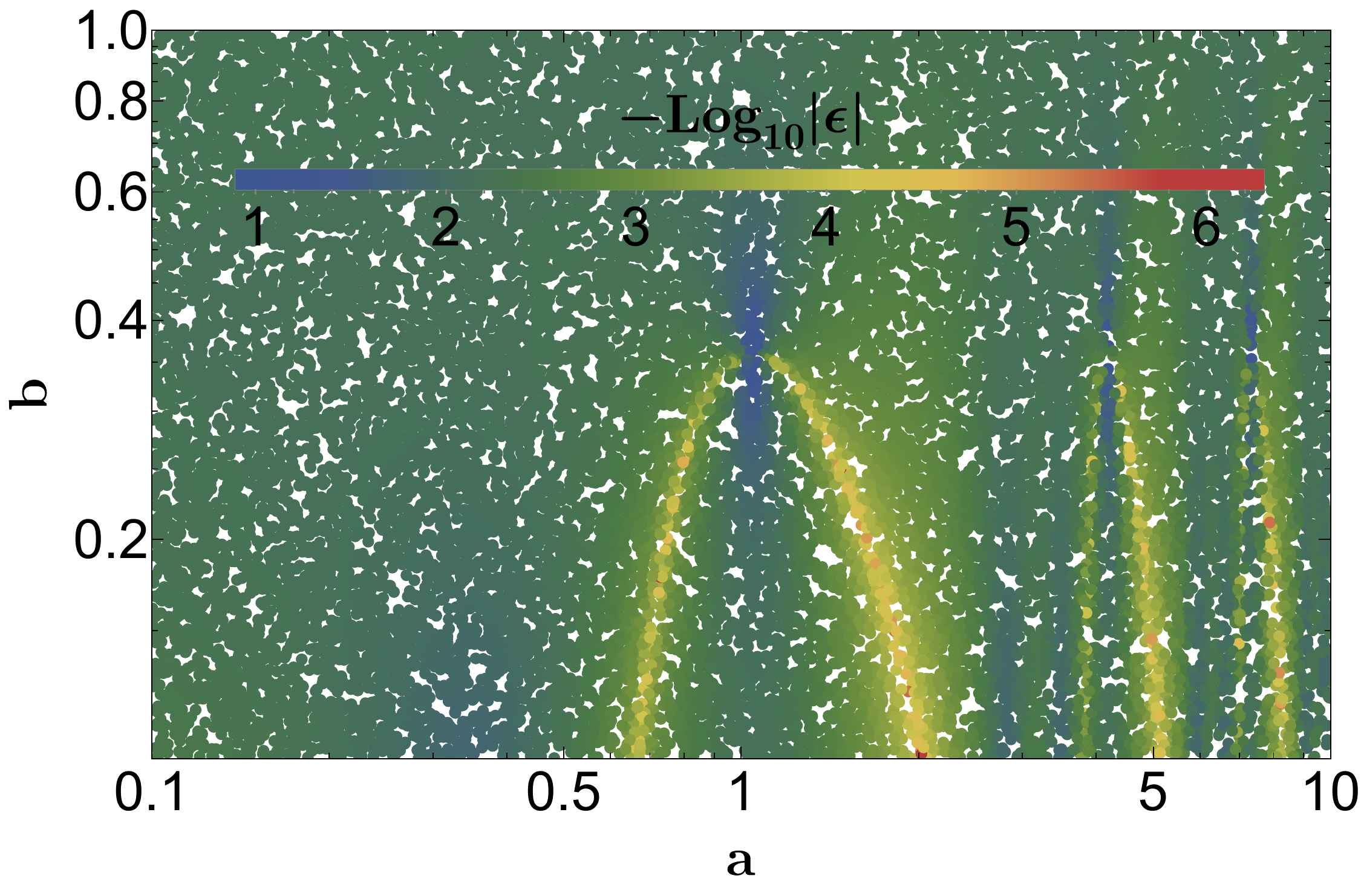}
\caption{This figure illustrates the dependency of the parameters \(a\) and \(b\) on the asymmetry (\(\varepsilon_{N_1}^{}\)) after accounting for the constraints imposed by neutrino masses. We are fixing other parameters as:
${\rm m_{N_1}^{}=2~TeV,}$
${\rm~m_{N_2}^{}=6~TeV,}$
${\rm~m_{\phi}=0.5~TeV,}$
${\rm~m_{\eta^0_{I_1}}^{}=0.4~TeV,}$
${\rm~m_{\eta^0_{I_2}}^{}=m_{\eta^0_{I_1}}^{}+m_{\phi}^{}+1~GeV,}$
${\rm~m_{\eta^0_{R_k}}^{}=m_{\eta^0_{I_k}}^{}+2~GeV},$
${\rm~m_{\eta_k^+}=m_{\eta^0_{I_k}}^{}+3~GeV,}$
${\rm~\mu_{12\phi}^{}=m_{\phi}^{},}$
${\rm~y_{12\phi}^{}=1},~\lambda_{kkH}=0.01$, with $k=1,~2$.}
\label{fig:ab}
\end{figure}
In fig\,.~\ref{fig:ab}, we illustrate the role of the rotation angle ${\rm z} = a + i~b $ in the asymmetry parameter $\varepsilon_{N_1}^{}$, taking into account the active neutrino masses. While \(a\) and \(b\) are unconstrained by theoretical or experimental limits, their influence on $\varepsilon_{N_1}^{}$ remains largely constant, with typical values around $\varepsilon_{N_1}^{} \sim 10^{-2.5}$. However, a notable decrease in $\varepsilon_{N_1}^{}$ is observed within the yellow-shaded region, where it drops to \(\lesssim 10^{-4}\). The maximum value of $\varepsilon_{N_1}^{}$ (\(\sim 10^{-1}\)) occurs for $\{a = 1,~b = 0.4\}$. Beyond this point, increasing \(a\) results in periodic variations in $\varepsilon_{N_1}^{}$.
\bibliographystyle{JHEP}
\bibliography{ref}
\end{document}